\documentclass[letterpaper,11pt]{article}
\pdfoutput=1 

\usepackage{jheppub} 

\usepackage[utf8]{inputenc}
\usepackage[T1]{fontenc}
\usepackage{blindtext}
\usepackage{lmodern}
\usepackage[mathscr]{euscript}
\usepackage{graphicx}
\graphicspath{ {figures/} }
\usepackage{array}
\usepackage{ytableau}
\usepackage{amsmath}
\numberwithin{equation}{section}
\usepackage{amssymb}
\usepackage{hyperref}
\usepackage{xcolor}
\usepackage{color}
\usepackage{dsfont}
\usepackage{setspace}
\usepackage{enumerate}
\usepackage{mathtools}
\usepackage[english]{babel}
\usepackage{fancyhdr}
\usepackage{braket}
\usepackage{stmaryrd}
\usepackage{xparse}

\ExplSyntaxOn
\NewDocumentCommand{\dslash}{O{}}
 {
  \str_case:nn { #1 }
   {
    {}{/\mkern-6mu/}
    {\big}{\big/\mkern-7mu\big/}
    {\Big}{\Big/\mkern-10mu\Big/}
    {\bigg}{\bigg/\mkern-14mu\bigg/}
    {\Bigg}{\Bigg/\mkern-18mu\Bigg/}
   }
 }
\ExplSyntaxOff

\usepackage{tikz}
\usetikzlibrary{calc,decorations.markings}
\usepackage{textcomp}
\usetikzlibrary{shapes.geometric}
 
\usepackage{extarrows} 
\definecolor{navyblue}{rgb}{0.0, 0.0, 0.5}
\definecolor{firebrick}{rgb}{0.7, 0.13, 0.13}
\usetikzlibrary{shapes.misc}

\makeatletter

\@addtoreset{equation}{section}
\makeatother
\let\refOld\ref
\renewcommand{\ref}[1]{(\refOld{#1})}

\DeclareFontFamily{U}{mathx}{\hyphenchar\font45}
\DeclareFontShape{U}{mathx}{m}{n}{
      <5> <6> <7> <8> <9> <10>
      <10.95> <12> <14.4> <17.28> <20.74> <24.88>
      mathx10
      }{}
\DeclareSymbolFont{mathx}{U}{mathx}{m}{n}
\DeclareFontSubstitution{U}{mathx}{m}{n}
\DeclareMathSymbol{\bigplus}        {1}{mathx}{"90}
\DeclareMathSymbol{\bigtimes}       {1}{mathx}{"91}

\DeclareFontFamily{OT1}{pzc}{}
\DeclareFontShape{OT1}{pzc}{m}{it}{<-> s * [1.10] pzcmi7t}{}
\DeclareMathAlphabet{\mathpzc}{OT1}{pzc}{m}{it}

\def\beq{\begin{equation}}
\def\eeq{\end{equation}}

\def\Tr{{\mathrm{Tr}}}

\newcommand{\qe}{\mathfrak{q}}

\newcommand{\BC}{{\mathbb{C}}}

\newcommand{\BP}{{\mathbb{P}}}

\newcommand{\ve}{{\varepsilon}}
\newcommand{\ii}{{\mathrm{i}}}

 \def\p{\partial}
 
 \def\a{\alpha}
 \def\b{\beta}
 \def\g{\gamma}
 \def\d{\delta}

 \def\th{\theta}

 \def\l{\lambda}

 \def\th{\theta}

 \def\o{\omega }

\usepackage{lingmacros}
\usepackage{tree-dvips}

\title{\boldmath
{Riemann-Hilbert correspondence \\ and blown up surface defects}}

\author[a]{Saebyeok Jeong}
\author[b]{and Nikita Nekrasov\footnote{on leave of absence from:
CAS Skoltech and IITP RAS, Moscow, Russia}}

\affiliation[a]{New High Energy Theory Center, Rutgers University, \\ 136 Frelinghuysen Road, Piscataway, New jersey 08854-8019, USA}
\affiliation[b]{Simons Center for Geometry and Physics, Stony Brook University,\\ C.N.~Yang Institute for Theoretical Physics, Stony Brook University\\
Stony Brook, NY 11794-3636, USA}

\emailAdd{saebyeok.jeong@physics.rutgers.edu} \emailAdd{nnekrasov@scgp.stonybrook.edu}

\abstract{The relationship of two dimensional quantum field theory and isomonodromic deformations of Fuchsian systems has a long history. Recently four-dimensional $\EuScript{N}=2$ gauge theories joined the party in a multitude of roles. In this paper we study the vacuum expectation values of  intersecting half-BPS surface defects in $SU(2)$ theory with $N_f=4$ fundamental hypermultiplets. We show they form a horizontal section of a Fuchsian system on a sphere with  $5$ regular singularities, calculate the monodromy, and define the associated isomonodromic tau-function. Using the blowup formula in the presence of half-BPS surface defects, initiated in the companion paper, we obtain the GIL formula, establishing an unexpected relation of the \emph{topological string/free fermion} regime of supersymmetric gauge theory to classical integrability.}

\begin{document} 
\maketitle
\flushbottom
\usetikzlibrary{calc,decorations.markings}

\section{Introduction}
The rich physics of four-dimensional $\EuScript{N}=2$ supersymmetric gauge theories is sometimes encoded in the intricate ways in the geometry of the moduli space of vacua. It is important to decipher this structure both for the understanding of less supersymmetric more realistic gauge theories, and for the unexpected applications. One recurring theme is the relation between the four dimensional physics and the two dimensional physics, such as the BPS/CFT correspondence \cite{NN2004, Losev:2003py}. 

Not only that the string/M-theory embedding \cite{Klemm:1996bj,Witten5} of such gauge theories enriched the physical intuition about these correspondences, the availability of exact localization computation \cite{Nekrasov:2002qd} of their partition functions have been largely utilized to precisely quantify them. One of the interesting discoveries was the duality of quantum and classical regimes. In particular, it was found that the problem of quantizing Hitchin integrable systems translates to the problem of holomorphic symplectic geometry of the moduli space of flat connections and its Lagrangian subvarieties \cite{NRS2011}. The physical understanding of such a duality was provided in \cite{ref:nekwit} by reducing the gauge theory to the sigma models with boundaries associated to certain branes, and a gauge theoretical derivation of the correspondence was provided in \cite{JN2018} at specific examples.

Another version of quantum/classical duality was presented in \cite{GIL2012,LLNZ2013}, where the $\EuScript{N}=2$ gauge theories are connected to the Riemann-Hilbert problem and isomonodromic deformations of Fuchsian systems on Riemann surfaces. To be precise, the correspondences were established in the language of Liouville conformal field theory in \cite{GIL2012,LLNZ2013}. The $\EuScript{N}=2$ gauge theories appear indirectly through their identification of their partition functions and the Liouville correlation functions \cite{agt}. It is actually better to state the correspondences in gauge theory context in some aspects, especially for the purpose of this work, as we shall see hereafter. In these works, some variants of the gauge theory partition functions, 
which purely lie in the quantum regime, are identified either with the
isomonodromic tau function \cite{GIL2012} or with the Hamilton-Jacobi potential for isomonodromic deformations of Fuchsian system \cite{LLNZ2013},
which belong to the quasi-classical regime\footnote{See also the related work \cite{Reshetikhin},\cite{Ribault:2005wp}}. 
In the case of the $\mathfrak{sl}(2)$ Fuchsian system on the Riemann sphere with four regular punctures, the main example throughout this work, the associated $\EuScript{N}=2$ gauge theory is the one with the $SU(2)$ gauge group and four fundamental hypermultiplets. It is known that the isomonodromic deformations of this Fuchsian system is described by Painlev\'{e} VI, the most general second-order non-linear ordinary differential equation with the Painlev\'{e} property.

{}A mystery is that even though the both results connect the 
$\EuScript{N}=2$ gauge theories to the Riemann-Hilbert problem and isomonodromic deformations of Fuchsian systems, the field theory settings in which the correspondence arises are rather different. The computation of the partition function of $\EuScript{N}=2$ gauge theories on the non-compact $\mathbb{C}^2$ involves a regularization implemented by the $\Omega$-background, weakly gauging the maximal torus of the spacetime isometry $U(1)_{\ve_1} \times U(1)_{\ve_2} \subset SO(4)$. In \cite{GIL2012}, the isomonodromic tau function for a Fuchsian system is expressed as an infinite sum of the gauge theory partition functions with shifted Coulomb moduli, subject to the \textit{self-dual limit} $\ve_2 = -\ve_1$ of the $\Omega$-background. The resulting sum is superficially similar to the \emph{dual magnetic partition function} in \cite{Nekrasov:2003rj}\footnote{Although it was pointed out in \cite{Nekrasov:2002qd} that the ${\ve}_{1} = - {\ve}_{2}$ partition function might be a tau-function of some version of KP-Toda hierarchy, since the latter can be expressed through free fermions using Sato Grassmanian.}. Meanwhile, in \cite{LLNZ2013}, the Hamilton-Jacobi potential for isomonodromic deformations of the same Fuchsian system is expressed as the \textit{free energy}, i.e., the asymptotics of the partition function in the \textit{NS limit} $\ve_2 \to 0$ of the $\Omega$-background. Recall that the time-derivative of the isomonodromic tau function \cite{SJM} is, by definition, the Hamiltonian for the isomonodromic flow, while the Hamilton-Jacobi equation equates the Hamiltonian to the time-derivative of the Hamilton-Jacobi potential. In this sense, the two approaches provide two seemingly different expressions for more or less the same mathematical quantity in two different limits of the equivariant parameters. The goal of this work
and the companion paper \cite{NikBlowup} is to reconcile this conflict between the self-dual limit and the NS limit, and to establish an explicit connection between the two approaches to the Riemann-Hilbert problem and the isomonodromic tau function.

The main character of the play is the blowup $\widehat{\mathbb{C}}^2$ which is essentially obtained by replacing the origin $0 \in \mathbb{C}^2$ of the spacetime by an exceptional divisor $\mathbb{P}^1$. The study of $\EuScript{N}=2$ gauge theories on the blowup was initiated in \cite{ny}. The partition function of $\EuScript{N}=2$ gauge theories on the blowup can also be computed by supersymmetric localization. It is an infinite sum of a product of two gauge theory partition functions on the ordinary $\mathbb{C}^2$, with shifted Coulomb moduli and $\Omega$-background parameters. In the limit of the blowing down, where the size of the exceptional divisor shrinks to zero, the spacetime reduces to the ordinary $\mathbb{C}^2$. Meanwhile, the physics should not depend on the size of the exceptional divisor, so that the partition function would not be affected by such a procedure. As a consequence, the gauge theory partition function satisfies a non-trivial relation which we refer to as the \textit{blowup formula}. The blowup formula contains rich analytic information on the gauge theory partition function, and in particular it was used in \cite{ny} to exactly prove that the asymptotics of the partition function in the limit $\ve_1,\ve_2 \to 0$ is identical to the Seiberg-Witten prepotential.

An interesting feature of the blowup formula is, as just mentioned, that it relates the gauge theory partition functions with shifted Coulomb moduli and $\Omega$-background parameters. More precisely, the blowup formula schematically looks like
\begin{align}
\mathcal{Z} (a,\mathbf{m},\ve_1,\ve_2;\qe) = \sum_{n\in \mathbb{Z}} \mathcal{Z}(a+n\ve_1, \mathbf{m}, \ve_1,\ve_2-\ve_1;\qe) \mathcal{Z} (a+n\ve_2, \mathbf{m}, \ve_1-\ve_2,\ve_2;\qe),
\end{align}
where $\mathbf{m}$ denotes the masses of hypermultiplets and $\qe$ denotes the gauge coupling. We immediately notice that the shift in the $\Omega$-background parameters occurs in such a way that it connects the self-dual limit and the NS limit of the $\Omega$-background. Thus, we may naturally expect, at least conceptually, that the blowup formula is the key to resolve our mystery.

The above blowup formula, however, does not directly lead to the solution as it is. The last important ingredients are half-BPS surface (co-dimension two) defects of $\EuScript{N}=2$ gauge theories. The half-BPS surface defects can be engineered in various ways, and the exact supersymmetric partition functions of the $\EuScript{N}=2$ gauge theory in the presence of these surface defects are also available. The surface defect can be viewed as a two-dimensional gauged linear sigma model on a surface coupled to the bulk four-dimensional gauge theory, whose local chiral operators thereby form non-trivial chiral ring relations. In the presence of the $\Omega$-background, such relations uplift to differential equations in the gauge coupling and the complexified FI parameter satisfied by the partition function, which can be regarded as \textit{double quantization} of the chiral ring relations. These differential equations can be exactly derived from the \textit{non-perturbative Dyson-Schwinger equations}, the constraints on the partition functions following from the regularity property of a special class of chiral observables called the $qq$-characters \cite{Nekrasov_BPS1,Nekrasov_BPS45,Jeong2017}. Now, the result of \cite{LLNZ2013} in fact can be re-phrased as stating that the free energy $S$ of the $\EuScript{N}=2$ gauge theories coupled to a surface defect on the $z_2$-plane,
\begin{align}
\Psi (a,\mathbf{m},\ve_1,\ve_2;\qe,z) = \exp \left( \frac{\ve_1}{\ve_2} S(a,\mathbf{m},\ve_1;\qe,z) + \mathcal{O}(1) \right)
\end{align}
is equivalent to the Hamilton-Jacobi potential. This is a consequence of taking the limit $\ve_2 \to 0$ to the mentioned differential equation satisfied by the surface defect partition function.

Equipped with non-local defects, we may wonder how the blowup formulas for their partition functions would work and what their consequences would be. In this paper, we suggest novel blowup formulas for the surface defect partition functions, which are schematically in the form
\begin{align}
\Psi (a,\mathbf{m},\ve_1,\ve_2;\qe,z) = \sum_{n\in \mathbb{Z}} \mathcal{Z}(a+n\ve_1,\mathbf{m}, \ve_1,\ve_2-\ve_1;\qe) \Psi (a+n\ve_2,\mathbf{m}, \ve_1-\ve_2,\ve_2;\qe,z),
\end{align}
where $z$ denotes the complexified FI parameter of the gauged linear sigma model of the defect on the $z_2$-plane. This blowup formula contains rich analytic information on the surface defect partition function, just as the previous one without the defect does for the ordinary partition function. Most importantly, we find that the result of \cite{GIL2012} can be derived from the result of \cite{LLNZ2013} by taking the NS limit $\ve_2 \to 0$ to this blowup formula. In this sense, the above blowup formula is a refinement of the relation in \cite{GIL2012} with the non-zero $\Omega$-background parameter $\ve_2 \neq 0$.

The derivation involves precise matching between the gauge theory parameters and the monodromy data of the associated Fuchsian system. For this, we also need to construct the horizontal section of the Fuchsian system in gauge theoretical language, from which we can explicitly compute the monodromy data in gauge theory parameters. It turns out that we need a further insertion of a half-BPS surface defect on the $z_1$-plane on top of the surface defect on the $z_2$-plane, so that the resulting configuration is \textit{intersecting surface defects} coupled to the bulk gauge theory. We show that the regular part of the partition function gives the horizontal section of the Fuchsian system in the limit $\ve_2 \to 0$. The intersecting surface defect partition functions are expressed as series in the gauge couplings and the complexified FI parameters, which are valid inside their own convergence domains only. To compute the monodromy data, we need the connection formulas between the horizontal sections lying in different convergence domains. By an investigation similar to the one in \cite{JN2018}, where the analytic continuation of the one-point function of a surface observable on the $z_1$-plane was discussed, we achieve such connection formulas by analytically continuing the intersecting surface defect partition functions from one domain to another. By properly concatenating the connection formulas, we finally express the monodromy data in gauge theoretical terms. The result verifies the expectations of \cite{GIL2012}.

Historically, the connection between the two dimensional quantum field theories, their lattice versions such as Ising model, and the Painlev{\'e} equations goes back to the works of \cite{ising1} and more recently, in the ${\mathcal{N}}=2$ $d=2$ supersymmetric context, to \cite{CV1992}. In some ways
our present work cements the link between the ${\mathcal{N}}=2$ $d=4$ physics and the two dimensional theories, by providing a natural habitat of the isomonodromic equations in the realm of correlation functions of four dimensional theories. For the works relating two-dimensional CFTs to the isomonodromic deformation problems, see \cite{ILT,Gav,Tes,LN2017,GavIL,CPT,BMGT}.

The paper is organized as follows. In section \ref{sec:pre}, we begin with the preliminaries of the Riemann-Hilbert correspondence and isomonodromic deformations of Fuchsian system on Riemann surfaces. The main purpose of this section is to give a minimal background for the conjectural relation of \cite{GIL2012} to the readers who are not familiar with it. Section \ref{sec:surfdef} provides constructions of half-BPS surface defects by orbifolding and partial higgsing. We also introduce construction of intersecting surface defects by partial higgsing. The partition functions of (intersecting) surface defects are written explicitly. Section \ref{sec:hjeq} explains how the free energy of the gauge theory coupled to a surface defect is identified with the Hamilton-Jacobi potential for isomonodromic deformations of Fuchsian systems. This is mainly re-phrasing the result of \cite{LLNZ2013} in purely gauge theoretical terms. In section \ref{sec:monodromy}, we show that the NS limit of the intersecting surface defect partition function provides the horizontal section of the Fuchsian system, and compute the monodromy data of the Fuchsian system in gauge theoretical terms. In particular, the analytic continuations of intersecting surface defect partition functions are studied to achieve the connection formulas between different convergence domains. In section \ref{sec:blowup}, we suggest new blowup formulas for surface defect partition functions. We also provide evidences to these formulas, including their consistency with analytic continuation along the flow of the surface defect parameter. Finally, we derive the result of \cite{GIL2012} by taking the NS limit of the blowup formula in section \ref{sec:taufunction}. We carefully take into account the action of the b\"{a}cklund transformations of Painlev\'{e} VI on the monodromy space, thereby recovering the exact expression written in \cite{GIL2012}. We conclude in section \ref{sec:discussion} with discussions. The appendices contain some computational details.

\acknowledgments
The authors are grateful to Misha Bershtein, Seok Kim, Igor Krichever, Barry McCoy, Gregory Moore,  Hiraku Nakajima, Andrei Okounkov, Alexander Zamolodchikov and Xinyu Zhang for discussions. SJ is also grateful to Giulio Bonelli, Alessandro Tanzini, Piljin Yi, Hee-Cheol Kim, and Jaewon Song for discussions and providing support during his visit to School and Workshop on Gauge Theories and Differential Invariants at ICTP, Korea Institute for Advanced Study, and Asian Pacific Center for Theoretical Physics. SJ also thanks Skolkovo Institute for Technology in Moscow for its hospitality during the Summer school in 2019 where part of the work was done. The work of SJ was supported by the US Department of Energy under grant DE-SC0010008. NN thanks V.~Mukhanov, A.~Vershik and M.~Zabzine for discussions and hospitality during his visits to Ludwig-Maximillian University (Munich), Steklov Mathematical Institute in Saint-Petersburg and Uppsala University, while this work was being prepared. 

\section{Preliminaries} \label{sec:pre}
This section is devoted to reviewing the generalities of Riemann-Hilbert correspondence and isomonodromic deformations of Fuchsian systems, in the view of their relations to the supersymmetric gauge theories. In particular, the main purpose of this section is to setup the conventions and to provide a background of the conjecture made in \cite{GIL2012} and its generalizations. Readers with expertise may safely skip to the next section.
\subsection{Painlev{\'e} VI and the GIL conjecture} \label{sec:pvi}
We closely follow the convention used in \cite{GIL2013}, unless specified. Painlev{\'e} equations were discovered as a result of the classification of the second-order first-degree nonlinear ordinary differential equations (ODEs) without movable critical points. The most general one, Painlev\'{e} VI (PVI), is written as follows:
\begin{align} \label{eq:pvi}
\begin{split}
\frac{d^2 w}{d \qe^2} &= \frac{1}{2} \left( \frac{1}{w}+ \frac{1}{w-1} + \frac{1}{w-\qe} \right) \left( \frac{d w}{d \qe} \right)^2 - \left( \frac{1}{\qe} +\frac{1}{\qe-1} +\frac{1}{w-\qe} \right) \frac{d w}{d \qe} \\
& + \frac{2 w (w-1)(w-\qe)}{\qe^2 (\qe-1)^2} \left( \left( \theta_\infty +\frac{1}{2} \right)^2 - \frac{\theta_0 ^2 \qe}{w^2} +\frac{\theta_{1}^2 (\qe-1)}{(w-1)^2} +\frac{\left( \frac{1}{4} -{\theta}_{\qe} ^2 \right) \qe (\qe-1)}{(w-\qe)^2} \right),
\end{split}
\end{align}
where $\theta_i$, $i=0,\qe,1,\infty$ are given parameters. The solution of this equation $w(\qe)$ is called the \textit{Painlev\'{e} transcendent}.

Painlev\'{e} equations admit Hamiltonian formulations, where the equations of motion are
\begin{align} \label{eq:eom}
\frac{dw}{d \qe} = \frac{\partial H}{\partial p_w} , \quad \frac{dp_w}{d\qe} = - \frac{\partial H}{\partial w}.
\end{align}
The relevant Hamiltonian is given by
\begin{align} \label{eq:hamw}
\begin{split}
H (w,p_w;\qe) &= \frac{w(w-\qe)(w-1)}{\qe (\qe-1)} p_w\left( p_w-\frac{2\th_0}{w} -\frac{2\th_\qe-1}{w-\qe}-\frac{2\th_1}{w-1} \right) \\
& +\frac{w(\th_0+\th_\qe+\th_1+\th_\infty)(\th_0+\th_\qe+\th_1-\th_\infty-1)}{\qe(\qe-1)}.
\end{split}
\end{align}
By solving for $p$ and substituting it back, we recover \eqref{eq:pvi} from the equations of motion \eqref{eq:eom}. We recognize $\qe$ plays the role of time in the Hamiltonian formulation, and thus refer to it (and, later, also its analogues) as \textit{time} from now on.

We define the Painlev\'{e} VI \textit{tau function} $\tau(\qe)$ as the generating function for the Hamiltonian. More precisely, it is defined to produce the Painlev\'{e} VI Hamiltonian under the derivative of its log \cite{GIL2012, GIL2013}:
\begin{align} \label{eq:dqder}
\begin{split}
\frac{d}{d\qe} \log \tau(\qe)&= H(w,p_w;\qe)-\frac{w (w-1)}{\qe(\qe-1)} p_w + \frac{w(\th_0+\th_\qe+\th_1+\th_\infty)}{\qe(\qe-1)} \\
& +\frac{(\th_0 + \th_\qe)^2}{\qe} - \frac{\th_\infty ^2 -\th_0 ^2 - \th_\qe ^2 -\th_1 ^2 -2\th_0 \th_\qe - 2 \th_0 \th_1 -2 \th_\qe \th_1}{1-\qe},
\end{split}
\end{align}
where the last two terms on the right hand side, which are only rational functions of time $\qe$, are not important in the sense that they could be absorbed into the Hamiltonian $H(w,p_w ;\qe)$ without affecting the dynamics \eqref{eq:eom}. The tau function $\tau(\qe)$ is always defined up to this ambiguity, and we fix this ambiguity by regarding the above equation as the definition of the tau function. The conjecture made in \cite{GIL2012} states that the Painlev\'{e} VI tau function, around the critical point $\qe=0$, can be expressed as an infinite sum of the partition functions of the four-dimensional $\EuScript{N}=2$ supersymmetric $SU(2)$ gauge theory with four fundamental hypermultiplets subject to the self-dual $\Omega$-background, with shifted arguments\footnote{The tau function $\tau(\qe)$ written in \eqref{eq:conj} differs from the tau function $\tau^{\text{GIL}} (\qe)$ written in \cite{GIL2012, GIL2013} by a simple function of $\qe$, namely \begin{align} \tau^{\text{GIL}} (\qe) = \qe^{-\th_0  ^2 -\th_\qe ^2} (1-\qe) ^{2\th_\qe \th_1} \tau (\qe). \nonumber \end{align} This difference is also reflected in the relation \eqref{eq:dqder}. The difference is of course non-essential and merely conventional. We find $\tau(\qe)$ more natural to consider in the gauge theory context, so we stick with our definition.}:
\begin{align} \label{eq:conj}
\begin{split}
\tau(\qe) &= \sum_{n \in \mathbb{Z}}  e^{n \beta} \mathcal{Z} (a + n \ve_1, \mathbf{m} ; \ve_1, -\ve_1;\qe) \\
&=\sum_{n \in \mathbb{Z}}  e^{n \beta} \mathcal{Z} (\alpha+n, \boldsymbol\th ; \qe).
\end{split}
\end{align}
In this equation, $\qe$ appears as the gauge coupling and is identified with the time in PVI \eqref{eq:pvi}. $a$, $\mathbf{m}=(m_i)_{i=1} ^4$, and $(\ve_1 , \ve_2=-\ve_1)$ are the equivariant parameters for the actions of the maximal tori of the global $SU(2)$ gauge symmetry, the $SO(8)$ flavor symmetry, and the $SO(4)$ Lorentz symmetry, respectively. They are also called the Coulomb modulus, the masses for the hypermultiplets, and the $\Omega$-background parameters, respectively (see appendix \ref{appA} for a brief review). As just mentioned, we have set the self-dual limit $\ve_2 = -\ve_1$ of the $\Omega$-background. Then the remaining $\Omega$-background parameter $\ve_1$ only plays the role of the mass scale, and we absorbed it in the second line into the definition of dimensionless parameters defined by
\begin{align} \label{eq:dimlessparm}
\alpha := \frac{a}{\ve_1}, \quad \theta_0:= \frac{m_3 -m_4}{2 \ve_1}, \quad \theta_\qe:= \frac{m_3 +m_4}{2 \ve_1}, \quad \theta_1:= \frac{m_1 +m_2}{2 \ve_1}, \quad  \theta_\infty:= \frac{m_1-m_2}{2 \ve_1},
\end{align}
relating the hypermultiplet masses with the $\theta$-parameters appearing in PVI \eqref{eq:pvi}. The parameters $\alpha$ and $\beta$ in \eqref{eq:conj} can be thought of as the integration constants for the equations of motion \eqref{eq:eom}. As we see here, $\a$ is identified with the Coulomb modulus of the gauge theory, while the gauge theoretical meaning of $\b$ is unclear yet and will be clarified later in section \ref{sec:monodromy}. The conjecture \eqref{eq:conj} was proven in \cite{ILT2014} by using the crossing symmetry of Liouville correlation functions. In this paper, we provide a gauge theoretical derivation of \eqref{eq:conj} by using half-BPS surface defects on the blowup.

\subsection{Riemann-Hilbert correspondence}
We begin by introducing Fuchsian systems on a Riemann surface. An $\mathfrak{sl} (2)$ Fuchsian system on the punctured sphere $\mathbb{P}^1 _{r+3}:=\mathbb{P}^1 \setminus \{ z_{-1} = \infty, z_0, \cdots, z_r, z_{r+1} \}$
 is defined by a matrix-valued linear differential equation,
\begin{align} \label{eq:fuch}
\frac{d \Phi}{d y} = A(y) \Phi := \sum_{i=0} ^{r+1} \frac{A_i (z)}{y-z_i} \Phi, \quad\quad\quad y \in \mathbb{P}^1 \setminus \{ z_{-1} = \infty, z_0, \cdots, z_r, z_{r+1} \}
\end{align}
where $z_i$'s are the positions of the $r+3$ punctures which are assumed to be distinct and
\begin{align}
(A_0, A_1, \cdots, A_{r+1}) \in \mathfrak{g} := \bigoplus_{i=0} ^{r+1} \mathfrak{sl}(2)
\end{align}
are matrix-valued functions. We also define
\begin{align} \label{eq:ainf}
A_\infty := -\sum_{i=0} ^{r+1} A_i \in \mathfrak{sl}(2).
\end{align}
We assume that the matrices $A_\infty$, $A_0, \cdots, A_{r+1}$ are diagonalizable and their eigenvalues are all distinct. 

In \eqref{eq:fuch}, $\Phi$ takes the value in $\mathbb{C}^2$. By placing two independent column solutions into a fundamental matrix, we can regard $\Phi$ as a $2 \times 2$ matrix. Now for each element in the fundamental group $\gamma \in \pi_1 \left( \mathbb{P}^1 _{r+3}\right)$, analytic continuation of the solution $\Phi$ along the loop $\gamma$ produces a new solution $\Phi'$, which is related to the original solution $\Phi$ by a monodromy $M_\gamma$:
\begin{align}
\Phi' = \Phi M_\gamma.
\end{align}
Hence the monodromies of the solution provides a representation of the fundamental group into $SL(2)$,
\begin{align}
\begin{split}
M: \pi_1 \left(\mathbb{P}^1_{r+3} \right) &\longrightarrow SL(2). \\
\gamma& \longmapsto M_\gamma
\end{split}
\end{align}
We did not fix the baespoint of the fundamental group, so that the monodromy $M$ is always defined up to an overall conjugation by $SL(2)$. The monodromy space $\EuScript{M}_{\mathbf{z}}$, where the monodromy data $M$ takes value, is thus given by 
\begin{align}
M \in \EuScript{M}_{\mathbf{z}} : = \text{Hom} \left(\pi_1 \left(\mathbb{P}^1_{r+3} \right) , SL(2) \right) /SL(2).
\end{align}
Hence a Fuchsian system is associated to a representation of the fundamental group by its monodromy data. We define the Riemann-Hilbert map by this monodromy representation,
\begin{align}
\text{RH}:\mathfrak{g} \longrightarrow \EuScript{M}_{\mathbf{z}}.
\end{align}

So far, we have seen that an $\mathfrak{sl}(2)$ Fuchsian system defines a representation of the fundamental group $\pi_1 (\mathbb{P}^1 _{r+3})$ into $SL(2)$ by the monodromies of the solution $\Phi$. The Riemann-Hilbert problem is the question about the converse: For a given monodromy data, can we reconstruct a Fuchsian system which exhibits this monodromy? The solution to this problem is not unique, and there are many solutions corresponding to the given monodromy data. In particular, the monodromy data do not depend on the positions $\mathbf{z}$ of the poles, while the Fuchsian system \eqref{eq:fuch} has an explicit dependence on them. Hence we are led to study the deformations of the Fuchsian system with respect to the positions of the poles, which preserve the monodromies so that the deformations lead to a family of solutions to the Riemann-Hilbert problem for the same monodromy data. Such deformations are called \textit{isomonodromic} deformations.

\subsection{Isomonodromic deformations of Fuchsian systems}
We study the deformations of the matrices $(A_i)_{i=0} ^{r+1}$ with respect to $(z_i)_{i=0} ^{r+1}$ which preserve the monodromies of the solution $\Phi$.  The deformations are isomonodromic when they can be compensated by a gauge transformation of the connection $\partial_y -A$, namely,
\begin{align}
-\frac{\partial A}{\partial z_j} = \left[ \partial_y-A, \epsilon_j \right].
\end{align}
Solving for the gauge variation parameter $\epsilon_j$ by equating the terms of order $(y-z_j)^{-2}$, we get $\epsilon_j = \frac{A_j}{y-z_j}$. Plugging it back and taking the residues in $y \to z_i$, we obtain
\begin{align} \label{eq:schlesinger}
\begin{split}
&\frac{\partial A_i}{\partial z_j} = \frac{ [A_i , A_j]} {z_i -z_j}, \quad\quad\quad i \neq j \\
&\frac{\partial A_i}{\partial z_i} = - \sum_{j \neq i} \frac{[A_i, A_j] }{z_i -z_j},
\end{split}
\end{align}
for $i,j=0, \cdots, r+1$. These equations are called the \textit{Schlesinger equations} for isomonodromic deformations of the Fuchsian system. Namely, the monodromies of the Fuchsian system \eqref{eq:fuch} do not depend on $(z_i)_{i=0} ^{r+1}$ if the matrices $(A_i (z))_{i=0} ^{r+1}$ satisfy \eqref{eq:schlesinger}. The converse, that any isomonodromic deformation is described by the Schlesinger equations, is generally not true unless we impose some reasonable assumptions. In this paper, we will be interested only in isomonodromic deformations generated by the Schlesinger equations.

The Schlesinger equations \eqref{eq:schlesinger} admit a canonical Hamiltonian formulation, as we now discuss. We consider the standard Lie-Poisson bracket on $\mathfrak{g}^* \sim \mathfrak{g}$ which is written as\footnote{Here $\mathfrak{g}$ is identified with its dual $\mathfrak{g}^*$ by the Killing form, $A \mapsto \Tr(A \; \cdot \;)$.}
\begin{align}
\Big\{ \left( A_i \right) ^a _b , \left( A_j \right) ^c _d \Big\} = \delta_{ij} \Big( \delta^a _d \left( A_i \right)^c _b - \delta^c _b \left( A_j \right) ^a _d \Big).
\end{align}
Then the time-dependent Hamiltonians defined by
\begin{align} \label{eq:hamiltonian}
H_i = \sum_{j \neq i} \frac{\text{Tr}\, A_i A_j }{z_i -z_j}, \quad\quad i=0, \cdots, r+1,
\end{align}
give the following equations of motion
\begin{align}
\frac{\partial A_i}{\partial z_j} = \{ A_i, H_j \},\quad\quad i,j=0,\cdots,r+1,
\end{align}
which precisely reproduce the Schlesinger equations \eqref{eq:schlesinger}. It is also straightforward to check that these Hamiltonians are mutually Poisson-commuting
\begin{align}
\{ H_i, H_j \} =0, \quad \quad i,j=0, \cdots, r+1,
\end{align}
and also satisfy the condition
\begin{align}
\frac{\partial H_i}{\partial z_j} = \frac{\partial H_j}{\partial z_i}, \quad\quad i,j=0, \cdots, r+1.
\end{align}
Due to these conditions, we can define the generating function $\tau (z)$ of the Hamiltonians, called the \textit{isomonodromic tau function}. Namely, 
\begin{align}
H_i = \frac{d}{d z_i} \log \tau (z), \quad\quad i=0, \cdots, r+1.
\end{align}
It is immediate that
\begin{align}
d \log \tau (z) = \sum_{i<j} \Tr A_i A_j \, d \log (z_i -z_j),
\end{align}
where the right hand side is a closed 1-form due to the Schlesinger equations \eqref{eq:schlesinger}.

The Hamiltonian system defined on $\mathfrak{g}$ can be reduced by a symplectic reduction. First we can restrict the system to a symplectic leaf
\begin{align}
\bigtimes_{i=1} ^{r+1} \mathcal{O}_i =\mathcal{O}_0 \times \cdots \times \mathcal{O}_{r+1} \subset \mathfrak{g},
\end{align}
obtained by choosing a conjugacy class (i.e., an adjoint orbit) $\mathcal{O}_i$ of $A_i$ by choosing $\det A_i = -\th_i ^2$ for each $i=0, \cdots, r+1$. Then we perform the symplectic quotient to get the reduced symplectic manifold
\begin{align}
\mathcal{E}_{\mathbf{z}} (\boldsymbol\th) :=\left( \bigtimes_{i=1} ^{r+1} \mathcal{O}_i  \right) \; \dslash[\bigg] \;SL(2),
\end{align}
where the double slash denotes the symplectic quotient. 
Note that $A_\infty$ is an integral of motion:
\begin{align}
\frac{\partial A_\infty}{\partial z_i} = \{ A_\infty, H_i \} =0 , \quad\quad i=0, \cdots, r+1,
\end{align}
which generates the action of $SL(2)$ on $\bigtimes_{i=1} ^{r+1} \mathcal{O}_i $ by simultanuous conjugations. Imposing the moment map equation for this action is equivalent to setting $A_\infty$ a fixed diagonal matrix. We can further take the quotient with respect to the residual symmetry of the conjugation of diagonal matrices, thereby performing a symplectic quotient. A simple dimension count shows that the reduced symplectic manifold is $2r$-dimensional, $\dim  \mathcal{E}_{\mathbf{z}} (\boldsymbol\th) =2r$. On the reduced symplectic leaves, the Hamiltonians \eqref{eq:hamiltonian} are redundant since the combinations
\begin{align}
\sum_{i=0} ^{r+1} H_i =0,\quad \sum_{i=0} ^{r+1} z_i H_i = \sum_{i<j} \Tr A_i A_j,
\end{align}
generate trivial dynamics. In particular,
\begin{align}
\sum_{i=0} ^{r+1} z_i \frac{\partial A_j}{\p z_i} = [A_j , A_\infty].
\end{align}
Thus the solutions to the Schlesinger equations are invariant under the reparametrizations of the times $z_0, \cdots, z_{r+1}$ by
\begin{align}
z_i \mapsto a z_i +b, \quad \quad i=0, \cdots, r+1, \quad a\neq 0.
\end{align}
Hence, without loss of generality, we can set $z_0 = 1$ and $z_{r+1} = 0$, only considering $r$ times $\left( z_i \right)_{i=1} ^{r}$ and corresponding $r$ Hamiltonians $(H_i)_{i=1} ^r$. Note that when the moduli space of Fuchsian system is restricted to $\mathcal{E}_{\mathbf{z}} (\boldsymbol\th) $, the monodromy space is also reduced to
\begin{align}
\EuScript{M}_{\mathbf{z}} (\boldsymbol\th) := \left\{\left. M \in \text{Hom} \left(\pi_1 \left(\mathbb{P}^1_{r+3} \right) , SL(2) \right)\; \right\vert\; \text{Tr}\, M_{\g_i} = 2 \cos 2\pi \th_i, \; i=-1,\cdots, r+1 \right\} \Big/ SL(2),
\end{align}
where $\g_i$ is a small loop around each puncture $z_i$, $i=-1, \cdots, r+1$. A simple dimension count shows that $\dim \EuScript{M}_{\mathbf{z}} (\boldsymbol\th)  = 2r$. The Riemann-Hilbert map is reduced to a symplectomorphism on the reduced moduli space,
\begin{align}
\text{RH}: \mathcal{E}_{\mathbf{z}} (\boldsymbol\th) \longrightarrow \EuScript{M}_{\mathbf{z}} (\boldsymbol\th).
\end{align} 

\subsection{Painlev\'{e} VI from isomonodromic deformation}
We end the section by explaining how Painlev\'{e} VI emerges in the simplest case $r=1$ of the isomonodromic deformations discussed so far. We choose the adjoint orbits $ \mathcal{O}_0 \times \mathcal{O}_\qe \times \mathcal{O}_1$ by imposing
\begin{align}
\text{det}\, A_i ^2 = -\th_i ^2 , \quad\quad i=0,\qe,1,
\end{align}
and restrict to a level set of the moment map
\begin{align} \label{eq:const}
A_\infty := \begin{pmatrix} -\th_\infty & 0 \\ 0 & \th_\infty \end{pmatrix} = -(A_0 + A_\qe + A_1).
\end{align}
We can parametrize the reduced symplectic leaf by
\begin{align} \label{eq:matrix}
A_i = \begin{pmatrix} u_i +\th_i & -u_i w_i \\ \frac{u_i +2\th_i}{w_i} & -u_i -\th_i \end{pmatrix}, \quad\quad i=0,\qe,1,
\end{align}
with the constraint \eqref{eq:const}. In particular, due to the constraint $\sum_{i=0,\qe,1} u_i w_i =0$, the component $A_{12}$ of the connection can be written as
\begin{align}
\sum_{i=0,\qe,1} -\frac{u_i w_i}{y-z_i} = \frac{k (y-w)}{y(y-\qe)(y-1)}.
\end{align}
It can be shown that the conjugate momentum for the variable $w$ is
\begin{align}
p_w:=  \sum_{i=0,\qe,1} \frac{u_i + 2\th_i}{w-z_i}.
\end{align}
We get rid of $k$ as a result of modding out the overall conjugation of $SL(2)$. It turns out that the remaining variables $(w,p_w)$ form a Darboux coordinate system on the two-dimensional reduced moduli space $\mathcal{E}_\qe (\boldsymbol\th)$. 

Painlev\'{e} VI arises precisely when we describe the isomonodromic flow of the Fuchsian system in the Darboux coordinates $(w,p_w)$. First we express the matrices $A_i$, $i=0,\qe,1$ in terms of $w$, $p_w$, and $k$. Then a straightforward computation shows that the Schlesinger equations
\begin{align}
\frac{dA_0}{d\qe} =- \frac{[A_0,A_\qe]}{\qe} , \quad\quad \frac{dA_1}{d\qe} = \frac{[A_1,A_\qe]}{1-\qe},
\end{align}
imply Painlev\'{e} VI \eqref{eq:pvi} satisfied by $w(\qe)$. Namely, Painlev\'{e} VI describes the isomonodromic flow of the $\mathfrak{sl}(2)$ Fuchsian system defined on the four-punctured sphere.

Let us consider the horizontal section $\Phi = \begin{pmatrix} \phi_1 \\ \phi_2 \end{pmatrix}$ of the Fuchsian system, $\left( \p_y - A(y)\right) \Phi(y)  =0$. We can convert this first-order differential equation into a second-order differential equation for $\phi_1$:
\begin{align}
0 = \left( \p_y ^2 - \left( \Tr A +\frac{\p_y A_{12}}{A_{12}} \right) \p_y + \det{A} -\p_y A_{11} +\frac{A_{11}}{A_{12}} \p_y A_{12} \right) \phi_1 (y).
\end{align}
By substituting \eqref{eq:matrix} into this equation, we get
\begin{align} \label{eq:horisecond}
\begin{split}
0&= \left[ \p_y ^2 + \left( \frac{1}{y} +\frac{1}{y-\qe}+\frac{1}{y-1} -\frac{1}{y-w}   \right) \p_y \right.\\
& \quad -\frac{\th_0^2}{y^2} -\frac{\th_\qe ^2}{(y-\qe)^2} -\frac{\th_1 ^2}{(y-1)^2} - \frac{\left( \th_\infty +\frac{1}{2} \right) ^2 -\th_0 ^2 -\th_\qe ^2 -\th_1 ^2 -\frac{1}{4}}{y(y-1)} \\
& \quad+\frac{w(w-1)}{y(y-w)(y-1)}\left( p_w  -\frac{\th_0}{w} -\frac{\th_\qe}{w-\qe} -\frac{\th_1}{w-1} \right) \\
&  \quad  - \frac{\qe(\qe-1)}{y(y-\qe)(y-1)}\left( H(w,p_w;\qe)  -\frac{\th_\qe}{z-\qe} \right. \\ 
&\left.\left. \quad\quad \quad\quad\quad\quad\quad+\frac{\th_0 +\th_\qe -2\th_0\th_\qe  -\qe\left( \th_0 ^2+ \th_\qe ^2 +\th_1 ^2 -\th_\infty ^2 -\th_\infty +\th_\qe +2\th_0 \th_1 \right)}{\qe (\qe-1)} \right) \right] \phi_1 (y),
\end{split}
\end{align}
where the Hamiltonian $H(w,p_w;\qe)$ is nothing but that of the Painlev\'{e} VI:
\begin{align} \label{eq:hamred}
\begin{split}
H(w,p_w;\qe) &= \frac{w(w-\qe)(w-1)}{\qe(\qe-1)} p_w \left( p_w - \frac{2\th_0}{w} -\frac{2\th_\qe-1}{w-\qe} - \frac{2\th_1}{w-1} \right) \\
& +\frac{w(\th_0+\th_\qe+\th_1+\th_\infty)(\th_0+\th_\qe+\th_1-\th_\infty-1)}{\qe(\qe-1)}.
\end{split}
\end{align}
Generically there are two independent solutions to the Fuchsian differential equation \eqref{eq:horisecond}, and we denote them as $\left(\phi_1 ^{(1)} (y), \phi_1 ^{(2)} (y) \right)$. As we have seen earlier, the solutions to the Schlesinger equation are given by the isomonodromic flow $(w,p_w)=(w(\qe),p_w (\qe))$. In other words, when the Fuchsian differential equation \eqref{eq:horisecond} is restricted to the isomonodromic flow $(w,p_w)=(w(\qe),p_w (\qe))$, the monodromies of the solution $\left(\phi_1 ^{(1)} (y), \phi_1 ^{(2)} (y) \right)$ define the monodromy data $M$ which is independent of $\qe$. This is precisely the image of $(w(\qe),p_w (\qe))$ in $\EuScript{M}_\qe (\boldsymbol\th)$ under the Riemann-Hilbert map.

To precisely describe the Riemann-Hilbert map, we need a coordinate system on the reduced monodromy space $\EuScript{M}_\qe (\boldsymbol\th)$. It is convenient to use Darboux coordinate systems to make the symplectomorphicity of the Riemann-Hilbert map manifest. The reduced monodromy space $\EuScript{M}_\qe (\boldsymbol\th)$ can be equipped with various kinds of Darboux coordinate systems, but the one which is relevant to its connection to the isomonodromic tau functions and supersymmetric gauge theories turns out to be the NRS coordinate system \cite{NRS2011}. The NRS coordinate system was introduced in \cite{NRS2011} for reduced moduli spaces of flat $SL(2)$-connections on generic Riemann surfaces. For our main example of the four-punctured sphere, it is a coordinate sysetm $(\a,\b)$ which simply parametrizes the monodromies $M_{A,B}$ along the two independent (in ${\pi}_{1}$) loops, which we denote as the $A$- and $B$-loops (see figure \ref{sphere}), by
\begin{align} \label{eq:nrsa}
&\Tr \, M_A  = -2 \cos 2\pi \a
\end{align}
and
\begin{align} \label{eq:nrsb}
\begin{split}
\text{Tr}\, M_B &= \frac{\left( \cos 2\pi \th_\infty +\cos 2\pi \th_1 \right) \left( \cos 2\pi \th_0 + \cos 2\pi \th_\qe \right)}{2 \sin ^2 \pi \a} \\
& + \frac{\left( \cos 2\pi \th_\infty -\cos 2\pi \th_1 \right) \left( \cos 2\pi \th_0 - \cos 2\pi \th_\qe \right)}{2 \cos ^2 \pi \a} \\
& - \sum_{\pm} 4 \frac{\prod_{\epsilon=\pm} \cos \pi (\mp \a - \th_\qe +\epsilon \th_0) \cos \pi (\mp \a -\th_1 +\epsilon \th_\infty) }{\sin ^2 2\pi \a}  e^{\pm\b} .
\end{split}
\end{align}
It can be shown that the coordinates $\a$ and $\b$ defined in this way form a Darboux coordinate system on $\EuScript{M}_\qe (\boldsymbol\th)$. The symplectomorphism of Riemann-Hilbert can be described in terms of Darboux coordinates
\begin{align}
\begin{split}
\text{RH}: \mathcal{E}_\qe (\boldsymbol\th) &\longrightarrow \EuScript{M}_\qe (\boldsymbol\th) \\
(w,p_w)& \longmapsto (\a,\b).
\end{split}
\end{align}
This implies the image $(\a,\b)$ is constant along the isomonodromic flow $(w(\qe),p_w (\qe))$. In this sense, $\a$ and $\b$ can be considered as the initial condition for Painlev\'{e} VI which is a second-order ODE.

\section{$\EuScript{N}=2$ supersymmetric gauge theories with surface defects} \label{sec:surfdef}
As we will see in later sections, half-BPS surface (codimension-two) defects play crucial roles in the correponsdence of four-dimensional $\EuScript{N}=2$ supersymmetric gauge theories and the isomonodromic deformations of Fuchsian systems. In particular, half-BPS surface defects can be used in the correspondence with the isomonodromic deformations to realize apparent singularities and horizontal sections of the associated Fuchsian system. In the M-theory perspective, we consider four-dimensional theories of class $\mathcal{S}$ realized as the worldvolume theory on M5-branes wrapping a Riemann surface \cite{gai1}. The relevant half-BPS surface defects are engineered by inserting M2-branes wrapping two-dimensional surfaces inside the four-dimensional spacetime. In the field theory limit, the bulk four-dimensional theory of class $\mathcal{S}$ gets coupled to the gauged linear sigma model living on a two-dimensional surface, which thereby realizes a surface defect. The position of the M2-brane insertion on the Riemann surface translates into the complexified FI parameter of this sigma model, and it provides an apparent singularity of the associated Fuchsian system on the Riemann surface.

In the presence of the $\Omega$-background, half-BPS surface defects may lie only on the $z_1$-plane or on the $z_2$-plane. These two choices are not equivalent, so that in particular they contribute to the partition function differently in the NS limit where one of the two $\Omega$-background parameters is taken to be zero. We choose our convention that the NS limit is always $\ve_2 \to 0$. Then we will see in later sections that the surface defects we need for the connection to the Riemann-Hilbert correspondence are the ones on the $z_2$-plane.\footnote{For the discussion on the surface defects on the $z_1$-plane and on the NS limit of their partition functions, see \cite{JN2018}.} To construct the horizontal section of the associated Fuchsian system, we need a further insertion of a surface defect on the $z_1$-plane, so that the resulting configuration is the intersecting surface defects coupled to the bulk theory. In this section, we discuss the constructions of these (intersecting) half-BPS surface defects and expressions of their partition functions.

Half-BPS surface defects can be constructed in various ways. Here, we introduce two constructions, orbifolding and partial higgsing, relevant to our discussion. We also introduce a construction of intersecting surface defects by partial higgsing. In particular, the main objects to be considered are the partition functions of the gauge theory in the presence of those (intersecting) surface defects. We compute them explicitly and discuss their properties. For preliminary discussions on the $\EuScript{N}=2$ partition functions and conventions used in this section, see appendix \ref{appA}. 

\subsection{Construction of surface defects} \label{subsec:surfacedef}

\subsubsection{Orbifold} \label{sec:orbdef}
We consider the $\EuScript{N}=2$ supersymmetric gauge theories on an orbifold defind by the following $\mathbb{Z}_p$-action on the flat spacetime $\mathbb{C}^2$,
\begin{align}
\zeta:(z_1, z_2)\longmapsto (\zeta z_1,  z_2), \quad\quad \zeta \in \mathbb{Z}_p.
\end{align}
Note that there is an orbifold singularity along the $z_2$-plane $\{ z_1 =0 \}$. This orbifold can be mapped to the ordinary $\mathbb{C}^2$ by
\begin{align}
(z_1, z_2) \longmapsto (\tilde{z}_1 = z_1 ^p, \tilde{z}_2 = z_2 ).
\end{align}
Then the $\EuScript{N}=2$ gauge theory on $\mathbb{C}^2 _{\tilde{z}_1, \tilde{z}_2}$ develops a surface defect on the $\tilde{z}_2$-plane by a singular boundary condition of the gauge field (see \cite{Nekrasov_BPS45, NikBlowup} for more details).

At the level of the partition functions, the $\mathbb{Z}_p$-action fractionalizes the contributions to the relevant equivariant integrations according to its irreducible representations. The partition function of the $\EuScript{N}=2$ gauge theory in the presence of the orbifold surface defect is, therefore, computed by keeping the $\mathbb{Z}_p$-invariant parts only. More precisely, the singular boundary condition breaks the global $U(N)$ gauge symmetry in general, and we have to specify the subgroup left to be preserved to fully characterize the surface defect. This is equivalent to the choice of the \textit{coloring function},
\begin{align}
c: N \longrightarrow \mathbb{Z}_p,
\end{align}
for which the preserved subgroup is assigned as $\bigtimes_{\o \in \mathbb{Z}_p} U( \vert c^{-1}(\o) \vert ) \subset U(N)$. We can turn on the magnetic fluxes on the support of the surface defect for each $U(1) \subset U(\vert c^{-1} (\o) \vert)$. The singularity in the gauge field and the magnetic flux combine into the fractionalized couplings $\qe_\o$, satisfying
\begin{align}
\qe = \prod_{\o \in \mathbb{Z}_p} \qe_\o,
\end{align}
where $\qe$ is the bulk instanton counting parameter. We can parametrize these couplings by
\begin{align}
\begin{split}
&\qe_0 = \qe \frac{z_1}{z_0} , \quad \qe_{p-1} = \frac{z_0}{z_{p-1}} \\
&\qe_\o = \frac{z_{\o+1}}{z_\o}, \quad \o =1, \cdots, p-2.
\end{split}
\end{align}

The partition function of the gauge theory in the presence of the orbifold surface defect is then computed by doing the path integral over the  $\mathbb{Z}_p$-invariant locus of fields. The path integral localizes to a finite-dimensional integral over the instanton moduli space, which admits the ADHM description. The ADHM construction of the moduli space $\EuScript{M}_{N,k}$ of $U(N)$-instantons with the instanton number $k$ involve the linear maps $(B_1,B_2,I,J)$, where $B_{1,2} \in \text{End} (K)$, $I \in \text{Hom}(N,K)$, and $J \in \text{Hom}(K,N)$, with the vector spaces $N = \mathbb{C}^N$ and $K = \mathbb{C}^k$. The instanton moduli space $\EuScript{M}_{N,k}$ is obtained by imposing the ADHM equation and the stability condition:
\begin{align}
\begin{split}
&[B_1, B_2] +IJ=0 \\
&K = \mathbb{C} [B_1, B_2] I(N)
\end{split}
\end{align}
modulo the $GL(K)$-action
\begin{align}
\left(B_1,B_2,I,J \right) \longrightarrow \left( g^{-1} B_1 g, g^{-1} B_2 g , g^{-1} I, J g \right), \quad \quad g \in GL(K).
\end{align}

Now upon the $\mathbb{Z}_p$-orbifolding, the space $N=\mathbb{C}^N$ and $K=\mathbb{C}^k$ are decomposed according to the $\mathbb{Z}_p$-representations as
\begin{align}
N = \bigoplus_{\o \in \mathbb{Z}_p} N_\o \otimes \EuScript{R}_\o , \quad\quad N_\o := \sum_{\a \in c^{-1} (\o)} e^{\b a_\a},
\end{align}
and
\begin{align}
K = \bigoplus_{\o \in \mathbb{Z}_p} K_\o \otimes \EuScript{R}_\o, \quad \quad K_\o := \sum_{\a =1}^N e^{\b a_\a} \sum_{j=1}^{ \l_1 ^{(\a)} } q_2 ^{j-1} \sum_{\substack{1 \leq i \leq \l_j ^{(\a)t}\\ c(\a) +i-1 \equiv \o \; \text{mod}\;p }} q_1 ^{i-1},
\end{align}
where $\EuScript{R}_\o$, $\o = 0, \cdots, p-1$ are one-dimensional irreducible representations of $\mathbb{Z}_p$ in which the generator $\zeta = e^{\frac{2\pi i}{p}}$ acts by $\zeta^\o$. Let $\Omega_N (\zeta)$ and $\Omega_K (\zeta)$ be the representation of the action of $\zeta$ on $N$ and $K$. Then we impose the constraint
\begin{align}
\zeta B_1 = \Omega_K ^{-1} B_1 \Omega_K  ,  \quad  B_2 = \Omega_K ^{-1} B_2 \Omega_K , \quad I = \Omega_K ^{-1} I \Omega_N, \quad \zeta J = \Omega_N ^{-1} J \Omega_K. 
\end{align}
These constraints imply the ADHM matrices are decomposed by
\begin{align}
\begin{split}
&B_{1,\o} : K_\o \longrightarrow K_{\o-1}, \quad B_{2,\o} : K_\o \longrightarrow K_{\o}, \\
&I_{\o} : N_\o \longrightarrow K_\o , \quad J_\o : K_\o \longrightarrow N_{\o-1}.
\end{split}
\end{align}
The ADHM equation is also decomposed into
\begin{align}
B_{1,\o} B_{2,\o} -B_{2,\o-1} B_{1,\o} +I_{\o-1} J_{\o} =0 , \quad \o \in \mathbb{Z}_p.
\end{align}
Let us define $\widetilde{B}_{1,2} \in \text{End} (\widetilde{K})$, $\widetilde{I} \in \text{Hom} (\widetilde{N},\widetilde{K})$, and $\widetilde{J} \in \text{Hom} (\widetilde{K} ,\widetilde{N})$ by
\begin{align}
\begin{split}
&\widetilde{K} = K_0, \quad \widetilde{N} = \bigoplus_{\o \in \mathbb{Z}_p} N_\o, \\
&\widetilde{B}_1 = B_{1,1} B_{1,2} \cdots B_{1,p-1} B_{1,0} , \quad \widetilde{B}_2 = B_{2,0} \\
& \widetilde{I} = \sum_{m=0} ^{p-1} B_{1,1} B_{1,2} \cdots B_{1,m-1} I_{m-1} \quad \widetilde{J} = \sum_{m=0} ^{p-1} J_m B_{1,m+1} B_{1,m+2} \cdots B_{1,p-1} B_{1,0}.
\end{split}
\end{align}
Then it is straightforward to show that we have the ordinary ADHM equation with the new matrices,
\begin{align}
[\widetilde{B}_1,\widetilde{B}_2 ] + \widetilde{I} \widetilde{J} =0.
\end{align}
In other words, we have constructed a projection $\EuScript{M}_{N,k} ^{\mathbb{Z}_p} \longrightarrow \EuScript{M}_{N, \widetilde{k}}$ of the moduli space of instantons on the $\mathbb{Z}_p$-orbifold to the moduli space of instantons on the ordinary $\mathbb{C}^2$. By integrating along the fiber of the projection, we produce a cohomology class of $\EuScript{M}_{N, \widetilde{k}}$ which we interpret as the surface defect observable.

At the level of the fixed points of the moduli spaces with respect to the global symmetry actions, the projection can be understood as a map
\begin{align} \label{eq:projy}
\rho : \boldsymbol\l \longmapsto \boldsymbol\Lambda
\end{align}
between two $N$-tuples of Young diagrams. Let us define
\begin{align}
\tilde{a}_\a = \begin{cases} a_{\a} ,\quad  c(\a) =0 \\ a_\a +(p-c(\a))\ve_1, \quad c(\a)=1,\cdots,p-1 \end{cases},
\end{align}
and also
\begin{align}
\widetilde{K}_\o : = \begin{cases} K_0 , \quad \o =0 \\ q_1 ^{p-\o} K_\o , \quad \o=1,\cdots, p-1 \end{cases}.
\end{align}
Then we see that
\begin{align}
\widetilde{K}=\widetilde{K}_0 = \sum_{\a =1} ^N e^{\b \tilde{a}_\a} \sum_{j=1} ^{\Lambda _1 ^{(\a)}} q_2 ^{j-1} \sum_{i=1} ^{\Lambda ^{(\a)t} _j } \tilde{q}_1 ^{i-1},
\end{align}
where the $N$-tuples of Young diagrams $\boldsymbol\Lambda$ is defined by $\Lambda_1 ^{(\a)} = \lambda_1 ^{(\a)} $ and
\begin{align}
\Lambda_j ^{(\a) t} = \begin{cases} 1+ \left[ \frac{\l _j ^{(\a)t} -1}{p} \right] , \quad c(\a) =0 \\ \left[ \frac{\l _j ^{(\a)t} c(\a) -1}{p} \right], \quad c(\a) = 1, \cdots p-1  \end{cases}.
\end{align}
Hence we obtain the projection $\rho$ \eqref{eq:projy} at the level of the fixed points with respect to the global symmetry action.

The surface defect observable can be obtained by first projecting to the $\mathbb{Z}_p$-invariant part and then re-expressing the projectetd partition function as expectation value of an observable. After the $\mathbb{Z}_p$-projection, the partition function is written as
\begin{align}
\Psi_{c,f} ^{\mathbb{Z}_2, \text{1-loop}} \Psi_{c,f} ^{\mathbb{Z}_2,\text{non-pert}}= \sum_{\boldsymbol\l} \prod_{\o \in \mathbb{Z}_p} \qe_\o ^{k_\o} E \left[ \mathcal{T} [\boldsymbol\l] ^{\mathbb{Z}_p} \right],
\end{align}
where
\begin{align} \label{eq:orbchar}
\mathcal{T}^{\mathbb{Z}_p} = \left[ - \frac{S S^*}{P_{12}^*} + \frac{M S^*}{P_{12}^*} \right] ^{\mathbb{Z}_p}.
\end{align}
We need to properly split this character into the bulk part and the surface defect part. For this, let us define
\begin{align}
\widetilde{S}_\o = \begin{cases} S_0, \quad \o = 0 \\ q_1 ^{p-\o} S_\o, \quad \o = 1, \cdots p-1 \end{cases}
\end{align}
and also $\widetilde{S}_{\o+p} = \widetilde{S}_\o$. Then we see that
\begin{align}
\widetilde{S} := \sum_{\o =0} ^{p-1} \widetilde{S}_\o = \widetilde{N} - \widetilde{P}_{12} \widetilde{K},
\end{align}
where $\tilde{q}_1 := q_1 ^p$, $\widetilde{P}_1 := 1-\tilde{q}_1$, and $\widetilde{P}_{12} := (1-\tilde{q}_1)(1-q_2)$. A straightforward computation shows that the first term in the character \eqref{eq:orbchar} becomes
\begin{align}
-\frac{\widetilde{S}\widetilde{S}^*}{\widetilde{P}_{12}^*} + \sum_{1\leq \o<\o' \leq p} \frac{\widetilde{S}_\o \widetilde{S}_{\o'} ^*}{P_2 ^*}.
\end{align}
The first term is precisely the bulk equivariant character. The second term should then be interpreted as the character for the surface defect observable.

When the gauge theory contains hypermultiplets, the equivariant parameters for the flavor symmetry group enter into the character as in the second term in \eqref{eq:orbchar}. We assign color $f$ to those hypermultiplet masses
\begin{align}
f:M \longrightarrow \mathbb{Z}_p,
\end{align}
and define $M_\o := f^{-1} (\o) $, $\o \in \mathbb{Z}_p$. Then we also modify the masses a bit by
\begin{align}
\widetilde{M}_\o = \begin{cases} M_0, \quad \o = 0 \\ q_1 ^{p-\o} M_\o, \quad \o = 1, \cdots p-1 \end{cases},
\end{align}
and define $\widetilde{M} = \sum_{\o=0} ^{p-1} \widetilde{M}_\o$. A straightforward computation shows that the second term in the character \eqref{eq:orbchar} becomes
\begin{align}
\frac{\widetilde{M} \widetilde{S}^*}{\widetilde{P}_{12}^*} - \sum_{1\leq \o<\o' \leq p} \frac{\widetilde{M}_\o \widetilde{S}_{\o'} ^*}{P_2 ^*}.
\end{align}
The first term is precisely the matter contribution to the bulk equivariant character. Thus, the second term should be interpreted as the character for the surface defect observable.

All in all, the partition function can be re-expressed as
\begin{align} \label{eq:orbpart}
\Psi_{c,f} ^{\mathbb{Z}_2, \text{1-loop}} \Psi_{c,f} ^{\mathbb{Z}_2,\text{non-pert}} = \sum_{\boldsymbol\Lambda} \qe^{\vert \boldsymbol\Lambda \vert} \EuScript{O}_{2,c,f} [\boldsymbol\Lambda] E\left[ - \frac{\widetilde{S}\widetilde{S}^*}{\widetilde{P}_{12} ^*} + \frac{\widetilde{M} \widetilde{S}^*}{\widetilde{P}_{12}^*} \right],
\end{align}
where the surface defect observable is
\begin{align}
\EuScript{O}_{2,c,f} [\boldsymbol\Lambda] := \sum_{\boldsymbol\l \in \rho^{-1} (\boldsymbol\Lambda)} \prod_{\o \in \mathbb{Z}_p} z_\o ^{k_{\o-1} -k_\o} E\left[ \sum_{1\leq\o'<\o''\leq p}  \frac{\left( \widetilde{S} _{\o'} - \widetilde{M}_{\o'} \right) \widetilde{S}_{\o''} ^*}{P_2 ^*} \right].
\end{align}
It was shown in  that the surface defect observable can be viewed as the partition function of the gauged linear sigma model on the $z_2$-plane and its coupling to the bulk gauge theory. The choice of coloring function $c$ corresponds to the choice of the vacuum of this gauged linear sigma model.

\subsubsection{Vortex string} \label{subsec:vortexstring}
We start from the superconformal $A_2$-quiver gauge theory. The instanton partition function of this theory can be written as
\begin{align}
\mathcal{Z}_{A_2} = \sum_{\boldsymbol\l} \prod_{\mathbf{i}=1,2} \qe^{\vert \boldsymbol\l^{(\mathbf{i})} \vert} E\left[ \mathcal{T}_{A_2} [\boldsymbol\l] \right],
\end{align}
where the character $\mathcal{T}_{A_2}$ is
\begin{align}
\begin{split}
\mathcal{T}_{A_2} =& \sum_{\mathbf{i}=1,2} (N_{\mathbf{i}} K_{\mathbf{i}}^* +q_{12} N_{\mathbf{i}}^* K_{\mathbf{i}} -P_{12} K_{\mathbf{i}} K_{\mathbf{i}}^*) -N_0 K_1 ^* -q_{12} N_3 ^* K_2 \\
&\quad\quad\quad\quad\quad\quad\quad\quad\quad\quad -N_1 K_2 ^* -q_{12} N_2 ^* K_1 +P_{12} K_1 K_2 ^*,
\end{split}
\end{align}
and the $2N$-tuple of Young diagrams $\boldsymbol\l = \left( \boldsymbol\l^{(1)}, \boldsymbol\l^{(2)} \right)$ enumerate fixed points of the instanton moduli space with respect to the global symmetry group.

We partially higgs the gauge group down to $U(N)$. The partial higgsing is initiated by the constraints
\begin{align} \label{eq:constvortex}
&a_{1,\a} = a_{0,\a} - \delta_{\a \g} \ve_1,
\end{align}
where we made a choice $\g \in\{1,\cdots, N\}$. These constraints make $N$ hypermultiplets nearly massless (exactly massless in the flat spacetime $\ve_1=0$). The massless hypermultiplet scalars may develop expectation values, higgsing the first $U(N)$ gauge node. Due to the $\ve_1$-mistach in the constraint, the $U(1)$ gauge group is restored along a codimension-two plane ($z_2$-plane) where the gauge field configuration is squeezed into a vortex. The net result is the $U(N)$ gauge theory with $2N$ hypermultiplets, coupled to a two-dimensional linear sigma model on the $z_2$-plane. The choice of $\g \in \{1,\cdots, N\} $ passes to the choice of the vacuum of this gauged linear sigma model. Hence we generate a surface defect coupled to the bulk gauge theory in this sense.

At the level of the fixed points of the instanton moduli space, the $N$-tuple of Young diagrams $\boldsymbol\l^{(1)}$ is restricted by the constraints \eqref{eq:constvortex} as
\begin{align}
\boldsymbol\l^{(1)} = \left(  \varnothing, \cdots ,\varnothing,\; \ytableausetup{mathmode,boxsize=1.5em, centertableaux}
\stackrel{\l ^{(1,\g)}}{\underbrace{\begin{ytableau}\scriptstyle &\scriptstyle  &\cdots& \scriptstyle \\
\end{ytableau}}_{k}} ,\; \varnothing , \cdots, \varnothing \right), \quad k \in \mathbb{Z}^{\geq 0}.
\end{align}
In terms of the vector space $K_1$, this implies
\begin{align}
K_1 = \mu \frac{1-q_2 ^k}{1-q_2},
\end{align}
with $\mu = e^{\b (a_{0,\g} -\ve_1)} $. The character $\mathcal{T}_{A_2}$ simplifies accordingly, and we can split it into the bulk and the surface defect contributions. The net result is
\begin{align}
\Psi_{\g}^{L,\text{non-pert}} = \sum_{\boldsymbol\l} \qe ^{\vert \boldsymbol\l  \vert} \EuScript{O}^{L} _{2,\g} \left[ \boldsymbol\l \right]  E \left[ \mathcal{T}_{A_1} \left[ \boldsymbol\l \right] \right] = \left\langle  \EuScript{O}^{L} _{2,\g} \right\rangle \mathcal{Z}_{A_1} ^{\text{inst}},
\end{align}
where $\mathcal{T}_{A_1}$ is the character for the $U(N)$ gauge theory with $2N$ hypermultiplets
\begin{align}
\mathcal{T}_{A_1} = NK^* + q_{12} N^* K -P_{12} KK^* -N_0 K^* -q_{12} N_3 ^* K,
\end{align}
and the surface defect observable $\EuScript{O}^{L} _{2,\g}$ is
\begin{align}
\EuScript{O}^{L }_{2,\g}  \left[ \boldsymbol\l  \right] := \sum_{k = 0 } ^\infty z ^{-k} \prod_{l=1} ^{k} \frac{\EuScript{Y}_2 (a_{0,\g} +l \ve_2) \left[ \boldsymbol\l \right]}{P_0 (a_{0,\g} +l \ve_2)} \prod_{\Box \in \boldsymbol\l^{(2)}} \frac{a_{0,\g} -\ve_1 -c_\Box}{a_{0,\g}-c_\Box}.
\end{align}
In the decoupling limit $\qe \to 0$, the surface defect observable reduces to the partition function of the gauged linear sigma model on the $z_2$-plane. When the gauge coupling is turned on, the non-trivial coupling between the degrees of freedom on the bulk and the surface defect start to contribute to the partition function, through the $\EuScript{Y}$-observable in the numerator. Note that the convergence domain for this partition function is 
$0 <\vert \qe\vert < 1 <\vert z\vert$.

We can similarly engineer the gauge theory coupled to the vortex string surface defect whose partition function converges in the domain $0 < \vert z \vert <\vert \qe\vert <1$. We start from the $A_2$-quiver gauge theory again. This time, we impose different constraints:
\begin{align}
a_{2,\a} = a_{3,\a} -\ve - \delta_{\a \g} \ve_1,
\end{align}
with a choice $\g \in \{1, \cdots, N\}$. The gauge group is partially higgsed as earlier, leaving the $U(N)$ gauge theory with $2N$ hypermultiplets coupled to a vortex string surface defect. Then, for the reasons to be clarified later, we would like to make the following re-definitions
\begin{align}
\begin{split}
&a_{0,\a} \longrightarrow -a_{0,\a} -\ve \\
&a_{1,\a} \longrightarrow -a_{1,\a}  \quad\quad\quad\quad \quad\quad  \a =1,\cdots,N. \\
&a_{3,\a} \longrightarrow -a_{3,\a} +2\ve 
\end{split}
\end{align}
The partition function can be written as
\begin{align}
\Psi_{\g}^{R, \text{non-pert}} = \sum_{\boldsymbol\l} \qe ^{\vert \boldsymbol\l  \vert}   E \left[ \mathcal{T}_{A_1} ' \left[ \boldsymbol\l \right] \right]  \sum_{k=0} ^\infty \left(\frac{z}{\qe}\right)^k \prod_{l=1} ^{k} \frac{\EuScript{Y}' (-a_{3,\g} +l\ve_2) \left[ \boldsymbol\l \right]}{P_3 ' (-a_{3,\g}  +2\ve +l \ve_2)} \prod_{\Box \in \boldsymbol\l} \frac{-a_{3,\g}  -\ve_1 -c_\Box '}{-a_{3,\g}  -c_\Box '}.
\end{align}
Here, we defined 
\begin{align}
\begin{split}
&\mathcal{T}_{A_1} ' [\boldsymbol\l] := \left. \left[ NK^* +q_{12} N^* K -P_{12} KK^* -N_0 K^* -q_{12}^2 N_3 ^* K \right]\right\vert_{\substack{\mathbf{a} \to -\mathbf{a} \\ \mathbf{a}_0 \to - \mathbf{a}_0 -\ve \\ \mathbf{a}_3 \to -\mathbf{a}_3 +2\ve }}, \\
&c_\Box ' : = \left. c_\Box \right\vert_{\mathbf{a} \to -\mathbf{a} }, \\
&P_3  ' (x) := \prod_{\a=1}^N (x+a_{3,\a} -2\ve), \\
&\EuScript{Y}' (x) := \left.\EuScript{Y}(x)\right\vert_{\mathbf{a} \to -\mathbf{a}}.
\end{split}
\end{align}

\subsection{Construction of intersecting surface defects} \label{sec:constint}
In this section, we construct $\EuScript{N}=2$ gauge theory with intersecting half-BPS surface defects. The surface defects would be lying on the $z_1$-plane and $z_2$-plane, so that they are intersecting with each other at the origin. Due to the presence of mutually intersecting defects, most of the $\EuScript{N}=2$ supersymmetry is broken but we still have the scalar supercharge preserved. Then, since $U(1)^2 \subset SO(4)$ isometry is still preserved, we can turn on the $\Omega$-background with respect to them. The partition function of the gauge theory in the presence of the intersecting defects is a path-integral which localizes onto the fixed points of this $\Omega$-deformed supercharge. This partition function can also be understood as the two-point function of the two surface defect observables. 

The configuration of intersecting surface defects were analyzed as a 4d/2d/0d coupled system in \cite{GFPP2016,PP2016}. In particular, their partition functions were conjecturally identified with the Liouville correlation functions with in the presence of arbitrary degenerate fields \cite{GFPP2016}, in the context of \cite{agt}. We may regard our analysis as giving a proof of this statement to relevant cases by using non-perturbative Dyson-Schwinger equations (see section \ref{subsec:intnonpert}). 

We would like to stress that the idea of using intersecting surface defects to construct solutions to KZ or BPZ type differential equations can be implemented in more general settings. In \cite{JLN}, the folded instanton configuration is considered on top of the $\mathbb{Z}_N$ regular orbifold surface defect, which allows a concrete connection between the gauge theory, WZW models, and spin chain systems. 

We provide a specific construction of such intersecting surface defects, which resembles the construction of a single surface defect by a partial Higging that we have seen in the last section. Here, we start from the linear $A_3$-quiver gauge theory. The partition function of the $A_3$-quiver gauge theory is written as
\begin{align} \label{eq:a3}
\mathcal{Z}_{A_3} = \sum_{\boldsymbol\l} \prod_{\mathbf{i} =1,2,3} \qe_{\mathbf{i}} ^{\vert \boldsymbol\l ^{(\mathbf{i})} \vert} E \left[ \mathcal{T}_{A_3} [\boldsymbol\l]\right],
\end{align}
where the character $\mathcal{T}_{A_3}$ is given by
\begin{align} \label{eq:chara3}
\begin{split}
\mathcal{T}_{A_3} &= \sum_{\mathbf{i}=1,2,3} \left( N_{\mathbf{i}} K_{\mathbf{i}} ^* +q_{12} N_{\mathbf{i}} ^* K_{\mathbf{i}} -P_{12} K_{\mathbf{i}} K_{\mathbf{i}} ^*\right) - N_0 K_{1} ^* - q_{12} N_4 ^* K_3   \\
& - N_1 K_2 ^* -q_{12} N_2 ^* K_1 + P_{12} K_1 K_2 ^* -N_2 K_3 ^* -q_{12} N_3 ^* K_2 +P_{12} K_2 K_3 ^*.
\end{split}
\end{align}
Let us initiate partial Higging for the first and the third gauge nodes by setting
\begin{align}
\begin{split}
&a_{1,\a} = a_{0,\a} - \delta_{\a, h} \ve_1 \\
&a_{3,\a} = a_{4,\a} -\ve - \delta_{\a, l} \ve_2,
\end{split}
\end{align}
for some chosen $h, l \in \{1, \cdots, N\}$, and define $\qe_2 = \qe$, $\qe_1 = \frac{1}{z}$, and $\qe_3 = \frac{y}{\qe}$. Then almost all the Young diagrams in $\boldsymbol\l^{(1)}$ become empty except a single-row $\l^{(1,h)}$, while almost all the Young diagrams in $\boldsymbol\l^{(3)}$ also become empty except a single-column $\l^{(3,l)}$. We have the following simplified expressions accordingly,
\begin{align}
K_1 = \mu \frac{1-q_2 ^{k_1}}{1-q_2}, \quad K_3 = \mu' \frac{1-q_1 ^{k_3}}{1-q_1},
\end{align}
where we defined $\mu = e^{a_{1,h}} = e^{a_{0,h} -\ve_1}$ and $\mu' = e^{a_{3,l}} =e^{a_{4,l} - \ve -\ve_2}$. Then the character \eqref{eq:chara3} simplifies to
\begin{align}
\begin{split}
\mathcal{T}_{A_3} =& N_2 K_2 ^* +q_{12} N_2 ^* K_2 -P_{12} K_2 K_2 ^* - N_0 K_2 ^* -q_{12}^2 N_4 ^* K_2 \\
& + \mu P_1 q_2 ^{k_1} K_1 ^* +q_{12} K_1 (N_1 ^* -S_2 ^*) -\mu P_1 K_2 ^* \\
& +\mu' P_2 q_1 ^{k_3} K_3 ^* +q_{12} K_3 (N_3 ^* -q_{12} ^{-1} S_2 ^* ) - q_{12} \mu'^{-1} P_2 ^{-1} K_2
\end{split}
\end{align}
Note that the first line is precisely the character for the $A_1$-quiver gauge theory, namely, the $U(N)$ gauge theory with $2N$ fundamental hypermultiplets (with some shifts in the masses that we will be more precise soon):
\begin{align}
\mathcal{T}_{A_1} =  N_2 K_2 ^* +q_{12} N_2 ^* K_2 -P_{12} K_2 K_2 ^* - N_0 K_2 ^* -q_{12}^2 N_4 ^* K_2.
\end{align}
The second line and the third line can be interpreted as the surface defect observables lying on $z_2$-plane and $z_1$-plane, respectively. Accordingly, the partition function becomes the two-point function of these two observables in the $A_1$-quiver gauge theory,
\begin{align} \label{eq:intsurf0yq1z}
\Upsilon_{l,h} ^{0<\vert y \vert <\vert \qe \vert <1 <\vert z \vert}  = \sum_{\boldsymbol\l } \qe ^{\vert \boldsymbol\l  \vert} \EuScript{O}_{1,l} ^{0<\vert y\vert<\vert \qe \vert} \left[ \boldsymbol\l \right] \EuScript{O}_{2,h} ^{1<\vert z \vert} \left[ \boldsymbol\l \right]  E \left[ \mathcal{T}_{A_1} \left[ \boldsymbol\l  \right] \right] = \left\langle \EuScript{O}_{1,l} ^{0<\vert y\vert<\vert \qe \vert} \EuScript{O}_{2,h} ^{1<\vert z \vert} \right\rangle \mathcal{Z}_{A_1},
\end{align}
whose convergence domain is $0<\vert y \vert <\vert \qe \vert <1 <\vert z \vert$. Here we have defined the instanton partition function for the $A_1$-quiver gauge theory
\begin{align}
\mathcal{Z}_{A_1} := \sum_{\boldsymbol\l}\qe ^{\vert \boldsymbol\l  \vert} E \left[ \mathcal{T}_{A_1} \left[ \boldsymbol\l\right] \right],
\end{align}
and the surface defect observables
\begin{align}
\begin{split}
&\EuScript{O}_{1,l} ^{0<\vert y\vert<\vert \qe \vert}  \left[ \boldsymbol\l  \right] := \sum_{k_3 =0}^ \infty \left( \frac{y}{\qe} \right) ^{k_3} \prod_{m=1} ^{k_3} \frac{\EuScript{Y} (a_{4,l}-2\ve +m\ve_1) \left[ \boldsymbol\l\right]}{P_4 (a_{4,l} +m \ve_1)} \prod_{\Box \in \boldsymbol\l} \frac{a_{4,l} -2\ve -\ve_2 -c_\Box}{a_{4,l} -2\ve -c_\Box} \\
&\EuScript{O}_{2,h} ^{1<\vert z\vert}  \left[ \boldsymbol\l  \right] := \sum_{k_1 = 0 } ^\infty z ^{-k_1} \prod_{m=1} ^{k_1} \frac{\EuScript{Y} (a_{0,h} +m \ve_2) \left[ \boldsymbol\l \right]}{P_0 (a_{0,h} +m\ve_2)} \prod_{\Box \in \boldsymbol\l} \frac{a_{0,h} -\ve_1 -c_\Box}{a_{0,h}-c_\Box}.
\end{split} 
\end{align}
In the zero bulk instanton sector, $\vert \boldsymbol\l ^{(2)}\vert =0$, $\EuScript{Y}_2$-observable simplifies to a polynomial $\EuScript{Y}_2 (x) = \prod_{\alpha=1} ^N (x- a_{2,\a})$. Then the surface defect observables $\EuScript{O}_1$ and $\EuScript{O}_2$ reduce to generalized hypergeometric functions, which are partition functions of the two-dimensional gauged linear sigma model on the $\text{Hom} (\mathcal{O}(-1) ,\mathbb{C}^N )$-bundle over $\mathbb{P}^{N-1}$ whose K\"{a}hler modulus is $\qe_1$ and $\qe_3$, respectively. The choice of $l,h \in \{1,\cdots, N\}$ is exactly the choice of vacua of these two gauged linear sigma models. Therefore, the intersecting surface defect partition function is just the product of these two partition functions of gauged linear sigma models lying on the $z_1$-plane and the $z_2$-plane. In the non-zero bulk instanton sector, we start to get contributions from the coupling between the bulk and the two surface defects.

We can similarly construct intersecting surface defects whose partition function is given by the series convergent in the domain $0<\vert z \vert <\vert \qe \vert <1 <\vert y \vert$. We also start from the $A_3$-quiver gauge theory \eqref{eq:a3}, and impose the constraints
\begin{align}
\begin{split}
&a_{1,\a} = a_{0,\a} - \delta_{\a, l} \ve_2 \\
&a_{3,\a} = a_{4,\a} -\ve - \delta_{\a, h} \ve_1.
\end{split}
\end{align}
These constraints initiate higging of the gauge group, leaving $U(N)$ gauge theory with $2N$ hypermultiplets coupled to intersecting vortex string surface defects. For reasons to be clear later, we re-define the parameters as
\begin{align}
\begin{split}
&a_{0,\a} \longrightarrow -a_{0,\a} -\ve \\
&a_{2,\a} \longrightarrow -a_{2,\a} \\
&a_{4,\a} \longrightarrow -a_{4,\a} +3\ve. 
\end{split}
\end{align}
The partition function thus defined is the intersecting surface defect partition function $\Upsilon_{l,h} ^{0<\vert z \vert <\vert \qe \vert <1 <\vert y \vert}  $ lying in the domain $0<\vert z \vert <\vert \qe \vert <1 <\vert y \vert$:
\begin{align}
\Upsilon_{l,h} ^{0<\vert z \vert <\vert \qe \vert <1 <\vert y \vert} = \sum_{\boldsymbol\l} \qe ^{\vert \boldsymbol\l\vert} {\EuScript{O}'} ^{1<\vert y \vert} _{1,l} [\boldsymbol\l] {\EuScript{O}'}_{2,h} ^{0<\vert z \vert<\vert \qe \vert} [\boldsymbol\l] E \left[ \mathcal{T} '_{A_1} [\boldsymbol\l] \right],
\end{align}
where the defect observables are
\begin{align}
\begin{split}
& {\EuScript{O}'}_{2,h}  ^{0<\vert z \vert<\vert \qe \vert}  [\boldsymbol\l] = \sum_{k_3=0} ^\infty \left( \frac{z}{\qe} \right)^{k_3} \prod_{m=1} ^{k_3} \frac{ \EuScript{Y}' (-a_{4,h} +\ve +m\ve_2)}{P_4 '(-a_{4,h} +3\ve+m\ve_2)} \prod_{\Box \in \boldsymbol\l} \frac{-a_{4,h} +\ve-\ve_1 -c' _\Box}{-a_{4,h} +\ve -c' _\Box}\\
& {\EuScript{O}'}_{1,l}  ^{1<\vert y \vert} [\boldsymbol\l] = \sum_{k_1 =0}  ^\infty y^{-k_1} \prod_{m=1} ^{k_1} \frac{ \EuScript{Y}' (-a_{0,l} -\ve+m\ve_1)}{P_0 ' (-a_{0,h}-\ve+m\ve_1)} \prod_{\Box \in \boldsymbol\l} \frac{-a_{0,l} -\ve-\ve_2 - c' _\Box}{-a_{0,l} -\ve -c' _\Box} ,
\end{split}
\end{align}
where we have defined
\begin{align}
\begin{split}
&\mathcal{T}_{A_1} ' = \left. \left[ N K ^* +q_{12} N ^* K -P_{12} K K ^* - N_0 K ^* -q_{12}^2 N_4 ^* K \right] \right\vert_{\substack{\mathbf{a}_0 \to - \mathbf{a}_0 -\ve \\ \mathbf{a} \to -\mathbf{a} \\ \mathbf{a}_4 \to -\mathbf{a}_4 +3\ve}} \\
&P_0 '(x) = \left. P_0 (x) \right\vert_{\mathbf{a}_0 \to - \mathbf{a}_0 -\ve} \\
&P_4 '(x) = \left. P_4 (x) \right\vert_{\mathbf{a}_4 \to - \mathbf{a}_4 +3\ve} \\
&c' _\Box = \left. c_\Box \right\vert_{\mathbf{a} \to -\mathbf{a}}.
\end{split}
\end{align}

\section{Surface defects and Hamilton-Jacobi equations for isomonodromic deformations} \label{sec:hjeq}
The partition functions of four-dimensional $\EuScript{N}=2$ gauge theories subject to $\Omega$-background have explicit dependence on gauge couplings and equivariant parameters. The partition function manifests various correspondences which relate the gauge theory with interesting mathematical objects, encoded in its responses to the variations of those parameters. Such analytic properties of the partition functions can be extracted by using a special class of chiral observables, called the $qq$-characters \cite{Nekrasov_BPS1}. The crucial property of the $qq$-characters is the regularity of their vacuum expectation values, which encode non-trivial chiral ring relations in the presence of $\Omega$-background \cite{Nekrasov_BPS1,JZ2019}. 

In the presence of half-BPS surface defects, the $qq$-characters are especially powerful since they lead to closed differential equations satisfied by the partition functions in many cases \cite{Nekrasov_BPS45,Jeong2017}, which can be regarded as double quantization of the chiral ring relation of the coupled system. In this section, we discuss such differential equations satisfied by the surface defect partition functions. The NS limit of the differential equation reveals that the asymptotics of the surface defect partition function is the Hamilton-Jacobi potential for the isomonodromic flow. Moreover, we verify that the very asymptotics of the surface defect partition function can also be viewed as the generating function of the Riemann-Hilbert map.

\subsection{Non-perturbative Dyson-Schwinger equations}
We state the non-perturbative Dyson-Schwinger equations satisfied by the partition functions of the $\EuScript{N}=2$ $U(2)$ gauge theory with four fundamental hypermultiplets in the presence of the half-BPS surface defects introduced in the previous section. From these equations it can be verified that the partition function satisfies a differential equation in gauge couplings. We do not reproduce the derivation here, but only state the result.

\subsubsection{Orbifold} \label{sec:orbsurfd}
Let us consider the $U(N)$ gauge theory with matter multiplets on the $\mathbb{Z}_p$-orbifold, introduced in section \ref{sec:orbdef}. The fundamental $qq$-characters of this theory are
\begin{align}
\EuScript{X}_\o (x) = \EuScript{Y}_{\o+1} (x+\ve) + (-1)^{\vert N_\o \vert + \vert M_\o \vert} \qe_\o \frac{P_\o (x)}{\EuScript{Y}_\o (x)}, \quad \o \in \mathbb{Z}_p,
\end{align}
where $P_\o (x) = \prod_{i \in f^{-1} (\o)} (x-m_i)$. The non-perturbative Dyson-Schwinger equations are constraints following from the regularity of their expectation values:
\begin{align} \label{eq:npds}
0 =\left[ x^{-n} \right] \Big\langle \EuScript{X}_\o (x) \Big\rangle , \quad \o \in \mathbb{Z}_p, \; n\geq1.
\end{align}

Our main example is the $U(2)$ gauge theory with four hypermultiplets on the $\mathbb{Z}_2$-orbifold, for which the color functions are chosen as
\begin{align}
\begin{split}
&c^{-1}(0) = \g, \quad c^{-1} (1) = \bar\g, \\
&M_\o   = \sum_{i=1,2} e^{\b m_{\o,i}} , \quad \o \in \mathbb{Z}_2.
\end{split}
\end{align}
Then we have two fundamental $qq$-characters:
\begin{align}
\begin{split}
&\EuScript{X}_0 (x) = \EuScript{Y}_1 (x+\ve) - \frac{\qe}{z} \frac{P_0 (x)}{\EuScript{Y}_0 (x)} \\
&\EuScript{X}_1 (x) = \EuScript{Y}_0 (x+\ve) - z \frac{P_1 (x)}{\EuScript{Y}_1 (x)} .
\end{split}
\end{align}
The non-perturbative Dyson-Schwinger equations for them imply that the partition function annihilates a differential operator in gauge couplings. We do not reproduce the derivation, but only state the result. First, for notation convenience let us get rid of all the tilde above the equivariant parameters, and re-define the hypermultiplet masses $\left( m_f \right)_{f=1} ^4$ as
\begin{align}
m_i = m_{-,i}, \quad m_{2+i} = m_{+,i}, \quad i=1,2.
\end{align}
Then we define
\begin{align}
\begin{split}
\widetilde{\Psi}_\g ^{\mathbb{Z}_2} =& z^{\frac{-a_\g +a_{\bar\g} +\ve_2}{2\ve_2} +\frac{\ve_1}{\ve_2} (\th_0 +\th_\qe) } \qe^{-\frac{(a_\g -a_{\bar\g})^2}{4\ve_1 \ve_2} +\frac{\ve_1}{\ve_2} (\th_0 +\th_\qe)^2 +\th_0+\th_\qe -\frac{\ve_2}{4\ve_1}}  \\
& (1-\qe) ^{-\frac{\ve_1}{\ve_2} (\th_0+\th_\qe+\th_1+\th_\infty)(\th_0+\th_\qe+\th_1-\th_\infty) -2\th_0 -\th_\qe -\th_1 -\frac{\ve_2}{2\ve_1}} \Psi ^{\mathbb{Z}_2 , \text{non-pert}}_\g,
\end{split}
\end{align}
where we have used the dimensionless $\th$-parameters \eqref{eq:dimlessparm} for the hypermultiplet masses. We have used the notation $\bar{\b} \in \{1,2 \} \setminus \{\b\}$. Then the non-perturbative Dyson-Schwinger equation \eqref{eq:npds} implies that the modified partition function satisfies the following differential equation
\begin{align} 
\begin{split}
0&=\left[ \ve_2 ^2 \partial_z ^2 - \ve_1 \ve_2 \left( \frac{2 \th_0}{z} + \frac{2 \th_\qe}{z-\qe} + \frac{2 \th_1}{z-1} \right) \partial_z + \ve_1 \ve_2 \frac{\qe(\qe-1)}{z(z-\qe)(z-1)} \frac{\partial}{\partial \qe} \right. \\
& + \frac{\ve_1 ^2}{z(z-\qe)(z-1)} \left\{ z \left( \th_0 +\th_\qe +\th_1 +\th_\infty + \frac{\ve_2}{2\ve_1} \right) \left( \th_0 +\th_\qe +\th_1 -\th_\infty +\frac{\ve_2}{2\ve_1} \right)  \right. \\ 
& \left. \left. \quad\quad\quad\quad\quad\quad\quad\quad\quad+\frac{\ve_2}{\ve_1 z} \left( \th_0  + \frac{\ve_2}{4 \ve_1} \right)    \right\} \right]  \widetilde\Psi ^{\mathbb{Z}_2} (\mathbf{a},\mathbf{m},\ve_1,\ve_2;\qe,z).
\end{split}
\end{align}
Note that the solutions $\widetilde\Psi^{\mathbb{Z}_2} (\mathbf{a},\mathbf{m},\ve_1,\ve_2;\qe,z)$ lie in the domain $0<\vert \qe \vert <\vert z \vert<1$.

\subsubsection{Vortex string}
Let us consider the $A_2$-quiver gauge theory. The fundamental $qq$-characters of this theory are
\begin{align}
\begin{split}
&\EuScript{X}_1 (x) = \EuScript{Y}_1 (x+\ve) +\qe_1 \frac{\EuScript{Y}_0 (x) \EuScript{Y}_2 (x+\ve)}{\EuScript{Y}_1 (x)} +\qe_1 \qe_2 \frac{\EuScript{Y}_0 (x) \EuScript{Y}_3 (x+\ve)}{\EuScript{Y}_2 (x)} \\
&\EuScript{X}_2 (x) = \EuScript{Y}_2 (x+\ve) + \qe_2 \frac{\EuScript{Y}_1 (x)\EuScript{Y}_3 (x+\ve)}{\EuScript{Y}_2 (x)} +\qe_1 \qe_2 \frac{\EuScript{Y}_0 (x-\ve) \EuScript{Y}_3 (x+\ve)}{\EuScript{Y}_1 (x-\ve)}.
\end{split}
\end{align}
The non-perturbative Dyson-Schwinger equations are the regularity conditions for their expectation values:
\begin{align} \label{eq:npdsvortex}
0 = \left[ x^{-n} \right] \Big\langle \EuScript{X}_i (x) \Big\rangle, \quad i=1,2, \;\; n\geq 1.
\end{align}
Now we partially higgs the gauge group by imposing the constraints
\begin{align}
&a_{1,\a} = a_{0,\a} - \delta_{\a \g} \ve_1.
\end{align}
As we have seen in section \ref{subsec:vortexstring}, the resulting theory is the $U(2)$ gauge theory with four fundamental hypermultiplets coupled to a vortex string surface defect. The non-perturbative Dyson-Schwinger equations \eqref{eq:npdsvortex} imply that the partition function of this theory satisfies a differential equation in gauge couplings. We do not reproduce the derivation here, but only state the result. First, let us re-define the hypermultiplet masses as
\begin{align}
m_{\a} = a_{0,\a}, \quad m_{2+\a} = a_{3,\a} -\ve, \quad \a=1,2,
\end{align}
so that we have the masses $\left( m_{f} \right)_{f=1}^4$ for four fundamentals. We also modify the partition function by
\begin{align}
\begin{split}
\widetilde\Psi_\g ^{L} =& z^{\frac{-m_\g + m_{\bar\g} +\ve_2}{2\ve_2} +\frac{\ve_1}{\ve_2} (\th_0 +\th_\qe+\th_1)  } \left( 1-z^{-1} \right)^{2{\th}_1 \frac{\ve_1}{\ve_2}  +1} \qe ^{-\frac{(a_1-a_2)^2}{4\ve_1\ve_2} +\frac{\ve_1}{\ve_2} (\th_0+\th_\qe)^2 +\th_0 +\th_\qe -\frac{\ve_2}{4\ve_1}} \\
&(1-\qe)^{-\frac{\ve_1}{\ve_2} (\th_0+\th_\qe+\th_1 +\th_\infty)(\th_0+\th_\qe+\th_1 -\th_\infty) -2\th_0 -\th_\qe -\th_1 -\frac{\ve_2}{2\ve_1} } \Psi_\g ^{L, \text{non-pert}},
\end{split}
\end{align}
where we have used the dimensionless $\th$-parameters \eqref{eq:dimlessparm}. Then the modified partition function satisfies
\begin{align} \label{eq:heavyeq}
\begin{split}
0&=\left[ \ve_2 ^2 \partial_z ^2 - \ve_1 \ve_2 \left( \frac{2 \th_0}{z} + \frac{2 \th_\qe}{z-\qe} + \frac{2 \th_1}{z-1} \right) \partial_z + \ve_1 \ve_2 \frac{\qe(\qe-1)}{z(z-\qe)(z-1)} \frac{\partial}{\partial \qe} \right. \\
& + \frac{\ve_1 ^2}{z(z-\qe)(z-1)} \left\{ z \left( \th_0 +\th_\qe +\th_1 +\th_\infty + \frac{\ve_2}{2\ve_1} \right) \left( \th_0 +\th_\qe +\th_1 -\th_\infty +\frac{\ve_2}{2\ve_1} \right)  \right. \\ 
& \left. \left. \quad\quad\quad\quad\quad\quad\quad\quad\quad+\frac{\ve_2}{\ve_1 z} \left( \th_0  + \frac{\ve_2}{4 \ve_1} \right)    \right\} \right]  \widetilde\Psi^L (\mathbf{a},\mathbf{m},\ve_1,\ve_2;\qe,z).
\end{split}
\end{align}
Note that the solutions $\widetilde\Psi^L (\mathbf{a},\mathbf{m},\ve_1,\ve_2;\qe,z)$ lie in the domain $0<\vert \qe \vert<1<\vert z \vert$.

Similarly, we can start from the $A_2$-quiver gauge theory with the constraints
\begin{align}
a_{2,\a} = a_{3,\a} -\ve - \delta_{\a \g} \ve_1,
\end{align}
and make the re-definition
\begin{align}
\begin{split}
&a_{0,\a} \longrightarrow -a_{0,\a} -\ve \\
&a_{1,\a} \longrightarrow -a_{1,\a}  \quad\quad\quad\quad \quad\quad  \a =1,2. \\
&a_{3,\a} \longrightarrow -a_{3,\a} +2\ve 
\end{split}
\end{align}
Then we make a shift in Coulomb moduli and modify the partition function by
\begin{align}
\begin{split}
&\widetilde\Psi_\g ^{R} (\mathbf{a} +\boldsymbol\d_\g \ve_1 )\\
 &= \left(\frac{z}{\qe} \right) ^{-\frac{m_{\g+2} -m_{\bar\g+2}}{2\ve_2}+ \frac{\ve_1}{\ve_2} \th_0 +\frac{1}{2}} (1-z)^{2{\th}_{1} \frac{\ve_1}{\ve_2}  +1} \qe^{-\frac{(a_{1,\g}- a_{1,\bar\g}+\ve_1)^2}{4\ve_1 \ve_2} +\frac{\ve_1}{\ve_2} \left( \th_0 + \th_\qe +\frac{1}{2} \right)^2 +\th_0 + \th_\qe +\frac{1}{2} -\frac{\ve_2}{4\ve_1}}\\
&(1-\qe)^{-\frac{\ve_1}{\ve_2} \left((\th_0 +\th_\qe+\th_1+\th_\infty)(\th_0 +\th_\qe+\th_1-\th_\infty) +2\th_\qe+2\th_1 +1  \right) -2\th_0 -3\th_\qe -3\th_1 -3 -\frac{5\ve_2}{2\ve_1}} \Psi_\g ^{R,\text{non-pert}} (\mathbf{a} +\boldsymbol\d_\g \ve_1),
\end{split}
\end{align}
where $\boldsymbol\d_\g = \left( \delta_{\alpha, \gamma} \right)_{\a =1,2}$. The modified partition function can be shown to satisfy the differential equation \eqref{eq:heavyeq}. Note that the solutions $\widetilde\Psi_\g ^{R} $ produced in this way are in the domain $0<\vert z \vert<\vert \qe \vert <1$.

\subsection{Hamilton-Jacobi equation for Painlev{\'e} VI} \label{subsec:HJeq}

The non-perturbative Dyson-Schwinger equations simplify in some limits in the space of parameters. In particular, for our study of isomonodromic deformations, we consider asymptotics of the surface defect expectation value in the NS limit $\ve_2 \to 0$, as well as the limit of the differential equation it obeys. The result is the Hamilton-Jacobi formulation  of the isomonodromic problem, as we recall now. Note that this observation was originally made in \cite{LLNZ2013}, albeit in the context of the Liouville conformal field theory which is related to our discussion through the identification of gauge theory partition functions and Liouville conformal blocks \cite{agt}. We emphasize that the non-perturbative Dyson-Schwinger equation allows us to re-establish the result of \cite{LLNZ2013} in purely gauge theoretical context, without resorting to the CFT arguments (see also \cite{NikBlowup}).

We have shown that the surface defect partition function satisfies
\begin{align} \label{eq:defeq}
\begin{split}
0&=\left[ \ve_2 ^2 \partial_z ^2 - \ve_1 \ve_2 \left( \frac{2 \th_0}{z} + \frac{2 \th_\qe}{z-\qe} + \frac{2 \th_1}{z-1} \right) \partial_z + \ve_1 \ve_2 \frac{\qe(\qe-1)}{z(z-\qe)(z-1)} \frac{\partial}{\partial \qe} \right. \\
& + \frac{\ve_1 ^2}{z(z-\qe)(z-1)} \left\{ z \left( \th_0 +\th_\qe +\th_1 +\th_\infty + \frac{\ve_2}{2\ve_1} \right) \left( \th_0 +\th_\qe +\th_1 -\th_\infty +\frac{\ve_2}{2\ve_1} \right)  \right. \\ 
& \left. \left. \quad\quad\quad\quad\quad\quad\quad\quad\quad+\frac{\ve_2}{\ve_1 z} \left( \th_0  + \frac{\ve_2}{4 \ve_1} \right)    \right\} \right]  \widetilde\Psi_\g (\mathbf{a},\mathbf{m},\ve_1,\ve_2;\qe,z)
\end{split}
\end{align}
In the NS limit $\ve_2 \to 0$, the partition function shows the following asymptotic behavior\footnote{Here, we assume that the Coulomb moduli is generic. It would be interesting to study special locus of Coulomb moduli for which the surface defect partition function splits into parts, each of which defines a twisted superpotential as its asymptotics \cite{Jeong2017}, in its correspondence to the Riemann-Hilbert problem.}
\begin{align} \label{eq:heavyasymp}
\widetilde\Psi_\g (\mathbf{a},\mathbf{m},\ve_1,\ve_2;\qe,z) = \exp \left( \frac{\ve_1}{\ve_2} \widetilde{S}_\g (\mathbf{a},\mathbf{m},\ve_1;\qe ,z) + \mathcal{O}(1) \right).
\end{align}
There is a choice of the vacuum $\g \in \{1,2\}$ of the gauged linear sigma model on the $z_2$-plane, which yield inequivalent asymptotics $\widetilde{S}_\g$. The analysis below on the asymptotics $\widetilde{S}$ applies for both choices of $\g$, so we omit the subscript from now on and consider the availability of two choices is always understood. Note that $\widetilde{S}$ is dimensionless, so that we can also write $\widetilde{S} (\mathbf{a},\mathbf{m},\ve_1;\qe,z) = \widetilde{S}  (\a,\boldsymbol\th ;\qe,z) $ by using the dimensionless parameters \eqref{eq:dimlessparm}. The crucial observation made in \cite{LLNZ2013} is that when the limit $\ve_2 \to 0$ is applied to \eqref{eq:defeq} it reduces to the Hamilton-Jacobi equation for Painlev\'{e} VI
\begin{align} \label{eq:HJeq}
H^+ \left( z, \frac{\partial \widetilde{S}}{\partial z} ; \qe \right) + \frac{\partial \widetilde{S}}{\partial \qe}=0,
\end{align}
where the Hamiltonian is given by
\begin{align} \label{eq:hamz}
\begin{split}
H^+(z,p;\qe) =& \frac{z(z-\qe)(z-1)}{\qe(\qe-1)} p \left( p- \frac{2\th_0}{z} - \frac{2\th_\qe}{z-\qe} -\frac{2\th_1}{z-1} \right) \\ &+\frac{z(\th_0+\th_\qe+\th_1+\th_\infty)(\th_0 + \th_\qe+\th_1-\th_\infty)}{\qe(\qe-1)}.
\end{split}
\end{align}
It is crucial to note that even though the equation of motion for this Hamiltonian implies $z(\qe)$ obeys Painlev\'{e} VI, it is not exactly $w(\qe)$ that we have seen in section \ref{sec:pvi}. By directly comparing their Hamiltonians $H^+(z,p;\qe)$ \eqref{eq:hamw} and $H(w,p_w;\qe)$ \eqref{eq:hamz}, we notice that there is following half-integer shift in $\theta$-parameters from $w(\qe)$ to $z(\qe)$,
\begin{align} \label{eq:shift}
\begin{split}
&\th_q \longrightarrow \th_q + \frac{1}{2} \\
&\th_\infty \longrightarrow \th_\infty -\frac{1}{2}.
\end{split}
\end{align}
This simple shift will actually play an important role in section \ref{sec:ok} in deriving the GIL relation and exactly identifying the monodromies of the associated Fuchsian system in terms of the $\a$ and $\b$ parameters.

\subsection{Generating function of Riemann-Hilbert symplectomorphism} \label{sec:genfun}
In the previous section, we have seen that the asymptotics $\widetilde{S} (\a,\boldsymbol\th,z;\qe)$ of the surface defect partition function in the NS limit $\ve_2 \to 0$ is identified with the Hamilton-Jacobi potential for the Painlev\'{e} VI. In this section, we will verify that $\widetilde{S} (\a,\boldsymbol\th,z;\qe)$ can also be viewed as the generating function of the Riemann-Hilbert map, which is a symplectomorphism from the moduli space of $\mathfrak{sl}(2)$ Fuchsian system on a Riemann surface to the space of representations of its fundamental group into $SL(2)$ given by the monodromies.

We define the Riemann-Hilbert map
\begin{align} \label{eq:rhmap}
\begin{split}
\text{RH} : \mathcal{E}_\qe \left( \th_0, \th_\qe+\frac{1}{2} ,\th_1,\th_\infty -\frac{1}{2} \right) &\longrightarrow \EuScript{M}_\qe \left( \th_0, \th_\qe+\frac{1}{2} ,\th_1,\th_\infty -\frac{1}{2} \right) \\
(z,p)& \longmapsto (\a,\b)
\end{split}
\end{align}
by
\begin{align}
p = \frac{\p \widetilde{S}}{\p z}, \quad \beta = \frac{\p \widetilde{S}}{\p \a}.
\end{align}
The first equation implicitly determines $\a$ in terms of $z$ and $p$, and then the second equation determines $\b$ in terms of $z$ and $p$. We can show that $(\a, \b)$ provides a local coordinate system on the monodromy space $ \EuScript{M}_\qe \left( \th_0, \th_\qe+\frac{1}{2} ,\th_1,\th_\infty -\frac{1}{2} \right)$ as follows. By taking the $\qe$-derivatives of these equations, we get
\begin{align}
\begin{split}
\frac{d p(\qe)}{d\qe} &= \left. \left( \frac{\p}{\p z} \frac{\p \widetilde{S}}{\p \qe} \right) \right\vert_{z=z(\qe)} + \left. \frac{\p ^2 \widetilde{S}}{\p z^2} \right\vert_{z=z(\qe)} \frac{dz(\qe)}{d\qe} + \left. \frac{\p^2 \widetilde{S}}{\p \a \p z} \right\vert_{z=z(\qe)}  \frac{d\a}{d\qe}\\
&= - \left. \frac{\p H^+(z,p;\qe)}{\p z} \right\vert_{\substack{z=z(\qe) \\ p=p(\qe)}} + \left. \frac{\p^2 \widetilde{S}}{\p \a \p z} \right\vert_{z=z(\qe)}  \frac{d\a}{d\qe},
\end{split}
\end{align}
and
\begin{align}
\begin{split}
&\frac{d \b}{d \qe} =\left. \left(  \frac{\p}{\p \a} \frac{\p \widetilde{S}}{\p \qe} \right) \right\vert_{z=z(\qe)} + \left. \frac{\p ^2 \widetilde{S} }{\p \a \p z}\right\vert_{z=z(\qe)} \frac{dz(\qe)}{d \qe} + \left. \frac{\p^2 \widetilde{S}}{\p \a^2} \right\vert_{z=z(\qe)} \frac{d\a}{d\qe} \\ 
&= \left. \frac{\p^2 \widetilde{S}}{\p \a \p z} \right\vert_{z=z(\qe)}  \left( \left. -\frac{\p H^+(z,p;\qe)}{\p p}\right\vert_{\substack{z=z(\qe) \\ p=p(\qe)}} + \frac{dz(\qe)}{d\qe} \right) + \left. \frac{\p^2 \widetilde{S}}{\p \a^2} \right\vert_{z=z(\qe)} \frac{d\a}{d\qe}.
\end{split}
\end{align}
These two equations imply $\a$ and $\b$ are constants if and only if $(z(\qe),p(\qe))$ is a solution to the Hamiltonian equations of motion, namely, an isomonodromic flow. Therefore, $(\a,\b)$ indeed parametrizes the monodromies of the $\mathfrak{sl}(2)$ Fuchsian system on the four-punctured sphere, and forms a local coordinate system on the monodromy space $ \EuScript{M}_\qe \left( \th_0, \th_\qe+\frac{1}{2} ,\th_1,\th_\infty -\frac{1}{2} \right)$. Also it is straightforward to show that the map defined in this way preserves the symplectic structure. In this sense, $\widetilde{S}$ is the generating function of the Riemann-Hilbert map \eqref{eq:rhmap}.

It is not clear yet whether $\a$ and $\b$ are precisely the NRS coordinates on $ \EuScript{M}_\qe \left( \th_0, \th_\qe+\frac{1}{2} ,\th_1,\th_\infty -\frac{1}{2} \right)$, which parametrize the trace invariants of the monodromies by
\begin{align}
&\Tr \, M_A = - 2 \cos 2\pi \a
\end{align}
and
\begin{align}
\begin{split}
\text{Tr}\, M_B &= \frac{\left( -\cos 2\pi \th_\infty +\cos 2\pi \th_1 \right) \left( \cos 2\pi \th_0 - \cos 2\pi \th_\qe \right)}{2 \sin ^2 \pi \a} \\
& - \frac{\left( \cos 2\pi \th_\infty +\cos 2\pi \th_1 \right) \left( \cos 2\pi \th_0 + \cos 2\pi \th_\qe \right)}{2 \cos ^2 \pi \a} \\
& + \sum_{\pm} 4 \frac{\prod_{\epsilon=\pm} \sin \pi (\mp \a - \th_\qe +\epsilon \th_0) \sin \pi (\mp \a -\th_1 +\epsilon \th_\infty) }{\sin ^2 2\pi \a}  e^{\pm\b} .
\end{split}
\end{align}
Be aware of the half-integer shifts in the $\th$-parameters compared to \eqref{eq:nrsa} and \eqref{eq:nrsb}. To verify that $\a = \frac{a}{\ve_1}$ and $\b = \frac{\p \widetilde{S}}{\p \a}$ are indeed the NRS Darboux coordinates, we need an explicit construction of the flat section of the Fuchsian system in the gauge theory context, from which we can compute the monodromies along the loops in $\pi_1 \left( \mathbb{P}^1 \setminus \{0,\qe,1,\infty\}\right)$ and express them in terms of $\a =\frac{a}{\ve_1}$ and $\b = \frac{\p \widetilde{S}}{\p \a}$. We will perform such a construction in the next section.

\section{Intersecting surface defects and monodromy data} \label{sec:monodromy}
Recall that our goal is to describe the isomonodromic tau function in gauge theoretical language. Being the constants of isomonodromic deformations, the monodromy data of the Fuchsian system explicitly appear in the expression of the tau function. Hence, it is important to know how such monodromy data are encoded in gauge theoretical terms.

In turn, we first need to construct the horizontal section of the Fuchsian system, as well as the Hamilton-Jacobi action of the previous section, in gauge theoretical terms. The relevant gauge theory setting turns out to be the intersecting surface defects that we have constructed in section \ref{sec:constint}.

In this section, we study the non-perturbative Dyson-Schwinger equations for the intersecting surface defect partition function and their implications. In particular, it is shown that the partition function satisfies a differential equation which reduces to the Fuchsian differential equation associated to the Fuchsian system in the NS limit $\ve_2 \to 0$. Next, we study how the intersecting surface defects partition functions analytically continue to each other. The monodromy data of the Fuchsian system are finally obtained in gauge theoretical terms by by concaternating such analytic continuations and taking the limit $\ve_2 \to 0$. 

\subsection{Intersecting surface defects and non-perturbative Dyson-Schwinger equations} \label{subsec:intnonpert}
Recall that we constructed the intersecting surface defects by starting from the $A_3$-quiver gauge theory and then partially Higgsing the gauge group. The fundamental $qq$-characters for the $A_3$-quiver gauge theory read as follows:
\begin{align} \label{eq:fundqqa3}
\begin{split}
&\EuScript{X}_1 (x) = \EuScript{Y}_1 (x+\ve) +\qe_1 \frac{\EuScript{Y}_0 (x) \EuScript{Y}_2 (x+\ve)}{\EuScript{Y}_1 (x)} +\qe_1 \qe_2 \frac{\EuScript{Y}_0 (x) \EuScript{Y}_3 (x+\ve)}{\EuScript{Y}_2 (x)} + \qe_1 \qe_2 \qe_3 \frac{ \EuScript{Y}_0 (x) \EuScript{Y}_4 (x+\ve)}{\EuScript{Y}_3 (x)} \\
&\EuScript{X}_2 (x) = \EuScript{Y}_2 (x+\ve) + \qe_2 \frac{\EuScript{Y}_1 (x) \EuScript{Y}_3 (x+\ve)}{\EuScript{Y}_2 (x)} +\qe_1 \qe_2 \frac{\EuScript{Y}_ 0 (x-\ve) \EuScript{Y}_3 (x+\ve)}{\EuScript{Y}_1 (x-\ve)} +\qe_2 \qe_3 \frac{\EuScript{Y}_1 (x) \EuScript{Y}_4 (x+\ve)}{\EuScript{Y}_3 (x)} \\
&\quad\quad\quad\quad+\qe_1 \qe_2 \qe_3 \frac{\EuScript{Y}_0 (x-\ve) \EuScript{Y}_2 (x) \EuScript{Y}_4 (x+\ve)}{\EuScript{Y}_1 (x-\ve) \EuScript{Y}_3 (x)} + \qe_1 \qe_2 ^2 \qe_3 \frac{\EuScript{Y}_0 (x-\ve) \EuScript{Y}_4(x+\ve)}{\EuScript{Y}_2(x-\ve)} \\
&\EuScript{X}_3 (x) = \EuScript{Y}_3 (x+\ve) +\qe_3 \frac{\EuScript{Y}_2 (x) \EuScript{Y}_4 (x+\ve)}{\EuScript{Y}_3 (x)} +\qe_2 \qe_3 \frac{ \EuScript{Y}_1 (x-\ve) \EuScript{Y}_4 (x+\ve)}{\EuScript{Y}_2 (x-\ve)} +\qe_1 \qe_2 \qe_3 \frac{\EuScript{Y}_0 (x-2\ve) \EuScript{Y}_4 (x+\ve)}{\EuScript{Y}_1 (x-2\ve)}.
\end{split}
\end{align}
The non-perturbative Dyson-Schwinger equations follow from the regularity of their expectation values. More precisely, we have
\begin{align} \label{eq:npdsa3}
\left[ x^{-n} \right] \Big\langle \EuScript{X}_i (x) \Big\rangle =0, \quad i=1,2,3,\;\; n\geq 1.
\end{align}
Now to construct the intersecting surface defects, we specialize some of the Coulomb moduli in the $A_3$-quiver gauge theory to initiate partial higging, as we have seen in section \ref{sec:constint},
\begin{align} \label{eq:higgscons}
\begin{split}
&a_{1,\a} = a_{0,\a} - \delta_{\a, h} \ve_1 \\
&a_{3,\a} = a_{4,\a} -\ve - \delta_{\a, l} \ve_2.
\end{split}
\end{align}
We have seen that these constraints partially higgs the gauge group and produce intersecting surface defects coupled to the bulk $U(2)$ gauge theory with four hypermultiplets. The intersecting surface defect partition function $\Upsilon_{l,h} ^{0<\vert y \vert <\vert \qe \vert <1 <\vert z \vert}  $ \eqref{eq:intsurf0yq1z} obtained in this way converges in the domain $0<\vert y \vert <\vert \qe\vert<1<\vert z \vert$. Now with these constraints imposed, we have simplified $\EuScript{Y}$-observables,
\begin{align}
\begin{split}
&\EuScript{Y}_1 (x) = \EuScript{Y}_0 (x) \frac{x-a_{0,h} +\ve_1 -k_1 \ve_2}{x-a_{0,h} -k_1 \ve_2}\\
&\EuScript{Y}_3 (x) = \EuScript{Y}_4 (x+\ve) \frac{x-a_{3,l} -k_3 \ve_1}{x-a_{3,l} -\ve_2 -k_3 \ve_1}.
\end{split}
\end{align}
Substituting these to the fundamental $qq$-characters \eqref{eq:fundqqa3}, we can obtain non-trivial identities satisfied by the partition function from the non-perturbative Dyson-Schwinger equations \eqref{eq:npdsa3}. More precisely, the intersecting partition function is a particular specialization of the case considered in \cite{Nekrasov_BPS45,Jeong2017}, where it was proven that the surface defect partition function solves the null-vector decoupling equation for the corresponding degenerate Liouville conformal block \cite{BPZ}. This differential equation was investigated in the conformal field theory point of view in \cite{LN2017}, but it is important not to resort to any CFT argument in our approach. We do not reproduce the derivation here, only the result is stated.

First, let us re-define the intersecting surface defect partition function with the following perturbative prefactors:
\begin{align}
\widehat{\Upsilon}_{l,h} ^{0<\vert y \vert <\vert \qe \vert <1 <\vert z \vert}  = \prod_{i=0} ^3 z_i ^{L_i} \prod_{0\leq i<j\leq 3} \left( 1-\frac{z_j}{z_i} \right) ^{T_{ij}} \Upsilon_{l,h} ^{0<\vert y \vert <\vert \qe \vert <1 <\vert z \vert} ,
\end{align}
where $z_0= z$, $z_1=1$, $z_2=\qe$, $z_3=y$, and
\begin{align} \label{eq:n2expnents}
\begin{split}
&L_i \equiv \frac{(a_{i+1,1}-a_{i+1,2})^2 - (a_{i,1}-a_{i,2})^2}{4 \varepsilon_1 \varepsilon_2} + \frac{ (\bar{a}_i -\bar{a}_{i+1} +\varepsilon )(\bar{a}_i -\bar{a}_{i+1})}{\varepsilon_1 \varepsilon_2} , \quad i=0,1,2,3 \\
&T_{ij} = \frac{2(\bar{a}_j -\bar{a}_{j+1} + \varepsilon) (\bar{a}_i -\bar{a}_{i+1} ) }{\varepsilon_1 \varepsilon_2}, \quad i,j=0,1,2,3.
\end{split}
\end{align}
With the higgsing constraint \eqref{eq:higgscons}, the prefactors simply so we can write
\begin{align}
\begin{split}
\widehat{\Upsilon}_{l,h} ^{0<\vert y \vert <\vert \qe \vert <1 <\vert z \vert} &= z^{\frac{-a_{0,h}+a_{0,\bar{h}} +2\ve_1+\ve_2}{2\ve_2}} \left( \frac{y}{\qe} \right)^{\frac{a_{4,l}-a_{4,\bar{l}} +\ve}{2\ve_1}} \qe ^{-\Delta_0 -\Delta_\qe +\frac{\ve^2 -(a_{2,1}-a_{2,2})^2}{4\ve_1 \ve_2} +\frac{2\ve_1 +3\ve_2}{4\ve_1} }\\
& \left( 1-\frac{1}{z} \right)^{\frac{2\bar{a}_0 -2\bar{a}_2 +\ve_1+2\ve_2}{2\ve_2}} \left( 1-\frac{\qe}{z} \right)^{\frac{2\bar{a}_0 -2\bar{a}_4 +4\ve_1 +5\ve_2}{2\ve_2}} \left( 1-\frac{y}{z} \right)^{-\frac 1 2} (1-y)^{\frac{-2\bar{a}_0 +2\bar{a}_2 +\ve_1}{2\ve_1}} \\
& (1-\qe)^{\frac{(2\bar{a}_0 -2\bar{a}_2 -\ve_1)(2\bar{a}_2 -2\bar{a}_4 +4\ve_1 +5\ve_2)}{2\ve_1 \ve_2}} \left( 1- \frac y \qe \right)^{\frac{-2\bar{a}_2+2\bar{a}_4 -2\ve_1-3\ve_2}{2\ve_1}} \Upsilon_{l,h} ^{0<\vert y \vert <\vert \qe \vert <1 <\vert z \vert}.
\end{split}
\end{align}
Here, we used the notation $\bar{a}_i = \frac{a_{i,1}+a_{i,2}}{2}$ and also defined
\begin{align} \label{eq:deltas}
\begin{split}
&\Delta_0 = \frac{\ve^2 - (a_{4,1}-a_{4,2})^2}{4\ve_1 \ve_2} \, , \ \Delta_\infty =\frac{\ve^2 - (a_{0,1}-a_{0,2})^2}{4\ve_1 \ve_2} \\
&\Delta_\qe = -\frac{(2\bar{a}_2-2\bar{a}_4 +2\ve_1+3\ve_2)(2\bar{a}_2 -2\bar{a}_4 +4\ve_1+5\ve_2)}{4\ve_1\ve_2} \\ 
&\Delta_1 = -\frac{(2\bar{a}_0-2\bar{a}_2-\ve_1)(2\bar{a}_0 -2\bar{a}_2 +\ve_1+2\ve_2)}{4\ve_1 \ve_2}\\
&\Delta_L = -\frac 1 2 -\frac{3\ve_2}{4\ve_1} \, , \  \Delta_H = -\frac 1 2 -\frac{3\ve_1}{4\ve_2}.
\end{split}
\end{align}
Then the modified partition function $\widehat{\Upsilon}_{l,h} ^{0<\vert y \vert <\vert \qe \vert <1 <\vert z \vert}$ satisfies the following differential equations
\begin{subequations} \label{eq:bpz}
\begin{align}
 0=&  \left[ \frac{\ve_2}{{\ve}_{1}} \partial ^2 _z -  \left( \frac{1}{z} +\frac{1}{z-1} \right) \partial_z +  \frac{\qe(\qe-1)}{z(z-\qe)(z-1)} \partial_\qe +  \frac{y(y-1)}{z(z-y)(z-1)} \partial_y \ + \right. \\
 & \left. \qquad \qquad + \ \frac{\Delta_\infty -\Delta_0-\Delta_\qe-\Delta_1-\Delta_L-\Delta_H}{z(z-1)}\  + \right. \nonumber\\
  & \left. \qquad \qquad \qquad \qquad  + \  \frac{\Delta_0}{z^2} +\frac{\Delta_\qe}{(z-\qe)^2} +\frac{\Delta_1}{(z-1)^2} +\frac{\Delta_L}{(z-y)^2}   \right] \ \widehat{\Upsilon}_{l,h} ^{0<\vert y \vert <\vert \qe \vert <1 <\vert z \vert},  \\
  0=  & \left[ \frac{\ve_1}{\ve_2} \partial ^2 _y -  \left( \frac{1}{y} +\frac{1}{y-1} \right) \partial_y  +  \frac{\qe(\qe-1)}{y(y-\qe)(y-1)} \partial_\qe + \frac{z(z-1)}{y(y-z)(y-1)} \partial_z + \right. \nonumber \\
    &\left.  \qquad \qquad + \ \frac{\Delta_\infty -\Delta_0-\Delta_\qe-\Delta_1-\Delta_L-\Delta_H}{y(y-1)} \ + \right. \\
    & \left.  \qquad\qquad\qquad\qquad + \frac{\Delta_0}{y^2} +\frac{\Delta_\qe}{(y-\qe)^2} +\frac{\Delta_1}{(y-1)^2} +\frac{\Delta_H}{(y-z)^2}   \right] \ \widehat{\Upsilon}_{l,h} ^{0<\vert y \vert <\vert \qe \vert <1 <\vert z \vert}.
\end{align}
\end{subequations}
In accordance with the correspondence proposed in \cite{agt} these equations are precisely the BPZ equations \cite{BPZ} satisfied by the Liouville conformal block with the insertion of two degenerate fields at $y$ and $z$. Since we are not resorting to any CFT argument throughout the paper, we shall not describe here the CFT context of these equations in detail. Note that the equations do not have explicit dependence on the choice of intersecting surface defects $l$ and $h$, so that we have four solutions in total for each $l,h \in \{1,2\}$.

Next, let us consider the other extremal domain, $0<\vert z \vert <\vert \qe\vert<1<\vert y \vert$. As we have seen in section \ref{sec:constint}, we construct the intersecting surface defect partition function $\Upsilon_{l,h} ^{0<\vert z \vert <\vert \qe \vert <1 <\vert y \vert} $ in this domain by the higgsing
\begin{align}
\begin{split}
&a_{1,\a} = a_{0,\a} - \delta_{\a, l} \ve_2 \\
&a_{3,\a} = a_{4,\a} -\ve - \delta_{\a, h} \ve_1,
\end{split}
\end{align}
followed by the re-definition of parameters,
\begin{align}
\begin{split}
&a_{0,\a} \longrightarrow -a_{0,\a} -\ve \\
&a_{2,\a} \longrightarrow -a_{2,\a} \\
&a_{4,\a} \longrightarrow -a_{4,\a} +3\ve. 
\end{split}
\end{align}
Then the regularity of the expectation values of the $qq$-characters \eqref{eq:fundqqa3} implies the following differential equation
\begin{subequations} \label{eq:bpz'}
\begin{align}
 0=&  \left[ \ve_2 ^2 \partial ^2 _z - \ve_1 \ve_2 \left( \frac{1}{z} +\frac{1}{z-1} \right) \partial_z +\ve_1 \ve_2 \frac{\qe(\qe-1)}{z(z-\qe)(z-1)} \partial_\qe +\ve_1 \ve_2 \frac{y(y-1)}{z(z-y)(z-1)} \partial_y \right. \nonumber\\
  & \left. +\ve_1 \ve_2 \left( \frac{\Delta_0}{z^2} +\frac{\Delta'_\qe}{(z-\qe)^2} +\frac{\Delta'_1}{(z-1)^2} +\frac{\Delta_L}{(z-y)^2} + \frac{\Delta_\infty -\Delta_0-\Delta'_\qe-\Delta'_1-\Delta_L-\Delta_H}{z(z-1)} \right) \right]  \nonumber \\
& \widehat{\Upsilon}_{l,h} ^{0<\vert z \vert <\vert \qe \vert <1 <\vert y \vert}, \\
  0=  & \left[\ve_1 ^2 \partial ^2 _y - \ve_1 \ve_2 \left( \frac{1}{y} +\frac{1}{y-1} \right) \partial_y +\ve_1 \ve_2 \frac{\qe(\qe-1)}{y(y-\qe)(y-1)} \partial_\qe +\ve_1 \ve_2 \frac{z(z-1)}{y(y-z)(y-1)} \partial_z \right. \nonumber \\
    &\left.+\ve_1 \ve_2 \left( \frac{\Delta_0}{y^2} +\frac{\Delta'_\qe}{(y-\qe)^2} +\frac{\Delta'_1}{(y-1)^2} +\frac{\Delta_H}{(y-z)^2} + \frac{\Delta_\infty -\Delta_0-\Delta'_\qe-\Delta'_1-\Delta_L-\Delta_H}{y(y-1)} \right) \right] \nonumber\\
&\widehat{\Upsilon}_{l,h} ^{0<\vert z \vert <\vert \qe \vert <1 <\vert y \vert},
\end{align}
\end{subequations}
satisfied by the modified partition function
\begin{align}
\begin{split}
\widehat{\Upsilon}_{l,h} ^{0<\vert z \vert <\vert \qe \vert <1 <\vert y \vert} &= y^{\frac{a_{0,l}-a_{0,\bar{l}} +\ve_1+2\ve_2}{2\ve_1}} \left( \frac{z}{\qe} \right)^{\frac{-a_{4,h}+a_{4,\bar{h}} +\ve}{2\ve_2}} \qe ^{-\Delta'_0 -\Delta'_\qe +\frac{\ve^2 -(a_{2,1}-a_{2,2})^2}{4\ve_1 \ve_2} +\frac{3\ve_1 +2\ve_2}{4\ve_2} }\\
& \left( 1-\frac{1}{y} \right)^{\frac{-2\bar{a}_0 +2\bar{a}_2 -\ve_2}{2\ve_1}} \left( 1-\frac{\qe}{y} \right)^{\frac{-2\bar{a}_0 +2\bar{a}_4 -3\ve_1 -4\ve_2}{2\ve_1}} \left( 1-\frac{z}{y} \right)^{-\frac 1 2} (1-z)^{\frac{2\bar{a}_0 -2\bar{a}_2 +2\ve_1+3\ve_2}{2\ve_2}} \\
& (1-\qe)^{\frac{(-2\bar{a}_0 +2\bar{a}_2 -2\ve_1-3\ve_2)(-2\bar{a}_2 +2\bar{a}_4 -\ve_1 -2\ve_2)}{2\ve_1 \ve_2}} \left( 1- \frac z \qe \right)^{\frac{2\bar{a}_2-2\bar{a}_4 +3\ve_1+4\ve_2}{2\ve_2}} \Upsilon_{l,h} ^{0<\vert z \vert <\vert \qe \vert <1 <\vert y \vert},
\end{split}
\end{align}
where we defined
\begin{align} \label{eq:deltas'}
\begin{split}
&\Delta' _\qe = -\frac{(2\bar{a}_2 -2\bar{a}_4 +\ve_1+2\ve_2)(2\bar{a}_2 -2\bar{a}_4 +3\ve_1+4\ve_2)}{4\ve_1\ve_2)} \\
&\Delta'_1 = -\frac{(2\bar{a}_0-2\bar{a}_2 +\ve_2)(2\bar{a}_0-2\bar{a}_2 +2\ve_1+3\ve_2)}{4\ve_1\ve_2}.
\end{split}
\end{align}
Note that the parameters \eqref{eq:deltas} become identical to \eqref{eq:deltas'} after the shift of the Coulomb moduli,
\begin{align} 
a_{2,\a} \longrightarrow a_{2,\a} -\d_{\a,l} \ve_2 -\d_{\a,h} \ve_1.
\end{align}
Hence, the differential equations \eqref{eq:bpz} become \eqref{eq:bpz'} after this shift. The intersecting surface defect partition functions $\widehat{\Upsilon}_{l,h} ^{0<\vert y \vert <\vert \qe \vert <1 <\vert z \vert}$ and $\widehat{\Upsilon}_{l,h} ^{0<\vert z \vert <\vert \qe \vert <1 <\vert y \vert}$ define solutions to these equations in the respective domain, up to this shift of the Coulomb moduli.

\subsection{Hamilton-Jacobi equation and Fuchsian differential equation}
In the previous section, we expressed the differential equations satisfied by the intersecting surface defect partition functions in the form of the BPZ equation for the corresponding degenerate Liouville conformal block. For the purpose of the present work, however, it is useful to properly re-define the partition function with extra multiplicative prefactor and re-write the differential equations accordingly. In particular, we show in this section that the NS limit $\ve_2 \to 0$ of those differential equations provide the Hamilton-Jacobi equation and the Fuchsian differential equation associated to the isomonodromy problem in section \ref{sec:pre}. We modify the intersecting surface defect partition function to set these equations into the form that we prefer. We emphasize that the modification is non-essential and merely conventional.

We modify the partition function by the following re-definition,
\begin{align}
\begin{split}
&\widetilde{\Upsilon}_{l,h} (\mathbf{a}_2,\mathbf{a}_0,\mathbf{a}_4,\ve_1,\ve_2;\qe,z,y) \\
&= \left( \frac{ y(y-\qe)(y-1)}{y-z} \right)^{\frac 1 2} z^{-\frac{a_{4,1}-a_{4,2} +\ve_1}{2\ve_2}} (1-z)^{-\frac{2\bar{a}_0 -2\bar{a}_2 -\ve_1}{2\ve_2}} (z-\qe)^{-\frac{\bar{a}_4 -\bar{a}_2 -2\ve}{\ve_2}} \\
&  \qe^{\frac{-2(a_{4,1}-a_{4,2})(\bar{a}_4 -\bar{a}_2 -2\ve_1-\ve_2) +2\ve_1 (\bar{a}_4 -\bar{a}_2 -2\ve)-\ve_2^2}{4\ve_1 \ve_2}} \\
& (1-\qe)^{\frac{2(a_{4,1}-\bar{a}_2 -2\ve)^2 +(a_{0,2}-\bar{a}_2) (a_{4,1}-a_{4,2}) +(a_{0,1}-\bar{a}_2) (2a_{0,2} -2\bar{a}_2+a_{4,1}-a_{4,2}) + 2\ve_1(\bar{a}_4 -\bar{a}_2 -2\ve)}{2\ve_1 \ve_2} + \frac{2(2\bar{a}_0 +2a_{4,1}-a_{4,2}-4\bar{a}_2 -2\ve) -\ve_1}{2\ve_1} +\frac{\ve_2}{\ve_1} } \\
& \times \widehat{\Upsilon}_{l,h} (\mathbf{a}_2,\mathbf{a}_0,\mathbf{a}_4,\ve_1,\ve_2;\qe,z,y).
\end{split}
\end{align}
Also, let us re-define the hypermultiplet masses so that they only appear as fundamentals:
\begin{align}
m_{\a}:= a_{0,\a}, \quad m_{\a+2}:= a_{4,\a}-2\ve, \quad \a=1,2,
\end{align}
while we omit the subscript of the Coulomb moduli:
\begin{align}
a_\a := a_{2,\a} , \quad\quad \a=1,2.
\end{align}
Then the differential equations \eqref{eq:bpz} are modified accordingly:
\begin{align} \label{eq:bpzheavy}
\begin{split}
0&=\left[ \ve_2 ^2 \partial_z ^2  -\left( \ve_1 \ve_2 \left( \frac{2\th_0}{z}+\frac{2\th_\qe}{z-\qe}+\frac{2\th_1}{z-1} \right) -\frac{\ve_2 ^2}{y-z} \right) \p_z +\ve_1 \ve_2 \frac{\th_\qe +\frac{1}{2} +\frac{\ve_2}{4\ve_1}}{(z-\qe)^2} \right. \\
&\quad +\ve_1 \ve_2 \frac{y(y-1)}{z(z-y)(z-1)} \left( \p_y +\frac{\th_0 }{y}+\frac{\th_\qe+\frac{1}{2}}{y-\qe} +\frac{\th_1 }{y-1} \right) + \ve_1 \ve_2 \frac{\qe(\qe-1)}{z(z-\qe)(z-1)}\left( \frac{\p}{\p \qe} -\frac{\th_\qe +\frac{1}{2}}{y-\qe} \right)   \\
&\quad+\frac{\ve_1 ^2}{z(z-\qe)(z-1)} \left\{ z\left( \th_0 +\th_\qe +\th_1+\th_\infty  \right) \left( \th_0 +\th_\qe +\th_1-\th_\infty \right)  + \frac{z \ve_2}{\ve_1} \left( 2\th_0 +\th_\qe+2\th_1 -\frac{1}{2} \right) \right. \\
&\left. \left. \quad\quad\quad\quad\quad\quad\quad\quad\quad -\frac{3\ve_2^2}{4\ve_1^2} z  +\frac{\qe \ve_2}{z\ve_1} \left( \th_0 +\frac{\ve_2}{4 \ve_1} \right) \right\} \right] \widetilde\Upsilon (\mathbf{a},\mathbf{m},\ve_1,\ve_2;\qe,z,y)
\end{split}
\end{align}
and
\begin{align} \label{eq:bpzlight}
\begin{split}
0&=\left[ \ve_1 ^2 \partial_y ^2 + \ve_1 ^2 \left( \frac{1}{y} +\frac{1}{y-\qe} +\frac{1}{y-1} -\frac{1}{y-z} - \frac{\ve_2}{\ve_1} \left( \frac{1}{y} + \frac{1}{y-1} \right) \right) \p_y \right. \\
&\quad -\ve_1 ^2 \left( \frac{\th_0 ^2-\frac{\ve_2 ^2}{4\ve_1 ^2}}{y^2} + \frac{\left(\th_\qe +\frac{1}{2}\right)^2 -\frac{\ve_2 ^2}{4\ve_1 ^2}}{(y-\qe)^2}  + \frac{\th_1 \left( \th_1 +\frac{\ve_2}{\ve_1} \right)}{(y-1)^2}  + \frac{\th_\infty ^2 -\th_0 ^2 -\left( \th_\qe +\frac{1}{2} \right)^2 -\th_1 ^2 -\frac{1}{4} +\frac{\ve_2}{\ve_1} (\th_1 -1) +\frac{\ve_2 ^2}{2 \ve_1 ^2} }{y(y-1)} \right) \\
&\quad  +\ve_1 ^2 \frac{z(z-1)}{y(y-z)(y-1)} \left( \frac{\ve_2}{\ve_1} \p_z -\frac{\th_0}{z} -\frac{\th_\qe +\frac{1}{2}}{z-\qe} -\frac{\th_1}{z-1} \right) \\
& \left.\quad  +\ve_1 ^2 \frac{\qe(\qe-1)}{y(y-\qe)(y-1)} \left(\frac{\ve_2}{\ve_1} \frac{\p}{\p \qe} + \frac{\th_\qe +\frac{1}{2}}{z-\qe} - \frac{\th_\qe +\frac{1}{2} - 2 \th_0 \th_\qe  -\qe \left( (\th_0 + \th_1)^2 +(\th_\qe+\th_\infty +1)(\th_\qe-\th_\infty+1) \right)}{\qe(\qe-1)} \right.\right. \\
& \left. \left. \quad\quad\quad\quad\quad\quad\quad\quad\quad\quad +\frac{\ve_2}{\ve_1} \frac{\th_0 + \qe\left( 2\th_0 +2\th_1 +\th_\qe -\frac{1}{2} \right)}{\qe(\qe-1)} + \frac{\ve_2 ^2}{\ve_1 ^2} \frac{1+ 3\qe}{4 \qe(\qe-1)} \right)  \right] \widetilde\Upsilon  (\mathbf{a},\mathbf{m},\ve_1,\ve_2;\qe,z,y),
\end{split}
\end{align}
where we have used the dimensionless $\th$-parameters \eqref{eq:dimlessparm}. 

Now, let us consider the NS limit $\ve_2 \to 0$ of these equations. The asymptotics of the partition function is governed by the twisted superpotential $\widetilde{S}_h$, which received contribution only from the defect on the $z_2$-plane, while the regular part $\chi_l$ gets contribution only from the defect on the $z_1$-plane. Here, $l,h\in \{1,2\}$ enumerates the choices of the surface defects on the $z_1$-plane and the $z_2$-plane, respectively. More explicitly, we can write
\begin{align}
 \widetilde{\Upsilon}_{l,h} (\mathbf{a},\mathbf{m},\ve_1,\ve_2;\qe,z,y) = \exp \left( \frac{\ve_1}{\ve_2} \widetilde{S}_h (\mathbf{a},\mathbf{m},\ve_1;\qe ,z) \right) \left( \chi_l ( \mathbf{a},\mathbf{m},\ve_1;\qe ,y) + \mathcal{O} (\ve_2) \right).
\end{align}
Note that the presence of the additional surface defect on the $z_1$-plane does not affect the singular part $ \widetilde{S}(\mathbf{a},\mathbf{m},\ve_1;\qe ,z)$, so that it is identical to the one \eqref{eq:heavyasymp} which appears in the asymptotics of the partition function with a single surface defect on the $z_2$-plane. However, the regular part is affected by the presence of the additional defect on the $z_1$-plane and provides a non-trivial function $\chi ( \mathbf{a},\mathbf{m},\ve_1;\qe ,y)$.

Taking the NS limit $\ve_2 \to 0$ to \eqref{eq:bpzheavy}, we recover the Hamilton-Jacobi equation \eqref{eq:HJeq} that we have seen in section \ref{subsec:HJeq},
\begin{align}
H^+\left( z, \frac{\partial \widetilde{S}}{\partial z} ; \qe \right) + \frac{\partial \widetilde{S}}{\partial \qe}=0,
\end{align}
where the Hamiltonian is given by
\begin{align} \label{eq:pviz}
\begin{split}
H^+(z,p;\qe) =& \frac{z(z-\qe)(z-1)}{\qe(\qe-1)} p \left( p- \frac{2\th_0}{z} - \frac{2\th_\qe}{z-\qe} -\frac{2\th_1}{z-1} \right) \\
&+\frac{z(\th_0+\th_\qe+\th_1+\th_\infty)(\th_0 + \th_\qe+\th_1-\th_\infty)}{\qe(\qe-1)}.
\end{split}
\end{align}
This is precisely the Painlev\'{e} VI Hamiltonian \eqref{eq:hamz} that we obtained in section \ref{subsec:HJeq} from a single surface defect on the $z_2$-plane. We emphasize again that this Hamiltonian defines an isomonodromic flow $z(\qe)$ which is different from the isomonodromic flow $w(\qe)$ defined by $H(w,p_w;\qe)$ in \eqref{eq:hamred}, since there is the half-integer shift \eqref{eq:shift} in $\th$-parameters in $H^+(z,p;\qe)$ compared to $H(w,p_w;\qe)$.

On the other hand, by taking the NS limit $\ve_2 \to 0$ to \eqref{eq:bpzlight} we get
\begin{align} 
\begin{split}
0&= \left[ \p_y ^2 + \left( \frac{1}{y} +\frac{1}{y-\qe}+\frac{1}{y-1} -\frac{1}{y-z}   \right) \p_y \right.\\
& \quad -\frac{\th_0^2}{y^2} -\frac{\left(\th_\qe +\frac{1}{2} \right)^2}{(y-\qe)^2} -\frac{\th_1 ^2}{(y-1)^2} - \frac{\th_\infty ^2 -\th_0 ^2 -\left(\th_\qe+\frac{1}{2}\right) ^2 -\th_1 ^2 -\frac{1}{4}}{y(y-1)} \\
& \quad+\frac{z(z-1)}{y(y-z)(y-1)}\left( \frac{\p \widetilde{S}}{\p z}  -\frac{\th_0}{z} -\frac{\th_\qe+\frac{1}{2}}{z-\qe} -\frac{\th_1}{z-1} \right) \\
&  \quad  - \frac{\qe(\qe-1)}{y(y-\qe)(y-1)}\left( -\frac{\p \widetilde{S}}{\p \qe}  -\frac{\th_\qe+\frac{1}{2}}{z-\qe} \right. \\ 
&\left.\left. \quad\quad \quad\quad\quad\quad\quad+\frac{\th_\qe +\frac{1}{2} - 2 \th_0 \th_\qe  -\qe \left( (\th_0 + \th_1)^2 +(\th_\qe+\th_\infty +1)(\th_\qe-\th_\infty+1) \right)}{\qe(\qe-1)} \right) \right] \chi(y),
\end{split}
\end{align}
or, recalling that $p = \frac{\p \widetilde{S}}{\p z}$ and $H^+\left( z,\frac{\p \widetilde{S}}{\p z};\qe \right) = - \frac{\p \widetilde{S}}{\p \qe}$,
\begin{align} \label{eq:horibpz}
\begin{split}
0&= \left[ \p_y ^2 + \left( \frac{1}{y} +\frac{1}{y-\qe}+\frac{1}{y-1} -\frac{1}{y-z}   \right) \p_y \right.\\
& \quad -\frac{\th_0^2}{y^2} -\frac{\left(\th_\qe +\frac{1}{2} \right)^2}{(y-\qe)^2} -\frac{\th_1 ^2}{(y-1)^2} - \frac{\th_\infty ^2 -\th_0 ^2 -\left(\th_\qe+\frac{1}{2}\right) ^2 -\th_1 ^2 -\frac{1}{4}}{y(y-1)} \\
& \quad+\frac{z(z-1)}{y(y-z)(y-1)}\left( p -\frac{\th_0}{z} -\frac{\th_\qe+\frac{1}{2}}{z-\qe} -\frac{\th_1}{z-1} \right) \\
&  \quad  - \frac{\qe(\qe-1)}{y(y-\qe)(y-1)}\left(H^+ (z,p;\qe) -\frac{\th_\qe+\frac{1}{2}}{z-\qe} \right. \\ 
&\left.\left. \quad\quad \quad\quad\quad\quad\quad+\frac{\th_\qe +\frac{1}{2} - 2 \th_0 \th_\qe  -\qe \left( (\th_0 + \th_1)^2 +(\th_\qe+\th_\infty +1)(\th_\qe-\th_\infty+1) \right)}{\qe(\qe-1)} \right) \right] \chi(y).
\end{split}
\end{align}
This Fuchsian differential equation is almost identical to \eqref{eq:horisecond}, which we directly got from the $\mathfrak{sl}(2)$ Fuchsian system on the four-punctured sphere with the local monodromies around the punctures fixed by the $\th$-parameters, but not exactly. Just as the Painlev\'{e} VI Hamiltonian \eqref{eq:pviz} has shifts \eqref{eq:shift} in $\th$-parameters compared to \eqref{eq:hamred}, the equations \eqref{eq:horisecond} and \eqref{eq:horibpz} differ by the same half-integer shift in $\th$-parameters \eqref{eq:shift}. Hence the solutions to these equations also define distinct monodromies along the loops in $\pi_1 (\mathbb{P}^1 \setminus \{ 0,\qe,1,\infty\})$. More precisely, the monodromy data of the Fuchsian differential equation \eqref{eq:horibpz} takes its value in $\EuScript{M}_\qe \left( \th_0, \th_\qe+\frac{1}{2}, \th_1,\th_\infty -\frac{1}{2} \right)$, while the monodromy data of \eqref{eq:horisecond} takes its value in $ \EuScript{M}_\qe  (\boldsymbol\th)=\EuScript{M}_\qe \left( \th_0, \th_\qe, \th_1,\th_\infty \right)$.

\subsection{Analytic continuation of the intersecting surface defects expectation value}
In the previous section, we have seen that the expectation value of the intersecting surface defects has a regular part in the NS limit $\ve_2 \to 0$, which solves the second-order Fuchsian differential equation associated to the $\mathfrak{sl}(2)$ Fuchsian system on a Riemann surface. Hence we can obtain the monodromy of the solutions to this equation by first computing the monodromy of the intersecting surface defect partition functions along given loops in the fundamental group and then taking the limit $\ve_2 \to 0$. This is analogous to the procedure in \cite{JN2018} of computing the monodromy data of opers.

The multi-valuedness of the surface defect expectation value is worth some discussion. The ``classical'' monodromy, under the simple transport $z \mapsto e^{2\pi \ii}z$, e.g. in the domain $0< \vert z \vert < \vert \qe\vert < 1 <\vert y \vert$, is due to the way we chose to normalize the surface defect operator in the $\Omega$-background. A typical normalization in the equivariant gauged  $A$-model on the $\Omega$-deformed disk (or cigar) contains the factor $\propto z^{{\sigma}/{\hbar}}$, where $\hbar$ is the $\Omega$-deformation parameter corresponding to the rotational symmetry of the cigar/disk, and $\sigma$ is the scalar in the vector multiplet corresponding to some gauge $U(1)$-symmetry, and $z$ is the exponentiated complex FI parameter corresponding to that symmetry. With boundary conditions at infinity (or the boundary of the disk) selecting a specific fixed point of the global symmetry action on the target space, the value of $\sigma$ becomes fixed in terms of the twisted masses -- the equivariant parameters of that global symmetry. In terms of the non-linear sigma model data, the role of the $\sigma$ is played by the value at the fixed point of the function, which completes the closed two-form, representing the K{\"a}hler class, to the equivariantly closed class of degree $2$.

To enable the computation of monodromy we need to understand how the expectation values of the intersecting surface defects analytically continue across the different convergence domains. In the $S$-class theory associated with the Riemann surface $\EuScript{C}$ the K\"{a}hler moduli of the sigma models living on the surface defect belongs to $\EuScript{C}$ \cite{gai1}. Strictly speaking, this statement has not been derived in quantum field theory. Our analysis is probably the best one hope for in the case of a Lagrangian $S$-class theory. We shall see that, indeed, for the $A_1$-type theory, the K{\"a}hler moduli of the surface defects belong to the $4$-punctured sphere $\EuScript{C} = {\BP}^{1} \backslash \{ 0, {\qe}, 1, \infty \}$. The Riemann surface $\EuScript{C}$ is divided into several convergence domains for the partition function, so that within each domain the partition function has an instanton expansion, both in the bulk gauge coupling and in the exponentiated complex FI parameters. The complexified FI parameters of both surface defects, the one on the $z_1$-plane and the one on the $z_2$-plane, can be adiabatically transported along $\EuScript{C}$. The corresponding non-abelian Berry phase is what we are after. 

\subsubsection{Adiabatic $z$-transport of the $z_1 = 0$ surface defect} \label{subsec:flowz2def}

\begin{figure}
\centering
\resizebox{4 in}{4 in}{
\begin{tikzpicture}
\path[black] (0,-9) edge[->] (0,9)  (-9,0) edge[->] (9,0);
\node[cross out,draw=black] at (0,0) {}; \node[cross out,draw=black] at (1,0) {}; \node[cross out,draw=black] at (2,0) {};\node[cross out,draw=black] at (3,0) {};\node[cross out,draw=black] at (4,0) {}; \node[cross out,draw=black] at (5,0) {};\node[cross out,draw=black] at (6,0) {};

\node[cross out,draw=black] at (3.2,5.6) {}; \node[cross out,draw=black] at (2.2,5.6) {};\node[cross out,draw=black] at (1.2,5.6) {};\node[cross out,draw=black] at (0.2,5.6) {};\node[cross out,draw=black] at (-0.8,5.6) {};\node[cross out,draw=black] at (-1.8,5.6) {};\node[cross out,draw=black] at (-2.8,5.6) {};\node[cross out,draw=black] at (-3.8,5.6) {};\node[cross out,draw=black] at (-4.8,5.6) {}; 
\node[circle,fill,inner sep=1pt] at (-2,4.6) {};
\node[circle,fill,inner sep=1pt] at (-2,3.8) {};
\node[circle,fill,inner sep=1pt] at (-2,3.0) {};
\node[cross out,draw=black] at (2.1,2) {};\node[cross out,draw=black] at (1.1,2) {};\node[cross out,draw=black] at (0.1,2) {}; \node[cross out,draw=black] at (-0.9,2) {}; \node[cross out,draw=black] at (-1.9,2) {};\node[cross out,draw=black] at (-2.9,2) {};\node[cross out,draw=black] at (-3.9,2) {}; \node[cross out,draw=black] at (-4.9,2) {};

\node[cross out,draw=black] at (4.6,-2.4) {}; \node[cross out,draw=black] at (3.6,-2.4) {};\node[cross out,draw=black] at (2.6,-2.4) {};\node[cross out,draw=black] at (1.6,-2.4) {}; \node[cross out,draw=black] at (0.6,-2.4) {}; \node[cross out,draw=black] at (-0.4,-2.4) {}; \node[cross out,draw=black] at (-1.4,-2.4) {};\node[cross out,draw=black] at (-2.4,-2.4) {};\node[cross out,draw=black] at (-3.4,-2.4) {};\node[cross out,draw=black] at (-4.4,-2.4) {};

\node[circle,fill,inner sep=1pt] at (-2,-3.2) {};
\node[circle,fill,inner sep=1pt] at (-2,-3.8) {};
\node[circle,fill,inner sep=1pt] at (-2,-4.4) {};
\node[cross out,draw=black] at (5.9,-5) {}; \node[cross out,draw=black] at (4.9,-5) {};\node[cross out,draw=black] at (3.9,-5) {};\node[cross out,draw=black] at (2.9,-5) {};\node[cross out,draw=black] at (1.9,-5) {};\node[cross out,draw=black] at (0.9,-5) {}; \node[cross out,draw=black] at (-0.1,-5) {}; \node[cross out,draw=black] at (-1.1,-5) {};\node[cross out,draw=black] at (-2.1,-5) {};\node[cross out,draw=black] at (-3.1,-5) {};\node[cross out,draw=black] at (-4.1,-5) {};

\draw[blue,very thick,decoration={markings, mark=at position 0.6 with {\arrow[scale=2]{latex}}}, postaction={decorate}] (0,-9) -- (0,-5.3); \draw[blue,very thick] (0,-5.3) -- (5.9,-5.3);
\draw[blue,very thick] (5.9,-5.3) arc (-90:90:0.3) -- (5.9,-4.7); \draw[blue,very thick] (5.9,-4.7) -- (0,-4.7);\draw[blue,very thick] (0,-4.7) -- (0,-4.2); 
\node[blue,circle,fill,inner sep=1pt] at (0,-4.0) {}; \node[blue,circle,fill,inner sep=1pt] at (0,-3.75) {};\node[blue,circle,fill,inner sep=1pt] at (0,-3.5) {};
\draw[blue,very thick] (0,-3.3) -- (0,-2.7); \draw[blue,very thick] (0,-2.7) -- (4.6,-2.7);
\draw[blue,very thick] (4.6,-2.7) arc (-90:90:0.3) -- (4.6,-2.1);\draw[blue,very thick] (4.6,-2.1) -- (0,-2.1);  \draw[blue,very thick,decoration={markings, mark=at position 0.6 with {\arrow[scale=2]{latex}}}, postaction={decorate}] (0,-2.1) -- (0,-0.3);  \draw[blue,very thick] (0,-0.3) arc (270:90:0.3) -- (0,0.3);  \draw[blue,very thick,decoration={markings, mark=at position 0.6 with {\arrow[scale=2]{latex}}}, postaction={decorate}] (0,0.3) -- (0,1.7); \draw[blue,very thick] (0,1.7) -- (2.1,1.7); 
\draw[blue, very thick] (2.1,1.7) arc (-90:90:0.3) -- (2.1,2.3); \draw[blue,very thick] (2.1,2.3) -- (0,2.3); \draw[blue,very thick] (0,2.3) -- (0,3);
\node[blue,circle,fill,inner sep=1pt] at (0,3.4) {}; \node[blue,circle,fill,inner sep=1pt] at (0,3.8) {};\node[blue,circle,fill,inner sep=1pt] at (0,4.2) {}; 
\draw[blue,very thick] (0,4.6) -- (0,5.3); \draw[blue,very thick] (0,5.3) -- (3.2,5.3);
\draw[blue,very thick] (3.2,5.3) arc (-90:90:0.3) -- (3.2,5.9);  \draw[blue,very thick] (3.2,5.9) -- (0,5.9); \draw[blue, very thick,decoration={markings, mark=at position 0.6 with {\arrow[scale=2]{latex}}}, postaction={decorate}] (0,5.9) -- (0,9); 

\draw[red, very thick,decoration={markings, mark=at position 0.3 with {\arrow[scale=2]{latex}}, mark=at position 0.7 with {\arrow[scale=2]{latex}}  }, postaction={decorate}] (0,9) arc (90:-90:9) -- (0,-9);
\node at (8,6) {\scalebox{2.5}{$\textcolor{red}{\mathcal{R}_+}$}};

\draw[teal,very thick,decoration={markings, mark=at position 0.3 with {\arrow[scale=2]{latex}}, mark=at position 0.7 with {\arrow[scale=2]{latex}}  }, postaction={decorate}] (0,9) arc (90:270:9) -- (0,-9);
\node at (-8,6) {\scalebox{2.5}{$\textcolor{teal}{\mathcal{R}_-}$}};

\node[circle,fill,inner sep=1pt] at (6.6,0) {};\node[circle,fill,inner sep=1pt] at (7.2,0) {};\node[circle,fill,inner sep=1pt] at (7.8,0) {};
\node[circle,fill,inner sep=1pt] at (-5.4,5.6) {};\node[circle,fill,inner sep=1pt] at (-5.8,5.6) {};\node[circle,fill,inner sep=1pt] at (-6.2,5.6) {};
\node[circle,fill,inner sep=1pt] at (-5.5,2) {};\node[circle,fill,inner sep=1pt] at (-5.9,2) {};\node[circle,fill,inner sep=1pt] at (-6.3,2) {};
\node[circle,fill,inner sep=1pt] at (-5,-2.4) {};\node[circle,fill,inner sep=1pt] at (-5.4,-2.4) {};\node[circle,fill,inner sep=1pt] at (-5.8,-2.4) {};
\node[circle,fill,inner sep=1pt] at (-4.7,-5) {};\node[circle,fill,inner sep=1pt] at (-5.1,-5) {};\node[circle,fill,inner sep=1pt] at (-5.5,-5) {};

\node at (-0.7,-1.2) {\scalebox{2.5}{$\textcolor{blue}{\mathcal{C}}$}};
\node at (4.8,4.9) {$\frac{a_{2,1}-a_{0,h}}{\varepsilon_2}+  \lambda^{(1)}_1 -1 $};
\node at (7.4,-5.8) {$\frac{a_{2,N}-a_{0,h}}{\varepsilon_1}+  \lambda^{(N)} _1  -1 $};
\node at (0,-0.5) {$0$}; \node at (1,-0.5) {$1$};\node at (2,-0.5) {$2$};
\end{tikzpicture}
}
\caption{The contour $\mathcal{C}$ on the $\frac{x-a_{0,h}}{\varepsilon_2}$-plane.} \label{fig1}
\end{figure}
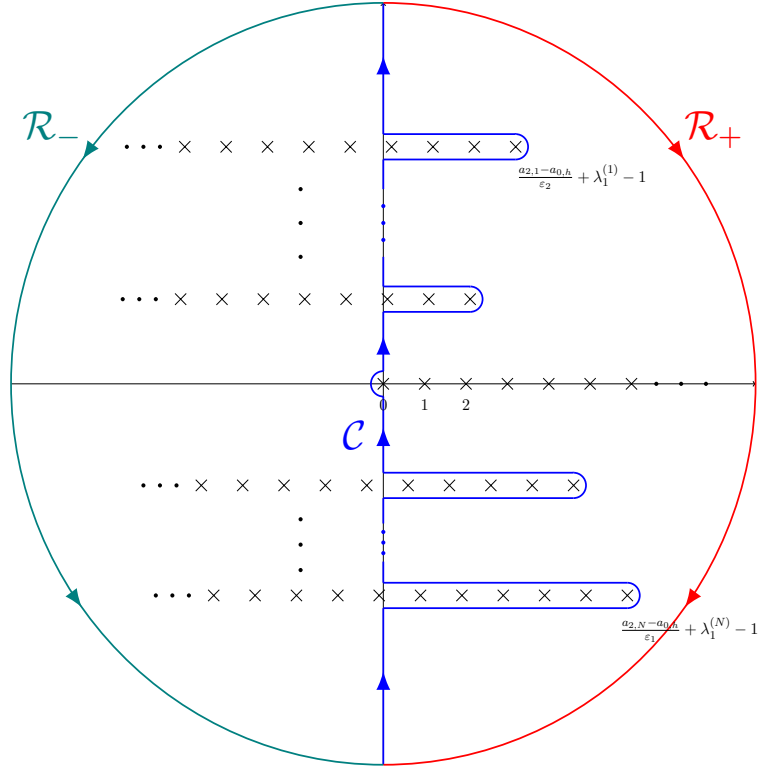

We start from the intersecting surface defect partition function in the domain $0<\vert y\vert<\vert\qe\vert<1<\vert z\vert$. Here, $z$ and $y$ are the complexified K{\"a}hler parameters of the effective two dimensional sigma models on the $z_2$-plane and the  $z_1$-plane, respectively. As we have seen in the section \ref{sec:constint}, the partition function can be viewed as the two-point function of the two surface defect observables $\EuScript{O}_1$ and $\EuScript{O}_2$. Then we can write out $\EuScript{O}_2$ to obtain
\begin{align} \label{eq:intpart1}
\begin{split}
&\Upsilon_{l,h} ^{0<\vert y\vert<\vert\qe\vert<1<\vert z\vert} \\
& = \sum_{\boldsymbol\l} \qe^{\vert \boldsymbol\l \vert} E\left[ \mathcal{T}_{A_1} [\boldsymbol\l] \right] \EuScript{O}_{1,l} ^{0<\vert y\vert <\vert \qe \vert} [\boldsymbol\l] \\
&\quad \sum_{k =0} ^\infty z^{-k} \prod_{\a =1 }^N \frac{\Gamma\left(k+1 +\frac{a_{0,h} -a_{2,\a}}{\ve_2} \right) \Gamma \left( 1+\frac{a_{0,h} -a_{0,\a}}{\ve_2} \right)}{\Gamma\left( 1+\frac{a_{0,h} -a_{2,\a}}{\ve_2} \right) \Gamma \left( k+1 + \frac{a_{0,h} -a_{0,\a}}{\ve_2} \right)} \prod_{\Box \in \boldsymbol\l} \frac{a_{0,h} +k\ve_2 -c_\Box -\ve_1}{a_{0,h} +k\ve_2 -c_\Box}\\
&=\sum_{\boldsymbol\l} \qe^{\vert \boldsymbol\l \vert} E\left[ \mathcal{T}_{A_1} [\boldsymbol\l] \right] \EuScript{O}_{1,l} ^{0<\vert y\vert <\vert \qe \vert} [\boldsymbol\l]\\
&\quad \sum_{k =0} ^\infty z^{-k} \prod_{\a =1 }^N \frac{\Gamma\left(k+1 -\l^{(\a)} _1 +\frac{a_{0,h} -a_{2,\a}}{\ve_2} \right) \Gamma \left( 1+\frac{a_{0,h} -a_{0,\a}}{\ve_2} \right)}{\Gamma\left( 1+\frac{a_{0,h} -a_{2,\a}}{\ve_2} \right) \Gamma \left( k+1 + \frac{a_{0,h} -a_{0,\a}}{\ve_2} \right)} \\
&\quad\quad\quad\quad\quad \prod_{ j=1 } ^{\l ^{(\a)}_1} \frac{a_{0,h}  -a_{2,\a} -\l^{(\a) t} _j \ve_1 +(k-j+1)\ve_2}{\ve_2}.
\end{split}
\end{align}
We can represent this partition function by the following contour integral,
\begin{align}
\begin{split}
&\Upsilon_{l,h} ^{0<\vert y\vert<\vert\qe\vert<1<\vert z\vert}\\
& = - \prod_{\a=1} ^N \frac{\Gamma \left( 1+\frac{a_{0,h} -a_{0,\a}}{\ve_2} \right)}{\Gamma \left( 1+\frac{a_{0,h} -a_{2,\a} }{\ve_2} \right)} (-z)^{\frac{a_{0,h}}{\ve_2}} \sum_{\boldsymbol\l} \qe^{\vert \boldsymbol\l \vert} E\left[ \mathcal{T}_{A_1}[\boldsymbol\l] \right] \EuScript{O}_{1,l} ^{0<\vert y\vert <\vert \qe \vert} [\boldsymbol\l] \\
& \int_{\mathcal{C}} dx \; (-z)^{-\frac{x}{\ve_2}} \frac{\Gamma \left( -\frac{x-a_{0,h}}{\ve_2} \right) \prod_{\a =1} ^N\Gamma \left( 1 - \l^{(\a)} _1+\frac{x-a_{2,\a}}{\ve_2} \right) }{\prod_{\a \neq h} \Gamma \left( 1+ \frac{x -a_{0,\a}}{\ve_2} \right)} \prod_{\a=1} ^N \prod_{j=1} ^{\l_1 ^{(\a)}} \frac{x-a_{2,\a} -\l_j ^{(\a)t} \ve_1 -(j-1)\ve_2}{\ve_2},
\end{split}
\end{align}
where the contour $\mathcal{C}$ is chosen as in figure \ref{fig1}. 
The series expansion \eqref{eq:intpart1} is recovered once we add to the contour a semicircle $\mathcal{R}_+$ closing to the right. It can be shown, for $\vert z \vert >1$,  that the integral along the semicircle goes to zero as the radius goes to infinity. Then we just pick up the residues from the simple poles on the right, recovering the series expansion \eqref{eq:intpart1} that we started with.

Now, let us adiabatically move $z$ while keeping $y$ fixed to get to the domain $\vert\qe\vert <\vert z \vert< 1$. Then it can be shown that the integral along the semicircle $\mathcal{R}_-$ on the left now converges to zero, so that we can close the contour in the opposite way. Then we pick up the residues from the simple poles on the left to obtain
\begin{align}
\Upsilon_{l,h} ^{0<\vert y\vert<\vert\qe\vert<1<\vert z\vert} = \sum_{h' =1 } ^N \prod_{\g \neq h} \frac{\Gamma \left( 1+\frac{a_{0,h} -a_{0,\g}}{\ve_2} \right)}{\Gamma \left( \frac{a_{2,h'}-a_{0,\g} }{\ve_2} \right)} \prod_{\a \neq h'} \frac{\Gamma \left( \frac{a_{2,h'} -a_{2,\a} }{\ve_2} \right)}{\Gamma \left( 1+\frac{a_{0,h} -a_{2,\a}}{\ve_2} \right)} z (-z)^{\frac{a_{0,h} -a_{2,h'}}{\ve_2}} \Upsilon_{l,h'} ^{0<\vert y\vert<\vert\qe\vert< \vert z \vert<1},
\end{align}
where we defined the instanton part of the intersecting surface defect partition function in the domain $0<\vert y \vert <\vert\qe\vert <\vert z \vert< 1$ by
\begin{align} 
\begin{split}
&\Upsilon_{l, h'} ^{0<\vert y \vert <\vert\qe\vert <\vert z \vert< 1} = \sum_{\boldsymbol\l} \qe ^{\vert \boldsymbol\l \vert} E[ \mathcal{T}_{A_1} [\boldsymbol\l] ] \EuScript{O}_{1,l} ^{0<\vert y\vert <\vert \qe \vert} [\boldsymbol\l] \EuScript{O}_{2,h'} ^{\vert \qe \vert <\vert z \vert <1},
\end{split}
\end{align}
where we defined the surface defect observable 
\begin{align} \label{eq:intsurf2}
\begin{split}
\EuScript{O}_{2,h'} ^{\vert \qe \vert <\vert z \vert <1} &= \sum_{k=0} ^\infty z^{k- \l^{(h')}_1} \frac{(-1)^k}{k!} \\
&\prod_{\a\neq h'} \frac{\Gamma \left( -k +\l ^{(h')} _1 -\l ^{(\a)} _1 + \frac{a_{2,h'} -a_{2,\a}}{\ve_2} \right)}{\Gamma \left( \frac{a_{2,h'} -a_{2,\a}}{\ve_2} \right)} \prod _{\d =1 }^N \frac{\Gamma \left( \frac{a_{2,h'} -a_{0,\d}}{\ve_2} \right)}{\Gamma \left( -k + \l ^{(h')}_1 + \frac{a_{2,h'} -a_{0,\d}}{\ve_2} \right)} \\
& \prod_{\delta=1} ^N \prod_{j=1} ^{\l^{(\d)} _1} \frac{a_{2,h'} -a_{2,\d} +\left(\l^{(h')}_1 -k -j \right) \ve_2 -\l^{(\d)t}_j \ve_1 }{\ve_2}.
\end{split}
\end{align}
We can also appropriately define the modified partition function $\Upsilon_{l,h} ^{0<\vert y \vert <\vert \qe\vert<\vert z \vert< 1}$ by
\begin{align}
\begin{split}
&\widehat{\Upsilon}_{l,h} ^{0<\vert y \vert <\vert \qe\vert<\vert z \vert< 1}\\
 &= z^{\frac{-a_{2,h} +a_{2,\bar{h}} +\ve}{2\ve_2}} \left(\frac y \qe \right)^{\frac{a_{4,l} -a_{4,\bar{l}} +\ve}{2\ve_1}} \qe^{-\Delta_0 -\Delta_\qe +\frac{\ve^2 -(a_{2,1}-a_{2,2})^2}{4\ve_1\ve_2} +\frac{2\ve_1+3\ve_2}{4\ve_1}}  \\
& (1-z)^{\frac{2\bar{a}_0-2\bar{a}_2 +\ve_1+2\ve_2}{2\ve_2}} \left( 1-\frac \qe z \right)^{\frac{2\bar{a}_2 -2\bar{a}_4 +4\ve_1+5\ve_2}{2\ve_2}} \left( 1- \frac y z \right) ^{-\frac 1 2} (1-y)^{\frac{-2\bar{a}_0 +2\bar{a}_2+\ve_1}{2\ve_1} } \\
& (1-\qe)^{\frac{(2\bar{a}_0 -2\bar{a}_2 -\ve_1)(2\bar{a}_2 -2\bar{a}_4 +4\ve_1+5\ve_2)}{2\ve_1\ve_2}} \left( 1-\frac y \qe \right)^{\frac{-2\bar{a}_2 +2\bar{a}_4 -2\ve_1 -3\ve_2}{2\ve_1}}  \Upsilon_{l,h} ^{0<\vert y \vert <\vert \qe\vert <\vert z \vert<1},
\end{split}
\end{align}
so that
\begin{align}
\widehat{\Upsilon}_{l,h} ^{0<\vert y\vert<\vert\qe\vert<1<\vert z\vert} = \sum_{h' =1 } ^N \prod_{\g \neq h} \frac{\Gamma \left( 1+\frac{a_{0,h} -a_{0,\g}}{\ve_2} \right)}{\Gamma \left( \frac{a_{2,h'}-a_{0,\g} }{\ve_2} \right)} \prod_{\a \neq h'} \frac{\Gamma \left( \frac{a_{2,h'} -a_{2,\a} }{\ve_2} \right)}{\Gamma \left( 1+\frac{a_{0,h} -a_{2,\a}}{\ve_2} \right)}  \widehat{\Upsilon}_{l,h'} ^{0<\vert y\vert<\vert\qe\vert< \vert z \vert<1}.
\end{align}
It is immediate that $\widehat{\Upsilon}_{l,h} ^{0<\vert y\vert<\vert\qe\vert< \vert z\vert<1}$ would solve differential equations in gauge couplings satisfied by $\widehat{\Upsilon}_{l,h} ^{0<\vert y\vert<\vert\qe\vert<1<\vert z\vert} $. In the case of $N=2$, these differential equations are \eqref{eq:bpzheavy} and \eqref{eq:bpzlight}. The above formula is nothing but the connection formula for the analytic continuation of solutions to these differential equations. Let us define the connection matrix
\begin{align}
\left( \mathbf{C} ^{(2)}_\infty \right)_{hh'} = \prod_{\g \neq h} \frac{\Gamma \left( 1+\frac{a_{0,h} -a_{0,\g}}{\ve_2} \right)}{\Gamma \left( \frac{a_{2,h'}-a_{0,\g} }{\ve_2} \right)} \prod_{\a \neq h'} \frac{\Gamma \left( \frac{a_{2,h'} -a_{2,\a} }{\ve_2} \right)}{\Gamma \left( 1+\frac{a_{0,h} -a_{2,\a}}{\ve_2} \right)},
\end{align}
where the superscript $(2)$ indicates that it is relevant to the adiabatic flow of the surface defect on the $z_2$-plane. Then the above connection formula can be written simply as
\begin{align}
\widehat{\Upsilon}_{l,h} ^{0<\vert y\vert<\vert\qe\vert<1<\vert z\vert} = \sum_{h' =1 } ^N \left( \mathbf{C} ^{(2)}_\infty \right)_{hh'} \widehat{\Upsilon}_{l,h'} ^{0<\vert y\vert<\vert\qe\vert< \vert z \vert<1}.
\end{align}

Next, we can initiate an adiabatic flow of $z$ starting from the domain $0<\vert z \vert <\vert \qe \vert < 1<\vert y \vert$. We have the intersecting surface defect partition function $\Upsilon ^{0<\vert z \vert <\vert \qe \vert < 1<\vert y \vert}$ lying in this domain. We may adiabatically flow $z$ to the domain $0<\vert \qe \vert<\vert z \vert  < 1<\vert y \vert$. By a procedure similar to the one described above, we get the connection formula
\begin{align}
\begin{split}
\Upsilon_{l,h} ^{0<\vert z \vert <\vert \qe \vert < 1<\vert y \vert} = \sum_{h'=1} ^N \prod_{\g \neq h} \frac{\Gamma \left( 1+\frac{a_{4,\g} -a_{4,h}}{\ve_2} \right)}{\Gamma\left( \frac{a_{4,\g} -a_{2,h'}-\ve}{\ve_2} \right)} \prod_{\a\neq h'} \frac{\Gamma \left( \frac{a_{2,\a}-a_{2,h'}}{\ve_2} \right)}{\Gamma \left( 1+\frac{a_{2,\a}-a_{4,h}+\ve}{\ve_2} \right)} \frac{\qe}{z}\left( -\frac{z}{\qe} \right)^{\frac{a_{4,h} -a_{2,h'}-\ve}{\ve_2}} \Upsilon_{l,h'} ^{0 <\vert \qe \vert <\vert z \vert< 1<\vert y \vert},
\end{split}
\end{align}
where the intersecting surface defect partition function in the domain $0 <\vert \qe \vert <\vert z \vert< 1<\vert y \vert$ is defined by
\begin{align}
\begin{split}
&\Upsilon_{l,h} ^{0 <\vert \qe \vert <\vert z \vert< 1<\vert y \vert}  =  \sum_{\boldsymbol\l} \qe ^{\vert \boldsymbol\l \vert} E[ \mathcal{T}_{A_1} ' [\boldsymbol\l] ] {\EuScript{O}'}_{1,l} ^{1<\vert y\vert} [\boldsymbol\l]  {\EuScript{O}'}_{2,h} ^{\vert \qe \vert < \vert z \vert<1} [\boldsymbol\l],
\end{split}
\end{align}
where the new surface defect observable is
\begin{align}
\begin{split}
{\EuScript{O}'}_{2,h} ^{\vert \qe \vert < \vert z \vert<1} [\boldsymbol\l] &=\sum_{k=0} ^\infty \left(\frac{\qe}{z}\right) ^{k-\l_1 ^{(h)}} \frac{(-1)^k}{k!} \\
&\prod_{\a \neq h} \frac{\Gamma\left(-k+\l_1 ^{(h)} -\l_1 ^{(\a)}+\frac{a_{2,\a}-a_{2,h}}{\ve_2}\right)}{\Gamma \left( \frac{a_{2,\a}-a_{2,h}}{\ve_2} \right)} \prod_{\d=1} ^N \frac{\Gamma\left( \frac{a_{4,\d}-a_{2,h}-\ve}{\ve_2} \right)}{\Gamma \left( -k +\l_1 ^{(h)} +\frac{a_{4,\d}-a_{2,h} -\ve}{\ve_2} \right)} \\
&\prod_{\d=1} ^N \prod_{j=1} ^{\l_1 ^{(\d)}} \frac{-a_{2,h} +a_{2,\d} -\l_j ^{(\d) t}\ve_1 -(k-j-\l_1 ^{(h)})\ve_2}{\ve_2}.
\end{split}
\end{align}
We can properly modify the partition function in the domain $0 <\vert \qe \vert <\vert z \vert< 1<\vert y \vert$ by 
\begin{align}
\begin{split}
\widehat{\Upsilon}_{l,h} ^{0 <\vert \qe \vert <\vert z \vert< 1<\vert y \vert} &= y^{\frac{a_{0,l}-a_{0,\bar{l}} +\ve_1+2\ve_2}{2\ve_1}} \left( \frac{\qe}{z} \right)^{\frac{a_{2,h}-a_{2,\bar{h}} -2\ve_1-\ve_2}{2\ve_2}} \qe ^{-\Delta'_0 -\Delta'_\qe +\frac{\ve^2 -(a_{2,1}-a_{2,2})^2}{4\ve_1 \ve_2} +\frac{3\ve_1 +2\ve_2}{4\ve_2} }\\
& \left( 1-\frac{1}{y} \right)^{\frac{-2\bar{a}_0 +2\bar{a}_2 -\ve_2}{2\ve_1}} \left( 1-\frac{\qe}{y} \right)^{\frac{-2\bar{a}_0 +2\bar{a}_4 -3\ve_1 -4\ve_2}{2\ve_1}} \left( 1-\frac{z}{y} \right)^{-\frac 1 2} (1-z)^{\frac{2\bar{a}_0 -2\bar{a}_2 +2\ve_1+3\ve_2}{2\ve_2}} \\
& (1-\qe)^{\frac{(-2\bar{a}_0 +2\bar{a}_2 -2\ve_1-3\ve_2)(-2\bar{a}_2 +2\bar{a}_4 -\ve_1 -2\ve_2)}{2\ve_1 \ve_2}} \left( 1- \frac \qe z \right)^{\frac{2\bar{a}_2-2\bar{a}_4 +3\ve_1+4\ve_2}{2\ve_2}} \Upsilon_{l,h} ^{0 <\vert \qe \vert <\vert z \vert< 1<\vert y \vert}.
\end{split}
\end{align}
Then the connection formula reads
\begin{align}
\Upsilon_{l,h} ^{0<\vert z \vert <\vert \qe \vert < 1<\vert y \vert} = \sum_{h'=1} ^N \left( \mathbf{C}_0 ^{(2)} \right)_{hh'} \widehat{\Upsilon}_{l,h'} ^{0 <\vert \qe \vert <\vert z \vert< 1<\vert y \vert},
\end{align}
where we defined the connection matrix
\begin{align}
\left( \mathbf{C}_0 ^{(2)} \right)_{hh'} = \prod_{\g \neq h} \frac{\Gamma \left( 1+\frac{a_{4,\g} -a_{4,h}}{\ve_2} \right)}{\Gamma\left( \frac{a_{4,\g} -a_{2,h'}-\ve}{\ve_2} \right)} \prod_{\a\neq h'} \frac{\Gamma \left( \frac{a_{2,\a}-a_{2,h'}}{\ve_2} \right)}{\Gamma \left( 1+\frac{a_{2,\a}-a_{4,h}+\ve}{\ve_2} \right)}.
\end{align}

\subsubsection{$y$-transporting the surface defect at $z_2 = 0$} \label{sec:ytrans}
Now we start from the intersecting surface defect partition function $\Upsilon_{l,h} ^{0<\vert y \vert <\vert \qe \vert < \vert z\vert<1} $ in the domain $0<\vert y \vert <\vert \qe \vert < \vert z\vert<1$ and move the complexified K{\"a}hler parameter $y$ of the effective sigma model on the $z_1$-plane to another domain $0<\vert \qe \vert <\vert y \vert<\vert z \vert<1$. We write out the observable $\EuScript{O}_{1,\b}$ in the partition sum as follows:
\begin{align}
\begin{split}
&\Upsilon_{l,h} ^{0<\vert y \vert <\vert \qe \vert < \vert z\vert<1} \\
&= \sum_{\boldsymbol\l} \qe^{\vert \boldsymbol\l\vert} E[\mathcal{T}_{A_1} [\boldsymbol\l]] \EuScript{O}_{2,h} ^{\vert \qe \vert< \vert z \vert <1} [\boldsymbol\l] \\
&\quad\quad\sum_{k=0} ^\infty \left( \frac{y}{\qe}  \right)^k \prod_{\a=1} ^N \frac{\Gamma \left( k+1 -l\left( \l^{(\a)} \right) +\frac{a_{4,l}-2\ve-a_{2,\a}}{\ve_1} \right) \Gamma \left( 1+\frac{a_{4,l}-a_{4,\a}}{\ve_1} \right)}{\Gamma\left( 1+\frac{a_{4,l}-2\ve-a_{2,\a}}{\ve_1} \right) \Gamma \left( k+1 +\frac{a_{4,l}-a_{4,\a}}{\ve_1} \right) }\\
&\quad\quad\quad\quad\quad \prod_{i=1}^{l\left( \l^{(\a)} \right)} \frac{a_{4,l}-2\ve -a_{2,\a} +(k-i+1)\ve_1 -\l_i ^{(\a)} \ve_2}{\ve_1}
\end{split}
\end{align}
By using a similar trick of contour integral, we can find the following analytic continuation formula
\begin{align}
\Upsilon_{l,h} ^{0<\vert y \vert <\vert \qe \vert < \vert z\vert<1}  = \sum_{\a =1 }^N \prod_{\b \neq l} \frac{\Gamma \left( 1+\frac{a_{4,l}-a_{4,\b}}{\ve_1} \right)}{\Gamma \left( \frac{a_{2,l'} +2\ve-a_{4,\b}}{\ve_1} \right)} \prod_{\a \neq l'} \frac{\Gamma \left( \frac{a_{2,l'}-a_{2,\a}}{\ve_1} \right)}{\Gamma \left(1+\frac{a_{4,l} -2\ve -a_{2,\a}}{\ve_1}\right)} \frac{\qe}{y} \left( -\frac{y}{\qe} \right)^{\frac{a_{2,l'}-a_{4,l}+2\ve}{\ve_1}} \Upsilon_{l',h} ^{0<\vert \qe \vert <\vert y \vert< \vert z\vert<1} ,
\end{align}
where we defined the analytically continued partition function by
\begin{align} \label{eq:intsurfint1}
\begin{split}
&\Upsilon_{l',h} ^{0<\vert \qe \vert <\vert y \vert< \vert z\vert<1} = \sum_{\boldsymbol\l} \qe^{\vert \boldsymbol\l\vert} E[\mathcal{T}_{A_1} [\boldsymbol\l]] \EuScript{O}_{2,h} ^{\vert \qe \vert < \vert z \vert <1} [\boldsymbol\l] \EuScript{O}_{1,l'} ^{\vert \qe \vert<\vert y \vert <1} [\boldsymbol\l],
\end{split}
\end{align}
with the new surface defect observable
\begin{align} \label{eq:intsurf1}
\begin{split}
\EuScript{O}_{1,l'} ^{\vert \qe \vert<\vert y \vert <1}  [\boldsymbol\l]&= \sum_{k=0 }^\infty \left(\frac{\qe}{y}\right)^{k-l\left( \l^{(l')}\right)}  \frac{(-1)^k}{k!}\\
&\prod_{\a \neq l'} \frac{\Gamma \left(-k +l \left( \l^{(l')} \right)-l \left( \l^{(\a)} \right)+\frac{a_{2,l'}-a_{2,\a}}{\ve_1} \right)}{\Gamma \left( \frac{a_{2,l'}-a_{2,\a}}{\ve_1} \right)} \prod_{\g=1} ^N \frac{\Gamma \left( \frac{a_{2,l'}+2\ve -a_{4,\g}}{\ve_1} \right)}{\Gamma \left( -k + l \left( \l^{(l')} \right) +\frac{a_{2,l'}+2\ve-a_{4,\g}}{\ve_1} \right)} \\
&\prod_{\g=1} ^N \prod_{i=1} ^{l \left( \l^{(\g)} \right)} \frac{a_{2,l'} -a_{2,\g} +\left( l \left( \l^{(l')} \right) -k-i \right)\ve_1 - \l_i ^{(\g)} \ve_2}{\ve_1}.
\end{split}
\end{align}
We can multiply a proper prefactor to $\Upsilon_{l,h} ^{0<\vert \qe \vert <\vert y \vert< \vert z\vert<1}$
\begin{align} \label{eq:intpartint1}
\begin{split}
&\widehat{\Upsilon}_{l,h} ^{0<\vert \qe \vert <\vert y\vert<\vert z \vert< 1}\\
 &= z^{\frac{-a_{2,h} +a_{2,\bar{h}} +\ve}{2\ve_2}} \left(\frac \qe y \right)^{\frac{-a_{2,l} +a_{2,\bar{l}} -\ve_1-2\ve_2}{2\ve_1}} \qe^{-\Delta_0 -\Delta_\qe +\frac{\ve^2 -(a_{2,1}-a_{2,2})^2}{4\ve_1\ve_2} +\frac{2\ve_1+3\ve_2}{4\ve_1}}  \\
& (1-z)^{\frac{2\bar{a}_0-2\bar{a}_2 +\ve_1+2\ve_2}{2\ve_2}} \left( 1-\frac \qe z \right)^{\frac{2\bar{a}_2 -2\bar{a}_4 +4\ve_1+5\ve_2}{2\ve_2}} \left( 1- \frac y z \right) ^{-\frac 1 2} (1-y)^{\frac{-2\bar{a}_0 +2\bar{a}_2+\ve_1}{2\ve_1} } \\
& (1-\qe)^{\frac{(2\bar{a}_0 -2\bar{a}_2 -\ve_1)(2\bar{a}_2 -2\bar{a}_4 +4\ve_1+5\ve_2)}{2\ve_1\ve_2}} \left( 1-\frac \qe y \right)^{\frac{-2\bar{a}_2 +2\bar{a}_4 -2\ve_1 -3\ve_2}{2\ve_1}}  \Upsilon_{l,h} ^{0<\vert \qe \vert <\vert y\vert <\vert z \vert<1},
\end{split}
\end{align}
so that the connection formula reads
\begin{align}
\widehat{\Upsilon}_{l,h} ^{0<\vert y \vert <\vert \qe \vert < \vert z\vert<1} = \sum_{\a =1 }^N \left( \mathbf{C}_0 ^{(1)} \right)_{l l'} \widehat{\Upsilon}_{l',h} ^{0<\vert \qe \vert <\vert y \vert< \vert z\vert<1}.
\end{align}
Here, we have defined the connection matrix
\begin{align}
\left( \mathbf{C}_0 ^{(1)} \right)_{l l'} = \prod_{\b \neq l} \frac{\Gamma \left( 1+\frac{a_{4,l}-a_{4,\b}}{\ve_1} \right)}{\Gamma \left( \frac{a_{2,l'} +2\ve-a_{4,\b}}{\ve_1} \right)} \prod_{\a \neq l'} \frac{\Gamma \left( \frac{a_{2,l'}-a_{2,\a}}{\ve_1} \right)}{\Gamma \left(1+\frac{a_{4,l} -2\ve -a_{2,\a}}{\ve_1}\right)},
\end{align}
where the superscript $(1)$ indicates the adiabatic flow of the surface defect on the $z_1$-plane.

Similarly, we can start from the domain $0<\vert \qe \vert <\vert z \vert<1<\vert y \vert$ and flow $y$ to another domain $0<\vert \qe \vert<\vert z \vert < \vert y \vert <1$, as follows.
\begin{align}
\Upsilon_{l,h} ^{0<\vert \qe \vert <\vert z \vert<1<\vert y \vert} = \sum_{l' =1 } ^N \prod_{\b \neq l} \frac{\Gamma \left( 1+\frac{a_{0,\b} -a_{0,l}}{\ve_1} \right)}{\Gamma \left( \frac{a_{0,\b} -a_{2,l'} +\ve }{\ve_1} \right)} \prod_{\a \neq l'} \frac{\Gamma \left( \frac{a_{2,\a} -a_{2,l'} }{\ve_1} \right)}{\Gamma \left( 1+\frac{a_{2,\a}-a_{0,l}-\ve}{\ve_1} \right)} y (-y)^{\frac{a_{2,l'}-a_{0,l}-\ve}{\ve_1}} \Upsilon_{l',h} ^{0<\vert \qe \vert<\vert z \vert < \vert y \vert <1} ,
\end{align}
where we have defined the intersecting surface defect partition function in the domain $0<\vert \qe \vert<\vert z \vert < \vert y \vert <1$ by
\begin{align}
\begin{split}
&\Upsilon_{l,h} ^{0<\vert \qe \vert <\vert z \vert<\vert y \vert<1} = \sum_{\boldsymbol\l} \qe ^{\vert \boldsymbol\l \vert} E[ \mathcal{T}'_{A_1} [\boldsymbol\l] ] {\EuScript{O}'}_{2,h} ^{\vert \qe \vert <\vert z \vert<1} [\boldsymbol\l] {\EuScript{O}'}_{1,l} ^{\vert \qe \vert <\vert y \vert<1} [\boldsymbol\l],
\end{split}
\end{align}
with the new surface defect observable
\begin{align}
\begin{split}
{\EuScript{O}'}_{1,l} ^{\vert \qe \vert <\vert y \vert<1} [\boldsymbol\l]&= \sum_{k=0} ^\infty y^{k- l\left(\l^{(l)}\right)} \frac{(-1)^k}{k!} \\
&\prod_{\a\neq l} \frac{\Gamma \left( -k +l \left( \l ^{(l)} \right) -l\left(\l ^{(\a')} \right) + \frac{a_{2,\a} -a_{2,l}}{\ve_1} \right)}{\Gamma \left( \frac{a_{2,\a} -a_{2,l}}{\ve_1} \right)} \prod _{\d =1 }^N \frac{\Gamma \left( \frac{a_{0,\d}-a_{2,l} +\ve}{\ve_1} \right)}{\Gamma \left( -k + l\left(\l ^{(l)}\right) + \frac{a_{0,\d}-a_{2,l} +\ve}{\ve_1} \right)} \\
& \prod_{\delta=1} ^N \prod_{i=1} ^{l\left(\l^{(\d)} \right)} \frac{-a_{2,l} +a_{2,\d} +\left(l\left(\l^{(l)}\right) -k -i \right) \ve_1 -\l^{(\d)}_i \ve_2 }{\ve_1}.
\end{split}
\end{align}
We may modify this partition function by
\begin{align} \label{eq:intsurfpartint2}
\begin{split}
\widehat{\Upsilon}_{l,h} ^{0 <\vert \qe \vert <\vert z \vert<\vert y \vert< 1} &= y^{\frac{a_{2,l}-a_{2,\bar{l}} +\ve}{2\ve_1}} \left( \frac{\qe}{z} \right)^{\frac{a_{2,h}-a_{2,\bar{h}} -2\ve_1-\ve_2}{2\ve_2}} \qe ^{-\Delta'_0 -\Delta'_\qe +\frac{\ve^2 -(a_{2,1}-a_{2,2})^2}{4\ve_1 \ve_2} +\frac{3\ve_1 +2\ve_2}{4\ve_2} }\\
& \left( 1- y \right)^{\frac{-2\bar{a}_0 +2\bar{a}_2 -\ve_2}{2\ve_1}} \left( 1-\frac{\qe}{y} \right)^{\frac{-2\bar{a}_0 +2\bar{a}_4 -3\ve_1 -4\ve_2}{2\ve_1}} \left( 1-\frac{z}{y} \right)^{-\frac 1 2} (1-z)^{\frac{2\bar{a}_0 -2\bar{a}_2 +2\ve_1+3\ve_2}{2\ve_2}} \\
& (1-\qe)^{\frac{(-2\bar{a}_0 +2\bar{a}_2 -2\ve_1-3\ve_2)(-2\bar{a}_2 +2\bar{a}_4 -\ve_1 -2\ve_2)}{2\ve_1 \ve_2}} \left( 1- \frac \qe z \right)^{\frac{2\bar{a}_2-2\bar{a}_4 +3\ve_1+4\ve_2}{2\ve_2}} \Upsilon_{l,h} ^{0 <\vert \qe \vert <\vert z \vert<\vert y \vert< 1}.
\end{split}
\end{align}
so that the connection formula reads
\begin{align}
\widehat{\Upsilon}_{l,h} ^{0<\vert \qe \vert <\vert z \vert<1<\vert y \vert} = \sum_{\a =1 }^N \left( \mathbf{C}_\infty ^{(1)} \right)_{l l'} \widehat{\Upsilon}_{l',h} ^{0<\vert \qe \vert<\vert z \vert < \vert y \vert <1},
\end{align}
where we have defined the connection matrix
\begin{align}
\left( \mathbf{C}_\infty ^{(1)} \right)_{ll' } = \prod_{\b \neq l} \frac{\Gamma \left( 1+\frac{a_{0,\b} -a_{0,l}}{\ve_1} \right)}{\Gamma \left( \frac{a_{0,\b} -a_{2,l'} +\ve }{\ve_1} \right)} \prod_{\a \neq l'} \frac{\Gamma \left( \frac{a_{2,\a} -a_{2,l'} }{\ve_1} \right)}{\Gamma \left( 1+\frac{a_{2,\a}-a_{0,l}-\ve}{\ve_1} \right)}.
\end{align}

\subsubsection{$y$-transporting the surface defect at $z_2=0$ across $z$}
Lastly, we need the connection formula between the solutions in the domain $0<\vert \qe \vert<\vert z \vert < \vert y \vert <1$ and $0<\vert \qe \vert< \vert y \vert <\vert z \vert  <1$. This amounts to resumming the intersecting surface defect partition function $\widehat{\Upsilon}_{l,h} ^{0<\vert \qe \vert <\vert y\vert<\vert z \vert< 1}$ as a series in $\frac y z$ and expanding it as a series in $\frac z y$ to glue it to $\widehat{\Upsilon}_{l,h} ^{0<\vert \qe \vert <\vert z\vert<\vert y \vert< 1}$. This resummation is trickier than the analytic continuations that we have obtained in the previous section, but can be accomplished in the following way.

Let us consider the degeneration limit $\qe \to 0$ of the partition function $\widehat{\Upsilon}_{l,h} ^{0<\vert \qe \vert <\vert y\vert<\vert z \vert< 1}$ \eqref{eq:intpartint1} in the domain $0<\vert \qe \vert <\vert y\vert<\vert z \vert< 1$ and the differential equations \eqref{eq:bpz} that it satisfies. We have to carefully decouple the $\qe$-dependent prefactor of $\widehat{\Upsilon}_{l,h} ^{0<\vert \qe \vert <\vert y\vert<\vert z \vert< 1}$ to have a well-defined limit:
\begin{align}
\widehat{\Upsilon}_{l,h} ^{0< \vert y\vert<\vert z \vert< 1} := \lim_{\qe \to 0} \qe^{\Delta_0 + \Delta_\qe -\frac{\ve^2 -(a_{2,1}-a_{2,2})^2}{4\ve_1\ve_2} +\frac{a_{2,l}-a_{2,\bar{l}}}{2\ve_1} +\frac{\ve_2}{4\ve_1} } \widehat{\Upsilon}_{l,h} ^{0<\vert \qe \vert <\vert y\vert<\vert z \vert< 1}.
\end{align}
By taking the limit to \eqref{eq:bpz} we have reduced differential equations,
\begin{subequations} \label{eq:bpzred}
\begin{align}
 0=&  \left[ \ve_2 ^2 \partial ^2 _z - \ve_1 \ve_2 \left( \frac{1}{z} +\frac{1}{z-1} \right) \partial_z +\ve_1 \ve_2 \frac{y(y-1)}{z(z-y)(z-1)} \partial_y \right. \nonumber\\
  & \left. +\ve_1 \ve_2 \left( \frac{\Delta_{0+\qe}}{z^2} +\frac{\Delta_1}{(z-1)^2} +\frac{\Delta_L}{(z-y)^2} + \frac{\Delta_\infty -\Delta_{0+\qe} -\Delta_1-\Delta_L-\Delta_H}{z(z-1)} \right) \right]  \widehat{\Upsilon}_{l,h} ^{0<\vert y \vert <\vert z \vert<1} , \\
  0=  & \left[\ve_1 ^2 \partial ^2 _y - \ve_1 \ve_2 \left( \frac{1}{y} +\frac{1}{y-1} \right) \partial_y +\ve_1 \ve_2 \frac{z(z-1)}{y(y-z)(y-1)} \partial_z \right. \nonumber \\
    &\left.+\ve_1 \ve_2 \left( \frac{\Delta_{0+\qe}}{y^2}  +\frac{\Delta_1}{(y-1)^2} +\frac{\Delta_H}{(y-z)^2} + \frac{\Delta_\infty -\Delta_{0+\qe}-\Delta_1-\Delta_L-\Delta_H}{y(y-1)} \right) \right]\widehat{\Upsilon}_{l,h} ^{0<\vert y \vert <\vert z \vert<1} \label{eq:bpzred1},
\end{align}
\end{subequations}
where we have defined
\begin{align}
\Delta_{0+\qe} = \frac{\ve^2 -(a_{2,l}-a_{2,\bar{l}} +\ve_2)^2}{4\ve_1\ve_2}.
\end{align}

Similarly, we carefully decouple the $\qe$-dependent prefactor in the solution $\widehat{\Upsilon}_{l,h} ^{0<\vert \qe \vert <\vert z\vert<\vert y \vert< 1}$ \eqref{eq:intsurfpartint2} in the other domain $0<\vert \qe \vert <\vert z\vert<\vert y \vert< 1$ by
\begin{align}
\widehat{\Upsilon}_{l,h} ^{0< \vert z\vert<\vert y \vert< 1} := \lim_{\qe \to 0} \qe^{\Delta'_0 + \Delta'_\qe -\frac{\ve^2 -(a_{2,1}-a_{2,2})^2}{4\ve_1\ve_2} -\frac{a_{2,h}-a_{2,\bar{h}}}{2\ve_2} +\frac{\ve_1}{4\ve_2} } \widehat{\Upsilon}_{l,h} ^{0<\vert \qe \vert <\vert z\vert<\vert y \vert< 1}.
\end{align}
Then the reduced differential equations obtained as the limit of \eqref{eq:bpz'} are
\begin{subequations} \label{eq:bpzred'}
\begin{align}
 0=&  \left[ \ve_2 ^2 \partial ^2 _z - \ve_1 \ve_2 \left( \frac{1}{z} +\frac{1}{z-1} \right) \partial_z +\ve_1 \ve_2 \frac{y(y-1)}{z(z-y)(z-1)} \partial_y \right. \nonumber\\
  & \left. +\ve_1 \ve_2 \left( \frac{\Delta'_{0+\qe}}{z^2} +\frac{\Delta'_1}{(z-1)^2} +\frac{\Delta_L}{(z-y)^2} + \frac{\Delta_\infty -\Delta'_{0+\qe} -\Delta'_1-\Delta_L-\Delta_H}{z(z-1)} \right) \right]  \widehat{\Upsilon}_{l,h} ^{0<\vert z \vert <\vert y \vert<1} , \\
  0=  & \left[\ve_1 ^2 \partial ^2 _y - \ve_1 \ve_2 \left( \frac{1}{y} +\frac{1}{y-1} \right) \partial_y +\ve_1 \ve_2 \frac{z(z-1)}{y(y-z)(y-1)} \partial_z \right. \nonumber \\
    &\left.+\ve_1 \ve_2 \left( \frac{\Delta'_{0+\qe}}{y^2}  +\frac{\Delta'_1}{(y-1)^2} +\frac{\Delta_H}{(y-z)^2} + \frac{\Delta_\infty -\Delta'_{0+\qe}-\Delta'_1-\Delta_L-\Delta_H}{y(y-1)} \right) \right] \widehat{\Upsilon}_{l,h} ^{0<\vert z \vert <\vert y \vert<1} \label{eq:bpzred1'},
\end{align}
\end{subequations}
where we have defined
\begin{align}
\Delta'_{0+\qe} = \frac{\ve^2-(a_{2,h}-a_{2,\bar{h}}-\ve_1)^2}{4\ve_1\ve_2}.
\end{align}
Effectively, what we have done is to bring the puncture at $\qe$ close to $0$ and merge them together. This is equivalent to turning off the bulk gauge coupling in the gauge theory point of view. Hence we are left with the two-dimensional sigma models on the $z_1$-plane and the $z_2$-plane interacting at the origin.

Let us note that the equations \eqref{eq:bpzred} become identical to \eqref{eq:bpzred'} after the following re-definition of the Coulomb moduli,
\begin{align} \label{eq:shift}
a_{2,\a} \longrightarrow a_{2,\a} -\d_{\a,l} \ve_2 -\d_{\a,h} \ve_1.
\end{align}
The reduced partition functions $\widehat{\Upsilon}_{l,h} ^{0<\vert y \vert <\vert z \vert<1}$ and $\widehat{\Upsilon}_{l,h} ^{0<\vert z \vert <\vert y \vert<1}$ provide the solutions to the equation in the respective domain. To accomplish the connection formua between these solutions, we need to re-expand $\widehat{\Upsilon}_{l,h} ^{0<\vert y \vert <\vert z \vert<1}$, which is a series in $\frac y z$ and $z$, as a series in $\frac z y$ and $y$. Then we can compare the re-expanded series with $\widehat{\Upsilon}_{l,h} ^{0<\vert z \vert <\vert y \vert<1}$ to achieve the connection formula.

For this, we need to explicitly write the series expansion of the reduced partition function. The non-perturbative part $\Upsilon_{l,h} ^{0<\vert y\vert<\vert z \vert< 1}$ of the reduced partition function is given by the limit $\qe \to 0$ to the non-perturbative part $\Upsilon_{l,h} ^{0<\vert \qe \vert <\vert y\vert<\vert z \vert< 1}$ of the full partition function given in \eqref{eq:intsurfint1}, with the surface observables given by \eqref{eq:intsurf1} and \eqref{eq:intsurf2}:
\begin{align}
\begin{split}
\Upsilon_{l,h} ^{0<\vert y\vert<\vert z \vert< 1} &= \lim_{\qe \to 0}  \sum_{\boldsymbol\l} \qe ^{\vert \boldsymbol\l \vert} E[ \mathcal{T}_{A_1} [\boldsymbol\l] ] \EuScript{O}_{1,l}  ^{\vert \qe \vert <\vert y \vert <1}  [\boldsymbol\l] \EuScript{O}_{2,h} ^{\vert \qe \vert <\vert z \vert <1}  [\boldsymbol\l] \\
&=\lim_{\qe \to 0 } \sum_{\boldsymbol\l} y^{\vert \boldsymbol\l \vert }E[ \mathcal{T}_{A_1} [\boldsymbol\l] ] \\
&\sum_{k=0} ^\infty \left( \frac \qe y \right)^{k+\vert \boldsymbol\l\vert - l\left( \l^{(l)} \right)} \frac{(-1)^k}{k!}\\
&\prod_{\a \neq l} \frac{\Gamma \left(-k +l \left( \l^{(l)} \right)-l \left( \l^{(\a)} \right)+\frac{a_{2,l}-a_{2,\a}}{\ve_1} \right)}{\Gamma \left( \frac{a_{2,l}-a_{2,\a}}{\ve_1} \right)} \prod_{\g=1} ^N \frac{\Gamma \left( \frac{a_{2,l}+2\ve -a_{4,\g}}{\ve_1} \right)}{\Gamma \left( -k + l \left( \l^{(l)} \right) +\frac{a_{2,l}+2\ve-a_{4,\g}}{\ve_1} \right)} \\
&\prod_{\g=1} ^N \prod_{i=1} ^{l \left( \l^{(\g)} \right)} \frac{a_{2,l} -a_{2,\g} +\left( l \left( \l^{(l)} \right) -k-i \right)\ve_1 - \l_i ^{(\g)} \ve_2}{\ve_1} \\
& \sum_{k'=0} ^\infty z^{k'- \l^{(h)}_1} \frac{(-1)^{k'}}{{k'}!} \\
&\prod_{\a\neq h} \frac{\Gamma \left( -k' +\l ^{(h)} _1 -\l ^{(\a)} _1 + \frac{a_{2,h} -a_{2,\a}}{\ve_2} \right)}{\Gamma \left( \frac{a_{2,h} -a_{2,\a}}{\ve_2} \right)} \prod _{\d =1 }^N \frac{\Gamma \left( \frac{a_{2,h} -a_{0,\d}}{\ve_2} \right)}{\Gamma \left( -k' + \l ^{(h)}_1 + \frac{a_{2,h} -a_{0,\d}}{\ve_2} \right)} \\
& \prod_{\delta=1} ^N \prod_{j=1} ^{\l^{(\d)} _1} \frac{a_{2,h} -a_{2,\d} +\left(\l^{(h)}_1 -k' -j \right) \ve_2 -\l^{(\d)t}_j \ve_1 }{\ve_2}.
\end{split}
\end{align}
Here, we have taken the rank of the bulk gauge group $N$ generic, but it is implicitly understood that we set $N=2$ whenever we restrict our attention to the solutions to the differential equations \eqref{eq:bpzred}. The summation over non-negative integers $k$ gets non-zero contribution only when $k+ \vert\boldsymbol\l\vert -l\left( \l^{(\l)} \right)=0$ due to the limit $\qe \to 0$. This implies $k=0$ and $\vert\boldsymbol\l  \vert= l\left( \l^{(l)} \right)$, namely, the Young diagrams $\left(\l^{(\a)} \right)_{\a=1} ^N$ are empty except $\l^{(l)}$ which is single-columned. The expression of the partition function is simplified accordingly. After many cancellations between the bulk contribution and the surface observable contribution, we obtain
\begin{align} \label{eq:redpartint11}
\begin{split}
\Upsilon_{l,h} ^{0<\vert y\vert<\vert z \vert< 1} =& \sum_{k=0} ^\infty  \frac{y^k}{k!} \prod_{\g=1} ^N \frac{\Gamma\left( k+\frac{a_{2,l} -a_{0,\g}}{\ve_1} \right)}{\Gamma\left( \frac{a_{2,l}-a_{0,\g}}{\ve_1} \right)} \prod_{\a \neq l} \frac{\Gamma\left( \frac{a_{2,l}-a_{2,\a}+\ve}{\ve_1} \right)}{\Gamma\left( k+\frac{a_{2,l} -a_{2,\a}+\ve}{\ve_1} \right)} \\
& \sum_{k'=0} ^\infty \frac{ z^{k' -(1-\d_{0,k})\d_{l,h}}}{k'!} \prod_{\g=1} ^N \frac{\Gamma\left( k'+1 +\frac{a_{0,\g}-a_{2,h}}{\ve_2} \right)}{\Gamma\left(  1+\frac{a_{0,\g}-a_{2,h}}{\ve_2}  \right) } \prod_{\a\neq l} \frac{\Gamma\left( 1+\frac{a_{2,\a}-a_{2,h}}{\ve_2} \right)}{\Gamma\left( k'+1 +\frac{a_{2,\a}-a_{2,h}}{\ve_2} \right)} \\
&  \prod_{\g=1} ^N \frac{\Gamma\left( -k'+ \frac{a_{2,h}-a_{0,\g}}{\ve_2} \right)}{\Gamma\left( -k' +(1-\d_{0,k})\d_{l,h} +\frac{a_{2,h}-a_{0,\g}}{\ve_2} \right)} \prod_{\a \neq h} \frac{\Gamma\left( -k' +(1-\d_{0,k})(\d_{l,h}-\d_{l,\a}) +\frac{a_{2,h}-a_{2,\a}}{\ve_2}  \right)}{\Gamma\left( -k' + \frac{a_{2,h}-a_{2,\a}}{\ve_2} \right)} \\
& \left(\frac{a_{2,h}-a_{2,l} +(\d_{l,h}-k'-1)\ve_2 -k\ve_1}{\ve_2}\right)^{1-\d_{0,k}}.
\end{split}
\end{align}
The first two lines indicate that this is the partition function of two-dimensional sigma models on the $z_1$-plane and the $z_2$-plane, whose target space is $\mathcal{O}(-1) \oplus \mathcal{O}(-1) \rightarrow \mathbb{P}^1$, respectively. It is not a simple product of the two, however, because of the interaction between the two models at the origin which contributes the third and the fourth lines to the partition function. Hence, we see that the bulk gauge theory decouples due to the decoupling limit $\qe \to 0$, and the surface defects becomes effectively a pair of two-dimensional sigma models, weakly interacting with each other via a bi-local observable. Here by the bi-local observable we mean an operator in the combined theory, which is local from the point of view of 
an observer, living on the worldsheet of one of these sigma models. In a sense, this configuration is a two dimensional analogue of the crossed instantons setup of \cite{Nekrasov_BPS1}.  

Note that the expression \eqref{eq:redpartint11} allows analytic continuation to the domain $0<\vert z \vert <\vert y \vert <1$ automatically. We just have to re-write $z= \frac z y y$ and expand. When $h=l$, we recognize that the power of $\frac z y$ begins from $-1$, when $k=1$ and $k'=0$. Hence the partition function is re-gathered as a series in $\frac z y$ and $y$ as
\begin{align} \label{eq:reexp1}
\begin{split}
&\Upsilon_{l,h=l} ^{0<\vert y\vert<\vert z \vert< 1} \\
&= -\frac y z  \prod_{\a \neq l} \frac{ a_{2,l}-a_{2,\a} }{ a_{2,l}-a_{2,\a}+\ve }  \\
&\left( \sum_{k,k'=0} ^\infty   y^{k+k' -1+\d_{0,k}} \left( \frac z y \right)^{k' +\d_{0,k}} \frac{(-1)^{k'}}{k!\, k'!} \prod_{\g=1} ^N \frac{\Gamma\left( k+\frac{a_{2,l} -a_{0,\g}}{\ve_1} \right)}{\Gamma\left( \frac{a_{2,l}-a_{0,\g}}{\ve_1} \right)} \frac{\Gamma\left( \frac{a_{2,h}-a_{0,\g}}{\ve_2} \right)}{\Gamma\left( -k' +1-\d_{0,k} +\frac{a_{2,h}-a_{0,\g}}{\ve_2} \right)} \right. \\
& \quad\quad \left. \prod_{\a \neq l}  \frac{\Gamma\left(1+ \frac{a_{2,l}-a_{2,\a}+\ve}{\ve_1} \right)}{\Gamma\left( k+\frac{a_{2,l} -a_{2,\a}+\ve}{\ve_1} \right)} \frac{\Gamma\left( -k' +1-\d_{0,k} +\frac{a_{2,l}-a_{2,\a}}{\ve_2}  \right)}{\Gamma\left(1+ \frac{a_{2,l}-a_{2,\a}}{\ve_2} \right)}  \times \left( \frac{ -k' \ve_2 -k\ve_1}{\ve_2} \right)^{1-\d_{0,k}} \times \left( -\frac{\ve_2}{\ve_1} \right) \right)  \\
&= -\frac y z  \prod_{\a \neq l} \frac{  a_{2,l}-a_{2,\a}  }{ a_{2,l}-a_{2,\a}+\ve } \times \left( 1+ \mathcal{O}\left( \frac z y, y \right) \right).
\end{split}
\end{align}

When $h \neq l$, on the other hand, it is straightforward from \eqref{eq:redpartint11} that the re-expansion as a series in $\frac z y$ and $y$ is trivial:
\begin{align} \label{eq:reexp2}
\Upsilon_{l,h(\neq l)} ^{0<\vert y\vert<\vert z \vert< 1} = 1 + \mathcal{O}\left(\frac z y, y\right).
\end{align}

As a result of the re-expansion \eqref{eq:reexp1} and \eqref{eq:reexp2} of $\Upsilon_{l,h} ^{0<\vert y\vert<\vert z \vert< 1} $, we produce the solutions to the reduced differential equations \eqref{eq:bpzred} in the domain $0<\vert z\vert<\vert y \vert< 1$. By explicitly solving the equations by a series expansion in $\frac z y $ and $y$, it follows that the solution is uniquely determined once the critical exponents of $y$ and $z$ are chosen. The uniqueness of the solution implies the identification of $\widehat{\Upsilon}_{l,h} ^{0<\vert y\vert<\vert z \vert< 1}$ and $\widehat{\Upsilon}_{l,h} ^{0<\vert z\vert<\vert y \vert< 1}$, up to the multiplicative prefactor in \eqref{eq:reexp1}. More precisely, after taking account of the shift \eqref{eq:shift} in the Coulomb moduli, we establish this connection formula as follows:
\begin{align} \label{eq:redanal}
\widehat{\Upsilon}_{l,h} ^{0<\vert y \vert <\vert z \vert<1}=\sum_{\a,\b,l' =1} ^N \left( \mathbf{C}_M ^{h} \right)_{l\a} \left( \mathbf{S}_2 \right)_{\a \b}  \left( \mathbf{S} ^h _1  \right)_{\b l'}  \widehat{\Upsilon}_{l',h} ^{0<\vert z \vert <\vert y \vert<1},
\end{align}
where the shift matrices are defined by
\begin{align}
\begin{split}
&\left(\mathbf{S}_2 \right)_{\a\b} = e^{\ve_2 \frac{\p}{\p a_\a}} \d_{\a\b}\\
&\left(\mathbf{S}_1 ^h \right)_{\a\b} = e^{\ve_1 \frac{\p}{\p a_h } } \d_{\a\b},
\end{split}
\end{align}
and the connection matrix is given by
\begin{align}
\left(\mathbf{C}_M ^{h} \right)_{ll'} = \left( \prod_{\a \neq l} \frac{a_{2,l} - a_{2,\a}}{a_{2,l}-a_{2 ,\a}+\ve} \right)^{\delta_{l,h}} \d_{l l'}.
\end{align}
In the case of $N=2$, we can explicitly write this $2\times 2$ matrix as
\begin{align}
\mathbf{C}_M ^{+} = \begin{pmatrix} \frac{a_{2,1}-a_{2,2}}{a_{2,1}-a_{2,2}+\ve} & 0 \\ 0 & 1 \end{pmatrix}, \quad\quad\mathbf{C}_M ^{-} = \begin{pmatrix} 1 & 0 \\ 0 & \frac{a_{2,2}-a_{2,1}}{a_{2,2}-a_{2,1}+\ve} \end{pmatrix}.
\end{align}
Here we are abusing the notation and use $h \in  \{ 1,2\}$ and $(-1)^{h-1} \in \{ \pm\}$ interchangeably.

Having established the analytic continuation \eqref{eq:redanal} of the reduced partition functions, we now investigate the analytic continuation of the full partition functions, $\widehat{\Upsilon}_{l,h} ^{0<\vert \qe\vert <\vert y \vert <\vert z \vert<1}$ and $\widehat{\Upsilon}_{l,h} ^{0<\vert \qe\vert <\vert z \vert <\vert y \vert<1}$, which solve \eqref{eq:bpz} and \eqref{eq:bpz'} in the respective domain. We can view the reduced partition functions $\widehat{\Upsilon}_{l,h} ^{0<\vert y \vert <\vert z \vert<1}$ and $\widehat{\Upsilon}_{l,h} ^{0<\vert z \vert <\vert y \vert<1}$ as the initial condition of the solutions at $\qe=0$. Then the solutions as series expansions in $\qe$ are uniquely determined by the differential equations. Hence, the connection formula \eqref{eq:redanal} immediately uplifts to the connection formula for the full partition functions as it is:
\begin{align} 
\widehat{\Upsilon}_{l,h} ^{0<\vert \qe \vert<\vert y \vert <\vert z \vert<1}=\sum_{\a,\b,l' =1} ^N \left( \mathbf{C}_M ^{h} \right)_{l\a} \left( \mathbf{S}_2 \right)_{\a \b}  \left( \mathbf{S} ^h _1  \right)_{\b l'}  \widehat{\Upsilon}_{l',h} ^{0<\vert \qe \vert<\vert z \vert <\vert y \vert<1}.
\end{align}
With this last piece of the puzzle, we have achieved all the analytic continuations of the intersecting surface defect partition functions that we need in computing the monodromy data of the associated Fuchsian system.

\subsection{Monodromy data} \label{sec:monod}
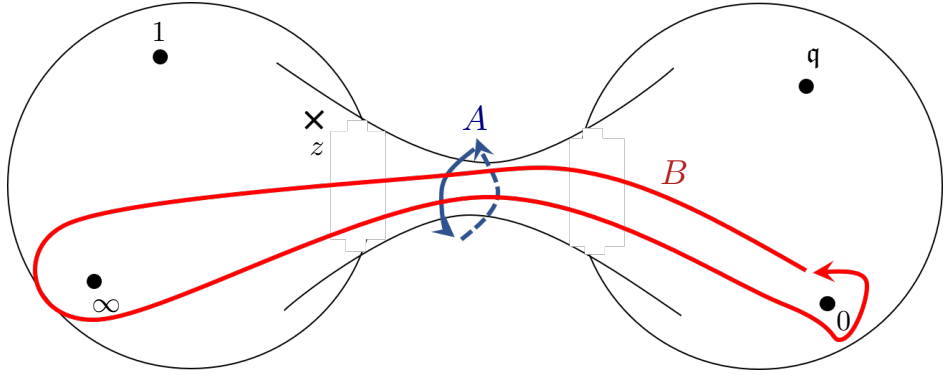
\begin{figure} 
\centering
\begin{tikzpicture}
\node[inner sep=0pt] (figure) at (0,0) {\includegraphics[width=0.8\textwidth]{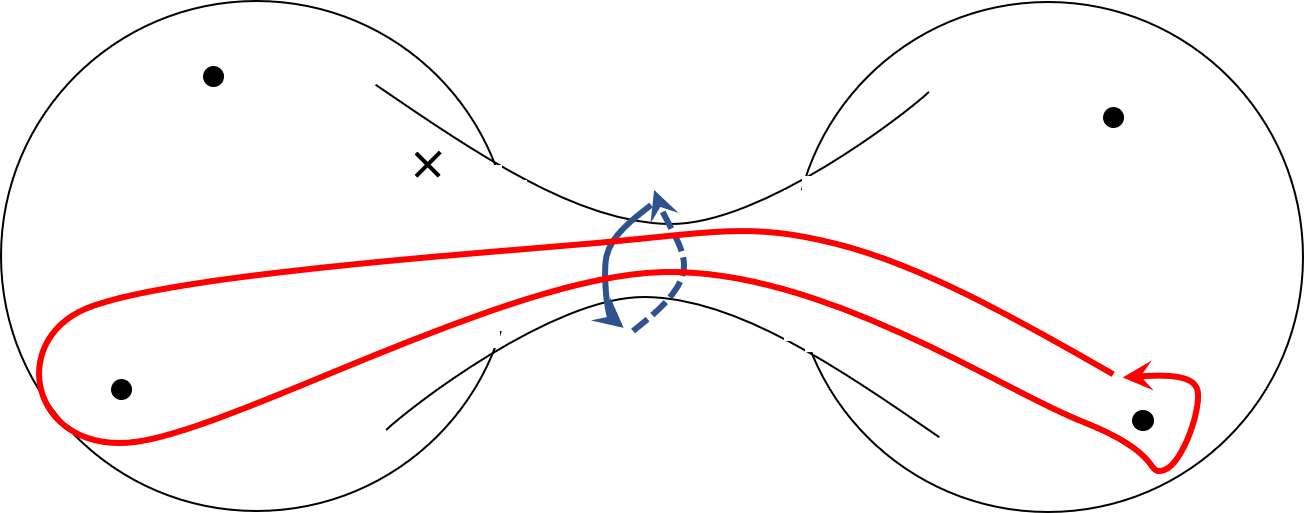}};
\node at (-4.9,-1.6) {\scalebox{1.0}{$\infty$}};
\node at (-4.2,2.05) {\scalebox{1.0}{$1$}};
\node at (4.5,1.7) {\scalebox{1.0}{$\mathfrak{q}$}};
\node at (4.9,-1.8) {\scalebox{1.0}{$0$}};
\node at (-2.1,0.5) {\scalebox{1.0}{$z$}};

\node at (0,0.9) {\scalebox{1.2}{$\textcolor{navyblue}{A}$}};
\node at (2.65,0.18) {\scalebox{1.2}{$\textcolor{firebrick}{B}$}};
\end{tikzpicture} 
\caption{The $A$-loop and the $B$-loop on the four-punctured sphere $\mathbb{P}^1 \backslash \{0,\mathfrak{q},1, \infty \}$. Here, $z$ is the position of the apparent singularity, around which the monodromy of the associated Fuchsian system is trivial. The $A$-loop is depicted as the blue line, while the $B$-loop is drawn as the red line. } \label{sphere}
\end{figure}

We have seen the $\EuScript{N}=2$ gauge theory with (intersecting) surface defects is associated to the Riemann-Hilbert correspondence and isomonodromic deformation of Fuchsian system through their partition functions. Their NS limits provide the generating function of the Riemann-Hilbert map, the Hamilton-Jacobi potential for the isomonodromic flow, and in particular the solutions to the Fuchsian differential equation. Based upon these relations, the monodromy data of the associated Fuchsian system can be expressed in gauge theoretical terms, as we describe now for our main example of the $\mathfrak{sl}(2)$ Fuchsian system on the four-punctured sphere (see figure \ref{sphere}).

Recall that the intersecting surface defect partition function $\widetilde\Upsilon$ annihilates a differential operator in coupling constants, which we denote as $\widehat{\widehat{\mathfrak{D}}}$ here, whose NS limit is the Fuchsian differential equation $\widehat{\mathfrak{D}} \chi =0$. In other words, the solutions to the associated Fuchsian differential equation are obtained as the regular part of the intersecting surface defect partition function in the NS limit $\ve_2 \to 0$. Hence, the strategy to compute the monodromy data of the Fuchsian system is simply to compute the monodromy of the intersecting surface defect partition functions first and then to take the NS limit of it. The Riemann surface $\EuScript{C}$ is divided into several convergence domains of the intersecting surface defect partition function, and a given loop $\gamma \in \pi_1 (\EuScript{C}) $ may remain in a single convergence domain or may pass through several convergence domains. In the former case, the computation of the monodromy $\mathbf{M}_\g$ is straightforward, while in the latter case the monodromy $\mathbf{M}_\g$ 
is a bit more involoved, being given by a concatenation of analytic continuations.

For our main example of the four-punctured sphere, $\EuScript{C} = \mathbb{P}^1 \setminus \{0,\qe,1,\infty\}$, there are six independent loops. Four of them are small loops encircling the four punctures and the remaining two, which we refer to as the $A$-loop and the $B$-loop, are \textit{non-local} loops depicted in figure \ref{sphere}. Let us begin with the small loop around the puncture at $0$. This loop is entirely contained in the domain $0< \vert y \vert < \vert \qe   \vert< \vert z \vert <1$, where the solution to $\widehat{\widehat{\mathfrak{D}}}$ is given by the intersecting surface defect partition function $\widetilde{\Upsilon}^{0< \vert y \vert < \vert \qe   \vert< \vert z \vert <1}$  that we obtained in section \ref{subsec:flowz2def}. Thus we simply continue this partition function along the path
\begin{align}
y \longrightarrow y \, e^{it} \quad \quad \text{with} \quad\quad 0 \leq t \leq 2\pi,
\end{align}
to enclose the punctures at $0$. Then the non-integral part of the exponent of $y$ produces a multiplicative factor as we move along $0 \leq t \leq 2\pi$, producing the monodromy $\mathbf{R}_0$. A straightforward computation shows that
\begin{align}
\mathbf{R}_0= \text{diag} \left( e^{\pi i \frac{m_1 - m_2 + \ve_2}{\ve_1}} ,  e^{\pi i\frac{-m_1 +m_2 +\ve_2}{\ve_1}}  \right).
\end{align}
The monodromy of the Fuchsian system is then computed by taking the limit $\ve_2 \to 0$,
\begin{align}
R_0 = \lim_{\ve_2 \to 0} \mathbf{R}_0 =  \text{diag} \left( e^{\pi i\frac{m_1 -m_2}{\ve_1}} ,  e^{-\pi i\frac{m_1 -m_2}{\ve_1}}  \right).
\end{align}
The monodromy itself is not an invariant notion since it depends on the basis in which it is expressed. Hence we take the trace invariant which is simply
\begin{align} \label{eq:r0}
\text{Tr}\, R_0 =  2 \cos 2\pi \th_0,
\end{align}
where we used the dimensionless $\th$-parameter. The outcome is not at all surprising, since it is immediate from the Fuchsian differential equation \eqref{eq:horibpz} that the the trace invariant of the monodromy $R_0$ should be as above.

We can deal with the monodromy around the puncture at $\infty$ in a similar manner. The loop encircling the infinity is entirely contained in the domain $0<\vert \qe \vert < \vert z \vert<1 <\vert y\vert$. The intersecting surface defect partition function $\widetilde{\Upsilon}^{0< \vert \qe   \vert< \vert z \vert <1<\vert y \vert  }$, also obtained in section \ref{subsec:flowz2def}, yields solutions to the differential operator $\widehat{\widehat{\mathfrak{D}}}$ in this domain. This time we should continue along the path which is clockwise from the origin
\begin{align}
y \longrightarrow y \, e^{-it} \quad \quad \text{with} \quad\quad 0 \leq t \leq 2\pi,
\end{align}
to enclose the infinity counterclockwise. Then we immediately compute the monodromy
\begin{align}
\mathbf{R}_\infty = \text{diag} \left( e^{\pi i\frac{-m_3+m_4 +\ve_1-2\ve_2}{\ve_1}} ,  e^{\pi i \frac{m_3-m_4 +\ve_1-2\ve_2}{\ve_1}}\right),
\end{align}
and also the trace invariant of the monodromy $R_\infty$,
\begin{align} \label{eq:rinf}
\text{Tr}\, R_\infty = \text{Tr}\, \lim_{\ve_2 \to 0} \mathbf{R}_\infty = -2\cos 2\pi \th_\infty.
\end{align}
Again, the result is expected from the Fuchsian differential equation \eqref{eq:horibpz}.

Before we consider the other two small loops around the punctures at $\qe$ and $1$, let us turn to the $A$-loop first. The $A$-loop only remains in the convergence domain in the middle, $0<\vert \qe \vert < \vert y \vert< \vert z \vert < 1$, so that the computation of the monodromy along it is immediate.\footnote{It is a matter of convention to choose the $A$-loop to be inside the domain $0<\vert \qe \vert < \vert y \vert< \vert z \vert < 1$ or the domain $0<\vert \qe \vert < \vert z \vert< \vert y \vert < 1$. As discussed earlier, $z$ is the position of the apparent singularity, and the monodromy around the small loop around $z$ is the identity.} As we have seen in section \ref{sec:ytrans}, the intersecting surface defect partition function $\widetilde{\Upsilon} ^{0<\vert \qe \vert < \vert y \vert< \vert z \vert < 1}$ provides the solutions to the $\widehat{\widehat{\mathfrak{D}}}$ in this domain. Thus we simply continue this partition function along the path
\begin{align}
y \longrightarrow y \, e^{it} \quad \quad \text{with} \quad\quad 0 \leq t \leq 2\pi,
\end{align}
to enclose the punctures at $0$ and $\qe$, thereby making the $A$-loop. A straightforward computation shows that
\begin{align}
\mathbf{M}_A = \text{diag} \left( e^{\pi i\frac{a_1 -a_2 -\ve_1+2\ve_2}{\ve_1}} ,  e^{\pi i\frac{-a_1 +a_2 -\ve_1+2\ve_2}{\ve_1}}  \right).
\end{align}
The monodromy of the Fuchsian system is then computed by taking the limit $\ve_2 \to 0$,
\begin{align}
M_A = \lim_{\ve_2 \to 0} \mathbf{M}_A =  \text{diag} \left( -e^{\pi i\frac{a_1 -a_2}{\ve_1}} ,  -e^{ \pi i\frac{-a_1 +a_2}{\ve_1}}  \right).
\end{align}
Hence the trace invariant is simply
\begin{align} \label{eq:acycle}
\text{Tr}\, M_A = - 2 \cos 2\pi \a,
\end{align}
where we used the dimensionless parameter $\a = \frac{a_1 -a_2}{2\ve_1} = \frac{a}{\ve_1}$.

Now, the small loops around the punctures $\qe$ and $1$ are not entirely contained in a single convergence domain. Rather, we can compose these loops out of the rotation matrices $\mathbf{R}_0$, $\mathbf{R}_\infty$ and the $A$-loop $\mathbf{M}_A$, which lie inside distinct convergence domains. These monodromy matrices are represented in different bases, so we need the connection matrices to account for the change of bases. These connection matrices are precisely what was computed in the section \ref{sec:ytrans}, by analytic continuations of intersecting surface defect partition functions. In particular, we can express
\begin{align}
\begin{split}
&\mathbf{R}_\qe = \mathbf{M}_A {\mathbf{C} ^{(1)} _0} ^{-1} \mathbf{R}_0 ^{-1} \mathbf{C} ^{(1) }_0 \\
&\mathbf{R}_1 = \mathbf{M}_A ^{-1} {\mathbf{C} ^ {(1)} _\infty} ^{-1} \mathbf{R}_\infty ^{-1} \mathbf{C} ^{(1) }_\infty.
\end{split}
\end{align}
Then the trace invariants of the monodromies of the Fuchsian system are
\begin{align} \label{eq:rq1}
\begin{split}
&\text{Tr}\, R_\qe = \text{Tr}\, \lim_{\ve_2 \to 0} \mathbf{R}_\qe = -2 \cos 2\pi \th_\qe \\
&\text{Tr}\, R_1 = \text{Tr}\, \lim_{\ve_2 \to 0} \mathbf{R}_1 = 2 \cos 2\pi \th_1,
\end{split}
\end{align}
where we re-expressed the hypermultiplet masses with the dimensionless $\th$-parameters. Once again, these trace invariants are indeed what we expect from the Fuchsian differential equation \eqref{eq:horibpz}. 

Finally, we are left with the $B$-loop which is  most complicated. By concatenating the connection matrices, the shift matrices, and the rotation matrices, we construct the following sequence of continuations of solutions along the $B$-loop,
\begin{align}
\begin{split}
&\widetilde{\Upsilon}^{0<\vert \qe \vert <\vert y \vert< \vert z\vert<1}  \xrightarrow{{\mathbf{C}_0 ^{(1)}} ^{-1}} \widetilde{\Upsilon} ^{0<\vert y \vert<\vert \qe \vert < \vert z\vert<1} \xrightarrow{\mathbf{R}_0 ^{-1}}  \widetilde{\Upsilon} ^{0<\vert y \vert<\vert \qe \vert < \vert z\vert<1}\xrightarrow{\mathbf{C}_0 ^{(1)} }\widetilde{\Upsilon} ^{0<\vert \qe \vert <\vert y \vert< \vert z\vert<1} \\
& \xrightarrow{ \mathbf{C}_M \mathbf{S}_2 \mathbf{S}_1 } \widetilde{\Upsilon} ^{0<\vert \qe \vert< \vert z\vert <\vert y \vert<1}  \xrightarrow{{\mathbf{C}_\infty ^{(1)}} ^{-1}} \widetilde{\Upsilon} ^{0<\vert \qe \vert< \vert z\vert <1<\vert y \vert}  \xrightarrow{\mathbf{R}_\infty ^{-1}}   \widetilde{\Upsilon} ^{0<\vert \qe \vert< \vert z\vert <1<\vert y \vert} \xrightarrow{\mathbf{C}_\infty ^{(1)} }   \widetilde{\Upsilon} ^{0<\vert \qe \vert< \vert z\vert <\vert y \vert<1}  \\
& \xrightarrow{\mathbf{S}_1 ^{-1}  \mathbf{S}_2  ^{-1} \mathbf{C}_M ^{-1}}\widetilde{\Upsilon}^{0<\vert \qe \vert <\vert y \vert< \vert z\vert<1}.
\end{split}
\end{align}
Hence the corresponding monodromy is
\begin{align}
\mathbf{M}_B ^\pm =  {\mathbf{C}_M ^\pm} \mathbf{S}_2  \mathbf{S}_1 ^\pm { \mathbf{C} ^{(1)}  _\infty}^{-1} \mathbf{R}_\infty \mathbf{C} ^{(1)} _\infty {\mathbf{S}_1^\pm} ^{-1}  \mathbf{S}_2 ^{-1} {\mathbf{C}_M ^\pm}^{-1} {\mathbf{C}_0 ^{(1)}} ^{-1} \mathbf{R}_0 \mathbf{C}_0 ^{(1)}.
\end{align}
Recalling the asymptotics of the intersecting surface defect partition function has the singular part, $\widetilde{\Upsilon}_{l,h} = \exp\left[\frac{\ve_1}{\ve_2} \widetilde{S}_h \right] (\chi_l+ \mathcal{O}(\ve_2) ) $, we compute the monodromy of the Fuchsian system along the $B$-loop by taking the limit $\ve_2 \to 0$. Also, note that we have a choice of the vacuum of the gauged linear sigma model on the $z_2$-plane, which is reflected in the superscript $\pm$ (again, we abuse the notation and use $h\in\{1,2\}$ and $(-1)^{h-1} \in \{\pm\}$ interchangeably). These two choices provide two different results for the monodromy computation, yielding
\begin{align}
M_B  ^\pm = \lim_{\ve_2 \to 0}\left( {\mathbf{C}_M ^\pm} \mathbf{S}_2  \mathbf{S}_1 ^\pm { \mathbf{C} ^{(1)}  _\infty}^{-1} \mathbf{R}_\infty \mathbf{C} ^{(1)} _\infty {\mathbf{S}_1^\pm} ^{-1}  \mathbf{S}_2 ^{-1} {\mathbf{C}_M ^\pm}^{-1} {\mathbf{C}_0 ^{(1)}} ^{-1} \mathbf{R}_0 \mathbf{C}_0 ^{(1)} e^{\frac{\ve_1}{\ve_2} \widetilde{S}_{\pm}}  \right).
\end{align}
A lengthy but straightforward computation gives
\begin{align}
\begin{split}
\text{Tr}\, M_B ^{+} &= \frac{\left( -\cos 2\pi \th_\infty +\cos 2\pi \th_1 \right) \left( \cos 2\pi \th_0 - \cos 2\pi \th_\qe \right)}{2 \sin ^2 \pi \a} \nonumber \\
& - \frac{\left( \cos 2\pi \th_\infty +\cos 2\pi \th_1 \right) \left( \cos 2\pi \th_0 + \cos 2\pi \th_\qe \right)}{2 \cos ^2 \pi \a} \nonumber\\
& + 4 \frac{\prod_{\pm} \sin \pi (-\a - \th_\qe \pm \th_0) \sin \pi (-\a -\th_1 \pm \th_\infty) }{\sin ^2 2\pi \a} \nonumber\\
&\quad \frac{\Gamma \left( 2\a \right)^2}{\Gamma \left( -2\a \right)^2} \prod_{\pm} \frac{\Gamma \left( -\a -\th_\qe \pm \th_0 \right) \Gamma \left( -\a -\th_1 \pm \th_\infty \right)}{\Gamma \left( \a -\th_\qe \pm \th_0 \right) \Gamma \left( \a -\th_1 \pm \th_\infty \right)} \frac{(2\a)^2}{\prod_{\pm} (\a -\th_1 \pm \th_\infty )} e^{\frac{\p \widetilde{S}_+}{\p \a}} \nonumber\\
& + 4 \frac{\prod_{\pm} \sin \pi (\a - \th_\qe \pm \th_0) \sin \pi (\a -\th_1 \pm \th_\infty) }{\sin ^2 2\pi \a} \nonumber\\
&\quad \left(\frac{\Gamma \left( 2\a \right)^2}{\Gamma \left( -2\a \right)^2} \prod_{\pm} \frac{\Gamma \left( -\a -\th_\qe \pm \th_0 \right) \Gamma \left( -\a -\th_1 \pm \th_\infty \right)}{\Gamma \left( \a -\th_\qe \pm \th_0 \right) \Gamma \left( \a -\th_1 \pm \th_\infty \right)} \frac{(2\a)^2}{\prod_{\pm} (\a -\th_1 \pm \th_\infty )} \right)^{-1} e^{-\frac{\p \widetilde{S}_+}{\p \a}},
\end{split}
\end{align}
and
\begin{align}
\begin{split}
\text{Tr}\, M_B ^{-} &= \frac{\left( -\cos 2\pi \th_\infty +\cos 2\pi \th_1 \right) \left( \cos 2\pi \th_0 - \cos 2\pi \th_\qe \right)}{2 \sin ^2 \pi \a} \nonumber \\
& - \frac{\left( \cos 2\pi \th_\infty +\cos 2\pi \th_1 \right) \left( \cos 2\pi \th_0 + \cos 2\pi \th_\qe \right)}{2 \cos ^2 \pi \a} \nonumber\\
& + 4 \frac{\prod_{\pm} \sin \pi (-\a - \th_\qe \pm \th_0) \sin \pi (-\a -\th_1 \pm \th_\infty) }{\sin ^2 2\pi \a} \nonumber\\
&\quad \frac{\Gamma \left( 2\a \right)^2}{\Gamma \left( -2\a \right)^2} \prod_{\pm} \frac{\Gamma \left( -\a -\th_\qe \pm \th_0 \right) \Gamma \left( -\a -\th_1 \pm \th_\infty \right)}{\Gamma \left( \a -\th_\qe \pm \th_0 \right) \Gamma \left( \a -\th_1 \pm \th_\infty \right)} \frac{\prod_{\pm} (-\a -\th_1 \pm \th_\infty )}{(2\a)^2} e^{\frac{\p \widetilde{S}_-}{\p \a}} \nonumber\\
& + 4 \frac{\prod_{\pm} \sin \pi (\a - \th_\qe \pm \th_0) \sin \pi (\a -\th_1 \pm \th_\infty) }{\sin ^2 2\pi \a} \nonumber\\
&\quad \left(\frac{\Gamma \left( 2\a \right)^2}{\Gamma \left( -2\a \right)^2} \prod_{\pm} \frac{\Gamma \left( -\a -\th_\qe \pm \th_0 \right) \Gamma \left( -\a -\th_1 \pm \th_\infty \right)}{\Gamma \left( \a -\th_\qe \pm \th_0 \right) \Gamma \left( \a -\th_1 \pm \th_\infty \right)} \frac{\prod_{\pm} (-\a -\th_1 \pm \th_\infty )}{(2\a)^2} \right)^{-1} e^{-\frac{\p \widetilde{S}_-}{\p \a}}.
\end{split}
\end{align}
It is crucial to note that the products of $\Gamma$-functions appearing in the third and the fifth lines are precisely the contributions from the 1-loop part of the asymptotics of the surface defect partition function, namely, the effective twisted superpotential $\widetilde{S}$. Note that the 1-loop part of the surface defect partition function is given by
\begin{align}
\begin{split}
S^{\text{1-loop}}_\pm =& \lim _{\ve_2 \to 0} \frac{\ve_2}{\ve_1} \log \left[ \prod_{\xi=\pm} \frac{\Gamma_2 (0;\ve_1,\ve_2) \Gamma_2 \left( \xi 2\a \ve_1 ;\ve_1, \ve_2 \right)}{\prod_{\xi' = \pm} \Gamma_2 \left( (\xi \a - \th_\qe + \xi' \th_0)\ve_1 ; \ve_1, \ve_2 \right)  \Gamma_2 \left( (\xi \a - \th_1 + \xi' \th_\infty)\ve_1 ; \ve_1, \ve_2 \right)} \right. \\
& \quad\quad\quad\quad\quad \left.  \frac{\prod_{\xi =\pm} \Gamma_1 ( (\pm \a -\th_1 +\xi \th_\infty )\ve_1 ; \ve_2)}{\Gamma_1 (\pm2\a \ve_1;\ve_2)} \right],
\end{split}
\end{align}
where the first line is the 1-loop contribution from the bulk, while the second line is the 1-loop contribution from the surface defect on $z_2$-plane. The subscript $\pm$ denotes the choice of the vacuum of the gauged linear sigma model on the $z_2$-plane. Then we see that
\begin{subequations} \label{eq:1loopdrv}
\begin{align}
&\frac{\p S_+ ^{\text{1-loop}}}{\p \a} = \log \left[  \frac{\Gamma \left( 2\a \right)^2}{\Gamma \left( -2\a \right)^2} \prod_{\pm} \frac{\Gamma \left( -\a -\th_\qe \pm \th_0 \right) \Gamma \left( -\a -\th_1 \pm \th_\infty \right)}{\Gamma \left( \a -\th_\qe \pm \th_0 \right) \Gamma \left( \a -\th_1 \pm \th_\infty \right)} \frac{(2\a)^2}{\prod_{\pm} (\a -\th_1 \pm \th_\infty )} \right] \\
&\frac{\p S_- ^{\text{1-loop}}}{\p \a} = \log \left[  \frac{\Gamma \left( 2\a \right)^2}{\Gamma \left( -2\a \right)^2} \prod_{\pm} \frac{\Gamma \left( -\a -\th_\qe \pm \th_0 \right) \Gamma \left( -\a -\th_1 \pm \th_\infty \right)}{\Gamma \left( \a -\th_\qe \pm \th_0 \right) \Gamma \left( \a -\th_1 \pm \th_\infty \right)} \frac{\prod_{\pm} (-\a -\th_1 \pm \th_\infty )}{(2\a)^2} \right].
\end{align}
\end{subequations}
See appendix \ref{app:1loop} for the detail of the computation. Therefore, this factor can be absorbed into the 1-loop part of the effective twisted superpotential $\widetilde{S}$, simplifying the expression of the monodromy as
\begin{align} \label{eq:bcycle}
\begin{split}
\text{Tr}\, M_B &= \frac{\left( -\cos 2\pi \th_\infty +\cos 2\pi \th_1 \right) \left( \cos 2\pi \th_0 - \cos 2\pi \th_\qe \right)}{2 \sin ^2 \pi \a} \\
& - \frac{\left( \cos 2\pi \th_\infty +\cos 2\pi \th_1 \right) \left( \cos 2\pi \th_0 + \cos 2\pi \th_\qe \right)}{2 \cos ^2 \pi \a} \\
& + \sum_{\pm} 4 \frac{\prod_{\epsilon=\pm} \sin \pi (\mp \a - \th_\qe +\epsilon \th_0) \sin \pi (\mp \a -\th_1 +\epsilon \th_\infty) }{\sin ^2 2\pi \a}  e^{\pm\frac{\p \widetilde{S}}{\p \a}},
\end{split}
\end{align}
for both choices of the surface defect on the $z_2$-plane.

To recapitulate, we have expressed the monodromy data of the Fuchsian system associated to the surface defect in four dimensional gauge theory, in terms of the gauge-theoretic physical quantities. The monodromy along the small loops encircling the punctures are fixed by the hypermultiplet masses, re-expressed in dimensionless $\th$-parameters as \eqref{eq:r0}, \eqref{eq:rinf}, and \eqref{eq:rq1}. This implies that we restrict our monodromy space to the reduced space $\EuScript{M}_\qe \left( \th_0, \th_\qe+\frac{1}{2} ,\th_1,\th_\infty -\frac{1}{2} \right)$, as expected from the Fuchsian differential equation \eqref{eq:horibpz}. We use the Darboux coordinates $(\a, \b)$  of \cite{NRS2011} to parametrize this space. In turn, they are determined in terms of the Coulomb modulus $a$ and the effective twisted superpotential $\widetilde{S}$ by \eqref{eq:acycle} and \eqref{eq:bcycle}. Namely,
\begin{align}
\a = \frac{a}{\ve_1}, \quad\quad \b = \frac{\p \widetilde{S} (\a, \boldsymbol\th ,z,\qe)}{\p \a}.
\end{align}
Thus we have understood how the image $(\a,\b)$ under the Riemann-Hilbert map parameterizes the monodromy data, so that $\a$ and $\b$ are indeed preserved along the isomonodromic flow $z(\qe)$.

\section{Surface defects on blowup} \label{sec:blowup}
$\EuScript{N}=2$ supersymmetric gauge theories placed on the blowup $\widehat{\mathbb{C}}^2$ are useful to unveil non-trivial properties of the ordinary $\EuScript{N}=2$ supersymmetric gauge theories on the $\mathbb{C}^2$. The blowup is defined by replacing the origin $ 0 \in \mathbb{C}^2$ by an exceptional divisor $\mathbb{P}^1$, namely,
\begin{align}
\widehat{\mathbb{C}}^2 = \left\{ \left((z_1,z_2), (w_1 :w_2) \right) \in \mathbb{C}^2 \times \mathbb{P}^1\; \vert \; z_1 w_2 = z_2 w_1 \right\}.
\end{align}
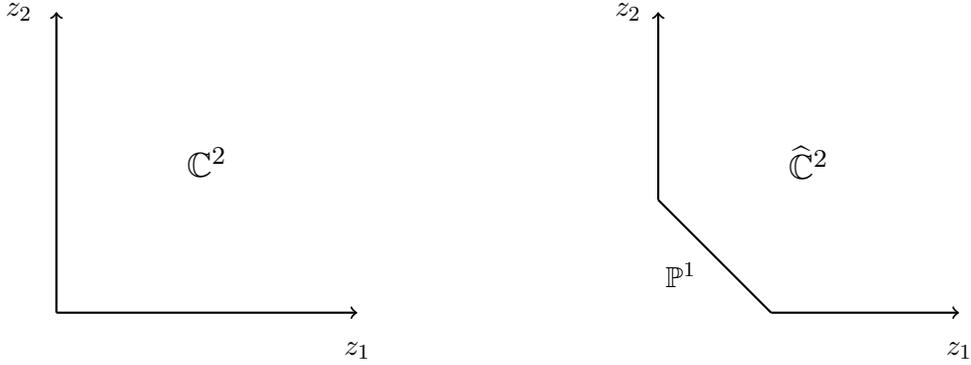
\begin{figure}
\centering
\begin{tikzpicture}
\path[black, thick] (-4,0) edge[->] (-4,4)  (-4,0) edge[->] (0,0);
\path[black, thick] (4,1.5) edge[->] (4,4)  (5.5,0) edge[->] (8,0);
\draw[black, thick] (4,1.5) -- (5.5,0);

\node at (-2,2) {\scalebox{1.2}{$\mathbb{C}^2$}};
\node at (6,2) {\scalebox{1.2}{$\widehat{\mathbb{C}}^2$}};
\node at (0,-0.5) {\scalebox{1.0}{$z_1$}};
\node at (-4.5,4) {\scalebox{1.0}{$z_2$}};
\node at (8,-0.5) {\scalebox{1.0}{$z_1$}};
\node at (3.6,4) {\scalebox{1.0}{$z_2$}};
\node at (4.3,0.5) {\scalebox{1.0}{$\mathbb{P}^1$}};
\end{tikzpicture}
\caption{The ordinary $\mathbb{C}^2$ and the blowup $\widehat{\mathbb{C}}^2$.} \label{fig:blowup}
\end{figure}The maximal torus of the spacetime isometry $U(1)_{\ve_1} \times U(1)_{\ve_2} \subset SO(4)$ of $\mathbb{C}^2$ uplifts to the isometry on $\widehat{\mathbb{C}}^2$ by
\begin{align}
\left( (z_1,z_2) , (w_1:w_2) \right) \longmapsto \left( (q_1 z_1, q_2 z_2), (q_1 w_1:q_2 w_2) \right), \quad (q_1,q_2) = \left(e^{\b \ve_1}, e^{\b \ve_2} \right) \in U(1)_{\ve_1} \times U(1)_{\ve_2}.
\end{align}
The fixed points of the isometry are the north and the south poles on the exceptional divisor. The spacetime locally looks like the ordinary $\mathbb{C}^2$ with different weights for the isometry action. Then the partition function of the $\EuScript{N}=2$ gauge theory on the blowup can be computed by properly multiplying the contributions from all the fixed points. Meanwhile, we may take the limit of the size of the exceptional divisor going to zero without affecting the physics, so that the partition function reduces to the one for the ordinary $\mathbb{C}^2$. The non-trivial identity satisfied by the partition function derived by this procedure essentially comprises what we call the \textit{blowup formula} for the $\EuScript{N}=2$ gauge theory partition functions. The non-trivial identity contains rich information on the gauge theory partition functions, and in particular it was used in \cite{ny} to exactly prove that the asymptotics of the partition function in the limit $\ve_1, \ve_2 \to 0$ is identical to the Seiberg-Witten prepotential.

An interesting question is how the blowup formula would work in the presence of non-local defects. For example, the insertion of Donaldson-type surface observables (more commonly known as two-observables of $\Tr \phi^2$) on the exceptional divisor was already discussed in \cite{ny}. In this section, we discuss the half-BPS surface defects that extend along one of the non-compact directions in $\widehat{\mathbb{C}}^2$, and suggest novel blowup formulas for their partition functions. We also discuss how those newly suggested blowup formulas are consistent with the analytic continuations of the complexified FI parameter of the gauged linear sigma model living on the surface defect, which have been explored in section \ref{subsec:flowz2def} and \cite{JN2018}. The application of the blowup formulas for the surface defect partition functions to the correspondence with the isomonodromic deformation of Fuchsian systems will be discussed in the next section.

\subsection{Blowup formula without surface defect}
We consider four-dimensional $\EuScript{N}=2$ supersymmetric $U(2)$ gauge theory with four fundamental hypermultiplets. The partition function depends on various equivariant parameters and the gauge coupling,
\begin{align}
\mathcal{Z} (\mathbf{a}, \mathbf{m}, \ve_1, \ve_2 ; \qe) = \mathcal{Z}^{\text{classical}} (\mathbf{a},  \ve_1, \ve_2 ; \qe) \mathcal{Z}^{\text{1-loop}} (\mathbf{a}, \mathbf{m}, \ve_1, \ve_2 ) \mathcal{Z}^{\text{inst}} (\mathbf{a}, \mathbf{m}, \ve_1, \ve_2 ; \qe).
\end{align}
Here, the perturbative parts are given by
\begin{align} \label{eq:partcl}
 \mathcal{Z}^{\text{classical}} (\mathbf{a}, \ve_1, \ve_2 ; \qe) = \qe^{-\frac{1}{2\ve_1 \ve_2} \sum_{\alpha=1} ^2 a_\alpha ^2}
\end{align}
and
\begin{align} \label{eq:part1loop}
\mathcal{Z}^{\text{1-loop}} (\mathbf{a}, \mathbf{m}, \ve_1, \ve_2 ) = \frac{\prod_{\alpha,\beta=1,2 }\Gamma_2 (a_\alpha -a_\beta ; \ve_1 ,\ve_2)}{\prod_{\substack{\alpha=1,2 \\ i=1,\cdots, 4}}\Gamma_2 (a_\alpha -m_{i} ;\ve_1, \ve_2)} .
\end{align}
Now we consider placing the theory on the blowup $\widehat{\mathbb{C}}^2$. The fixed points of the spacetime isometry are the north pole and the south pole of the exceptional divisor $\mathbb{P}^1$. Around these fixed points, the spacetime locally looks like a $\mathbb{C}^2$ with shifted weights of the isometry action. Hence when the size of the exceptional divisor is large, the partition function is computed as the sum over the fluxes on the exceptional divisor where the summand is a product of two ordinary partition functions with shifted arguments. Thus, we are led to the blowup formula for the partition function,
\begin{align}
\begin{split}
&\widehat{\mathcal{Z}}_{c_1 = kC} (\mathbf{a}, \mathbf{m}, \ve_1, \ve_2 ; \qe) \\ &= \sum_{\substack{\mathbf{n} = (n_1, n_2) \in \left( \mathbb{Z} + \frac{k}{2} \right)^2 \\ n_1 +n_2 =0}} \mathcal{Z} \left(\mathbf{a+n} \ve_1, \mathbf{m} + \frac{k}{2} \ve_1, \ve_1 , \ve_2-\ve_1 ; \qe\right) \mathcal{Z} \left(\mathbf{a+n}\ve_2, \mathbf{m} +\frac{k}{2} \ve_2, \ve_1-\ve_2, \ve_2 ; \qe\right),
\end{split}
\end{align}
where $C$ denotes the exceptional divisor and $c_1$ is the first Chern class of the torsion free sheave on the blowup $\widehat{\mathbb{C}}^2$ giving instanton. See \cite{ny} for more detail. We can determine the blowup partition function $\widehat{\mathcal{Z}}_{c_1 = kC} (\mathbf{a}, \mathbf{m}, \ve_1, \ve_2 ; \qe) $ by taking the opposite limit of the size of the exceptional divisor. When the total flux is zero ($k=0$), the gauge theory on the blowup reduces to the theory on $\mathbb{C}^2$ as the size of the exceptional divisor shrinks to zero, and we expect to recover the original partition function:
\begin{align} \label{eq:blowup1}
\begin{split}
&\mathcal{Z} (\mathbf{a}, \mathbf{m}, \ve_1, \ve_2 ; \qe)= \sum_{\substack{\mathbf{n} = (n_1, n_2) \in \mathbb{Z}^2 \\ n_1 +n_2 =0}} \mathcal{Z} (\mathbf{a+n} \ve_1, \mathbf{m}, \ve_1 , \ve_2-\ve_1 ; \qe) \mathcal{Z} (\mathbf{a+n}\ve_2, \mathbf{m} , \ve_1-\ve_2, \ve_2 ; \qe).
\end{split}
\end{align}
When the total flux is non-zero ($k=1$), the gauge theory on the blowup does not have a sensible limit as the size of the exceptional divisor shrinks to zero. Thus, we expect that the blowup partition function simply vanishes:
\begin{align} \label{eq:blowup2}
\begin{split}
0 = \sum_{\substack{\mathbf{n} = (n_1, n_2) \in \left( \mathbb{Z} + \frac{1}{2} \right)^2 \\ n_1 +n_2 =0}} \mathcal{Z} \left(\mathbf{a+n} \ve_1, \mathbf{m} + \frac{\ve_1}{2}, \ve_1 , \ve_2 -\ve_1; \qe\right) \mathcal{Z} \left(\mathbf{a+n}\ve_2, \mathbf{m} + \frac{\ve_2}{2}, \ve_1-\ve_2, \ve_2 ; \qe\right).
\end{split}
\end{align}

It is sometimes useful to write out the blowup formulas for the instanton part of the partition function. This can be accomplished by explicitly computing the ratio of perturbative parts \eqref{eq:partcl}-\eqref{eq:part1loop} appearing in these blowup formulas. For notational convenience, let us define the ratio of 1-loop part of the partition functions as follows,
\begin{align}
\begin{split}
L^{\mathbf{n},k} (\mathbf{a},\mathbf{m},\ve_1,\ve_2):= \frac{\mathcal{Z}^{\text{1-loop}} \left(\mathbf{a}+\mathbf{n} \ve_1 , \mathbf{m}+\frac{k}{2} \ve_1 , \ve_1,\ve_2-\ve_1\right) \mathcal{Z}^{\text{1-loop}} \left(\mathbf{a}+\mathbf{n} \ve_2 , \mathbf{m}+\frac{k}{2} \ve_2 , \ve_1-\ve_2,\ve_2\right)}{\mathcal{Z}^{\text{1-loop}} \left(\mathbf{a}  , \mathbf{m} , \ve_1,\ve_2\right)}.
\end{split}
\end{align}
Although this ratio of double gamma functions looks quite complicated, it is in fact a simple rational function in equivariant parameters (see appendix \ref{app:1loop}). Then the blowup formulas \eqref{eq:blowup1} and \eqref{eq:blowup2} become
\begin{align}
\begin{split}
&\mathcal{Z}^{\text{inst}} (\mathbf{a}, \mathbf{m}, \ve_1, \ve_2 ; \qe) \\
 =& \sum_{\substack{\mathbf{n} = (n_1, n_2) \in \mathbb{Z}^2 \\ n_1 +n_2 =0}} \qe^{\frac{1}{2} (n_1 ^2 + n_2 ^2)} L^{\mathbf{n},k=0} (\mathbf{a},\mathbf{m} ,\ve_1,\ve_2) \\
&\quad\quad\quad\quad\quad \mathcal{Z}^{\text{inst}} (\mathbf{a+n} \ve_1, \mathbf{m},  \ve_1 , \ve_2 -\ve_1 ; \qe)  \mathcal{Z}^{\text{inst}} (\mathbf{a+n}\ve_2, \mathbf{m}, , \ve_1-\ve_2, \ve_2 ; \qe),
\end{split}
\end{align}
and
\begin{align}
\begin{split}
0=& \sum_{\substack{\mathbf{n} = (n_1, n_2) \in \left( \mathbb{Z} + \frac{1}{2} \right)^2 \\ n_1 +n_2 =0}} \qe^{\frac{1}{2} (n_1 ^2 + n_2 ^2)} L^{\mathbf{n},k=1} (\mathbf{a},\mathbf{m} ,\ve_1,\ve_2) \\
&\quad\quad\quad\quad\quad \mathcal{Z}^{\text{inst}} \left(\mathbf{a+n} \ve_1, \mathbf{m} +\frac{\ve_1}{2},  \ve_1 , \ve_2 -\ve_1; \qe\right)  \mathcal{Z}^{\text{inst}} \left(\mathbf{a+n}\ve_2, \mathbf{m} +\frac{\ve_2}{2}, \ve_1-\ve_2, \ve_2 ; \qe\right).
\end{split}
\end{align}

\subsection{Blowup formula with surface defect} \label{subsec:blowupsurf}
Now we consider the $\EuScript{N}=2$ gauge theory with the insertion of a half-BPS surface defect. In particular, to make a contact with the system of isomonodromic deformation we need to insert a canonical surface defect which can be engineered by either a $\mathbb{Z}_2$-orbifold or a partial higgsing of a larger gauge group. The surface defect partition functions can be computed in both constructions. Since we can still put the theory on the blowup $\widehat{\mathbb{C}} ^2$, a natural question is what the blowup formula for this surface defect partition function would be.
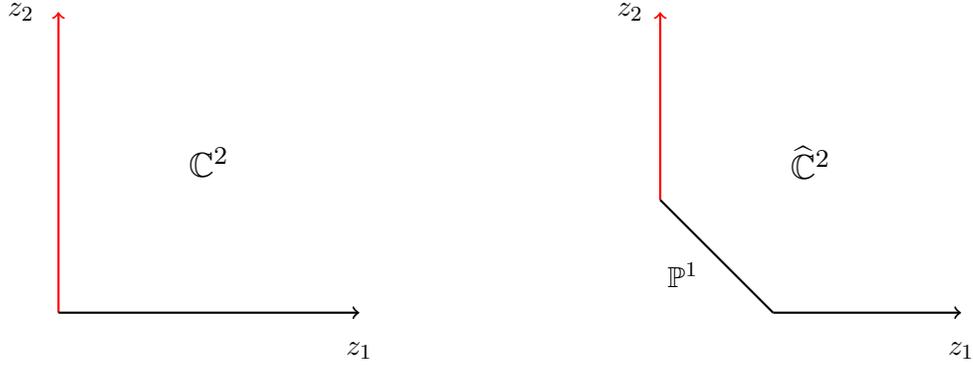
\begin{figure}
\centering
\begin{tikzpicture}
\path[red, thick] (-4,0) edge[->] (-4,4);
\path[black, thick] (-4,0) edge[->] (0,0);
\path[red, thick] (4,1.5) edge[->] (4,4);
\path[black, thick] (5.5,0) edge[->] (8,0);
\draw[black, thick] (4,1.5) -- (5.5,0);

\node at (-2,2) {\scalebox{1.2}{$\mathbb{C}^2$}};
\node at (6,2) {\scalebox{1.2}{$\widehat{\mathbb{C}}^2$}};
\node at (0,-0.5) {\scalebox{1.0}{$z_1$}};
\node at (-4.5,4) {\scalebox{1.0}{$z_2$}};
\node at (8,-0.5) {\scalebox{1.0}{$z_1$}};
\node at (3.6,4) {\scalebox{1.0}{$z_2$}};
\node at (4.3,0.5) {\scalebox{1.0}{$\mathbb{P}^1$}};
\end{tikzpicture}
\caption{A surface defect on the ordinary $\mathbb{C}^2$ and the blowup $\widehat{\mathbb{C}}^2$. The surface defect is extended along the non-compact $z_2$-plane, drawn in the red line. On the blowup $\widehat{\mathbb{C}}^2$, the surface defect is attached to the south pole of the exceptional divisor $\mathbb{P}^1$.} \label{fig:blowupsurf}
\end{figure}

There are two choices of the support of the surface defect which are consistent with the $\Omega$-background: the $z_1$-plane and the $z_2$-plane. Without loss of generality, we choose to insert the defect on the $z_2$-plane. Now when our theory is placed on the blowup $\widehat{\mathbb{C}}^2$, one of the two fixed points of the isometry, say, the north pole of the exceptional divisor is attached to the $z_1$-plane, while the south pole is attached to the $z_2$-plane. In turn, our theory looks very differently locally around those fixed points compared to the theory without a surface defect. Namely, on the local patch of $\mathbb{C}^2$ around the north pole we see our theory being absent of any defect insertion, while we recover the theory with the defect on the local patch of $\mathbb{C}^2$ around the south pole. Consequently, we expect the following blowup formula for the surface defect partition function to be satisfied,
\begin{align}
\begin{split}
&\widehat{\Psi}_{c_1 = kC} (\mathbf{a}, \mathbf{m}, \ve_1, \ve_2 ; \qe,z) \\ &= \sum_{\substack{\mathbf{n} = (n_1, n_2) \in \left( \mathbb{Z} + \frac{k}{2} \right)^2 \\ n_1 +n_2 =0}} \mathcal{Z} \left(\mathbf{a+n} \ve_1, \mathbf{m} + \frac{k}{2} \ve_1,  \ve_1 , \ve_2-\ve_1 ; \qe\right)  \Psi \left(\mathbf{a+n}\ve_2, \mathbf{m}  +\frac{k}{2} \ve_2,  \ve_1-\ve_2, \ve_2 ; \qe,z\right).
\end{split}
\end{align}
As in the case without the defect, we expect that the blowup partition function $\widehat{\Psi}_{c_1 = kC} (\mathbf{a}, \mathbf{m}, \ve_1, \ve_2 ; \qe,z)$ reduces to the original surface defect partition function if the total flux is zero ($k=0$). Therefore, we are led to
\begin{align} \label{eq:blowupsurf}
\begin{split}
&\Psi (\mathbf{a}, \mathbf{m}, \ve_1, \ve_2 ; \qe,z)= \sum_{\substack{\mathbf{n} = (n_1, n_2) \in \mathbb{Z}^2 \\ n_1 +n_2 =0}} \mathcal{Z} (\mathbf{a+n} \ve_1, \mathbf{m},  \ve_1 , \ve_2-\ve_1 ; \qe) \Psi (\mathbf{a+n}\ve_2, \mathbf{m},  \ve_1-\ve_2, \ve_2 ; \qe,z).
\end{split}
\end{align}
Now we can make an interesting expectation for the case when the total flux is non-zero ($k=1$). When there is no surface defect, the flux on any 2-cycle is always zero so that there is no sensible limit when the size of the exceptional divisor shrinks to zero, yieding zero for the blowup partition function. However, when we have a surface defect, there is already a non-zero flux along the support of the defect so that we can expect that the flux along the exceptional divisor gets absorbed into the support of the defect when we blow down the theory to $\mathbb{C}^2$. Hence, we expect that the blowup partition function is, instead of being zero, again proportional to the surface defect partition function $\mathbb{C}^2$,
\begin{align} \label{eq:blowhalf}
\begin{split}
&\Psi (\mathbf{a}, \mathbf{m}, \varepsilon_1, \varepsilon_2 ; \mathfrak{q},z) \\ \sim & \sum_{\substack{\mathbf{n} = (n_1, n_2) \in \left( \mathbb{Z} + \frac{1}{2} \right)^2 \\ n_1 +n_2 =0}} \mathcal{Z}\left(\mathbf{a+n} \ve_1,\mathbf{m}  + \frac{\ve_1}{2}, \varepsilon_1 , \varepsilon_2-\ve_1; \mathfrak{q}\right)  \Psi \left(\mathbf{a +n} \ve_2, \mathbf{m} +\frac{\ve_2}{2}, \varepsilon_1-\ve_2, \varepsilon_2  ; \mathfrak{q},z\right).
\end{split}
\end{align}
The precise multiplicative factor in front is to be determined (it cannot be $1$ because of the mismatch of the classical part at least), but it already exhibits a drastic difference from the blowup formula for the ordinary partition function without the defect, where the left hand side is simply zero.

\subsubsection{Orbifold}
In section \ref{sec:orbdef}, we have seen that the orbifolding of the spacetime can be used to engineer a surface defect in the gauge theory. In our main example of $U(2)$ gauge theory with four hypermultiplets on the $\mathbb{Z}_2$-orbifold, we can explicitly write down the partition function \eqref{eq:orbpart} as
\begin{align}
\Psi_\b ^{\mathbb{Z}_2} (\mathbf{a},\mathbf{m},\ve_1,\ve_2;\qe,z) = \Psi_\b^{\mathbb{Z}_2 ,\text{classical}} (\mathbf{a},\ve_1,\ve_2;\qe,z)\Psi_\b ^{\mathbb{Z}_2 ,\text{1-loop}} (\mathbf{a},\mathbf{m},\ve_1,\ve_2) \Psi_\b ^{\mathbb{Z}_2 ,\text{non-pert}} (\mathbf{a},\mathbf{m},\ve_1,\ve_2;\qe,z),
\end{align}
where $\b \in \{1,2\}$ is the choice of the vacuum of the gauged linear sigma model on the orbifold surface defect, realized by the $\mathbb{Z}_2$-coloring. We have seen that the exponent of the complexified FI parameter $z$ satisfies $0<\vert \qe \vert <\vert z \vert<1$. The classical part of the partition function is simply
\begin{align}
\Psi^{\mathbb{Z}_2 ,\text{classical}}_\beta (\mathbf{a},\ve_1,\ve_2 ; \qe,z) = z^{-\frac{a_{\beta}-a_{\bar\beta}}{2\ve_2}} \qe^{-\frac{a_\b ^2 +a_{\bar\beta} ^2}{2\ve_1 \ve_2}}
\end{align}
and the 1-loop part is
\begin{align}
\Psi^{\mathbb{Z}_2 ,\text{1-loop}}_\beta (\mathbf{a},\mathbf{m}, \ve_1,\ve_2) =   \frac{\prod_{\a,\a' =1,2} \Gamma_2 ( a_\a -a_{\a'}  ;\ve_1,\ve_2)  }{\prod_{\a=1,2}\prod_{i=1}^4 \Gamma_2 ( a_\a - m_i;\ve_1,\ve_2)} \frac{ \prod_{i=1,2} \Gamma_1 (a_\b - m_i ;\ve_2)}{\Gamma_1 ( a_\b -a_{\bar\b};\ve_2)}.
\end{align}
The instanton part is given by
\begin{align}
\begin{split}
&\Psi^{\mathbb{Z}_2 ,\text{non-pert}}_\beta (\mathbf{a},\mathbf{m},\ve_1,\ve_2;\qe,z) \\
&= \sum_{\boldsymbol\Lambda} \qe^{\vert\boldsymbol\Lambda \vert} E \left[ \mathcal{T}[\boldsymbol\Lambda] \right] \\
&\sum_{\boldsymbol\l \in \rho^{-1} (\boldsymbol\Lambda)} z^{k_1 -k_0} E\left[ (K_1 ^* -K_0 ^*) (N_1 -P_2 K_1 +q_1 P_2 K_0 -M_1) +q_2 N_0 ^* (K_1 -q_1 K_0) \right].
\end{split}
\end{align}
The blowup formula \eqref{eq:blowupsurf} in the absence of the flux $k=0$ is expected to be satisfied by the full surface defect partition function $\Psi_\b (\mathbf{a},\mathbf{m},\ve_1,\ve_2;\qe,z)$. By plugging the explicit forms of the perturbative parts into the formula \eqref{eq:blowupsurf}, we obtain the blowup formula for the non-perturbative part of the partition function,
\begin{align} \label{eq:blowuporbint}
\begin{split}
\Psi^{\mathbb{Z}_2 ,\text{non-pert}}_\beta &(\mathbf{a},\mathbf{m}, \ve_1, \ve_2; \qe, z ) =  \sum_{\substack{\mathbf{n} = (n_1, n_2) \in \mathbb{Z}^2 \\ n_1 +n_2 =0}} \qe^{\frac{1}{2} (n_1 ^2 + n_2 ^2 )} z^{\frac{1}{2} (n_{\bar\beta}-n_\beta)} L^{\mathbf{n},k=0} (\mathbf{a},\mathbf{m} ,\ve_1,\ve_2) \\
& \frac{\Gamma \left( \frac{a_\beta-a_{\bar\beta}}{\ve_2}+n_\beta-n_{\bar\beta} \right)}{\Gamma \left( \frac{a_\beta-a_{\bar\beta}}{\ve_2} \right)} \prod_{i=1,2} \frac{\Gamma \left( \frac{a_{\beta} - m_{i}}{\ve_2} \right)}{\Gamma \left( \frac{a_{\beta} -m_{i}}{\ve_2} + n_\beta  \right)} \\
& \mathcal{Z}^{\text{inst}} (\mathbf{a}+\mathbf{n} \ve_1, \mathbf{m}, \ve_1,\ve_2-\ve_1;\qe) \Psi_\beta ^{\mathbb{Z}_2 ,\text{non-pert}} (\mathbf{a} + \mathbf{n}\ve_2 , \mathbf{m} ,\ve_1-\ve_2,\ve_2;\qe,z).
\end{split}
\end{align}
This is indeed a non-trivial identity to be satisfied by the surface defect partition function. We checked that this identity holds in series expansion in $\frac{\qe}{z}$ and $z$.

It is possible to repeat this exercise to the case of non-zero flux $k=1$, \eqref{eq:blowhalf}. Remarkably, we find another non-trivial identity satisfied by the surface defect partition function:
\begin{align} \label{eq:blowuporbhalf}
\begin{split}
&\Psi_\b ^{\mathbb{Z}_2 } (\mathbf{a}, \mathbf{m}, \varepsilon_1, \varepsilon_2 ; \mathfrak{q},z) \\ = & -\qe^{-\frac{1}{4}} z^{\frac{1}{2}} \frac{1-\qe}{1-z}  \\ &\sum_{\substack{\mathbf{n} = (n_1, n_2) \in \left( \mathbb{Z} + \frac{1}{2} \right)^2 \\ n_1 +n_2 =0}} \mathcal{Z}\left(\mathbf{a+n} \ve_1,\mathbf{m}  + \frac{\ve_1}{2}, \varepsilon_1 , \varepsilon_2-\ve_1; \mathfrak{q}\right)  \Psi_\b ^{\mathbb{Z}_2 } \left(\mathbf{a +n} \ve_2, \mathbf{m} +\frac{\ve_2}{2}, \varepsilon_1-\ve_2, \varepsilon_2  ; \mathfrak{q},z\right).
\end{split}
\end{align}
Note that the left hand side is non-zero but proportional to the surface defect partition function even though the sum is taken over half-integers. We have checked this identitiy in series expansion in $\frac{\qe}{z}$ and $z$.

\subsubsection{Vortex string}

The vortex string surface defect in the $U(2)$ gauge theory with four hypermultiplets can be engineered by starting from the quiver $U(2) \times U(2)$ gauge theory and partially higgsing the gauge group \cite{Nekrasov_BPS45, JN2018}. The surface defect partition function can be decomposed into the classical part, the 1-loop part, and the non-perturbative part:
\begin{align}
\Psi_\b ^{L} (\mathbf{a},\mathbf{m},\ve_1,\ve_2;\qe,z) = \Psi_\b^{L, \text{classical}} (\mathbf{a},\mathbf{m},\ve_1,\ve_2;\qe,z)\Psi_\b ^{L,\text{1-loop}} (\mathbf{a},\mathbf{m},\ve_1,\ve_2) \Psi_\b ^{L,\text{non-pert}} (\mathbf{a},\mathbf{m},\ve_1,\ve_2;\qe,z),
\end{align}
where $\b \in \{1,2\}$ is the choice of the vacuum of the gauged linear sigma model on the $z_2$-plane. Depending on the contents of the gauged linear sigma model on the $z_2$-plane and its coupling to the bulk gauge theory, all the pieces of the partition function vary accordingly. Let us start from the vortex string surface defect, whose complexified FI parameter is in the domain $0 < \vert \qe\vert <1<\vert z \vert$. The classical part of the partition function is
\begin{align}
\Psi^{L,\text{classical}}_\beta (\mathbf{a}, \mathbf{m},\ve_1,\ve_2 ; \qe,z) = z^{-\frac{m_{\beta}-m_{\bar\beta}}{2\ve_2}} \qe^{-\frac{a_1 ^2 +a_2 ^2}{2\ve_1 \ve_2}},
\end{align}
and the 1-loop part is given by
\begin{align}
\Psi^{L,\text{1-loop}}_\beta (\mathbf{a},\mathbf{m}, \ve_1,\ve_2) =   \frac{\prod_{\a,\a' =1,2} \Gamma_2 ( a_\a -a_{\a'}  ;\ve_1,\ve_2)  }{\prod_{\a=1,2}\prod_{i=1}^4 \Gamma_2 ( a_\a - m_i;\ve_1,\ve_2)} \frac{ \prod_{\a=1,2} \Gamma_1 (a_\a - m_\b ;\ve_2)}{\Gamma_1 ( m_\b -m_{\bar\b};\ve_2)}.
\end{align}
The non-perturbative part can be written out as
\begin{align}
 \Psi_\b ^{L,\text{non-pert}} (\mathbf{a},\mathbf{m},\ve_1,\ve_2;\qe,z) = \sum_{\boldsymbol\l} \qe^{\vert \boldsymbol\l \vert} E\left[ \mathcal{T}[ \boldsymbol\l ]\right] \sum_{k=0} ^\infty z^{-k}  \prod_{l=1} ^{k} \frac{\EuScript{Y} (m_{\b} +l \ve_2) \left[ \boldsymbol\l \right]}{ l\ve_2 (m_\b -m_{\bar\b}+l\ve_2)} \prod_{\Box \in \boldsymbol\l} \frac{m_{\b} -\ve_1 -c_\Box}{m_{\b}-c_\Box}.
\end{align}
The blowup formula \eqref{eq:blowupsurf} in the absence of the flux $k=0$ is expected to be satisfied by the full partition function $\Psi_\b (\mathbf{a},\mathbf{m},\ve_1,\ve_2;\qe,z) $.
We can convert this equation to a non-trivial identity satisfied by the non-perturbative part, by explicitly substituting the perturbative part. The result is
\begin{align} \label{eq:blowupL}
\begin{split}
& \Psi^{L,\text{non-pert}}_\beta (\mathbf{a},\mathbf{m}, \ve_1, \ve_2; \qe, z ) \\
&=  \sum_{\substack{\mathbf{n} = (n_1, n_2) \in \mathbb{Z}^2 \\ n_1 +n_2 =0}} \qe^{\frac{1}{2} (n_1 ^2 + n_2 ^2 )}  L^{\mathbf{n},k=0} (\mathbf{a},\mathbf{m} ,\ve_1,\ve_2) \prod_{\g=1,2} \frac{\Gamma \left( \frac{a_{\g} - m_{\b}}{\ve_2} \right)}{\Gamma \left( \frac{a_{\g} -m_{\b}}{\ve_2} + n_\g  \right)} \\
&\quad\quad\quad\quad\quad \mathcal{Z}^{\text{inst}} (\mathbf{a}+\mathbf{n} \ve_1, \mathbf{m}, \ve_1,\ve_2-\ve_1;\qe) \Psi_\beta ^{L,\text{non-pert}} (\mathbf{a} + \mathbf{n}\ve_2 , \mathbf{m} ,\ve_1-\ve_2,\ve_2;\qe,z).
\end{split}
\end{align}
This is indeed a non-trivial identity to be satisfied by the surface defect partition function. We have checked this identity in series expansion in $\qe$ and $z^{-1}$.

We can do a similar exercise to the case of non-zero flux $k=1$. We find another non-trivial identity:
\begin{align} \label{eq:vortexhalf}
\begin{split}
&\Psi_\b ^{L} (\mathbf{a}, \mathbf{m}, \varepsilon_1, \varepsilon_2 ; \mathfrak{q},z) \\ = &\qe^{-\frac{1}{4}} (1-\qe) \\&  \sum_{\substack{\mathbf{n} = (n_1, n_2) \in \left( \mathbb{Z} + \frac{1}{2} \right)^2 \\ n_1 +n_2 =0}} \mathcal{Z}\left(\mathbf{a+n} \ve_1,\mathbf{m}  + \frac{\ve_1}{2}, \varepsilon_1 , \varepsilon_2-\ve_1; \mathfrak{q}\right)  \Psi_\b ^{L} \left(\mathbf{a +n} \ve_2, \mathbf{m} +\frac{\ve_2}{2}, \varepsilon_1-\ve_2, \varepsilon_2  ; \mathfrak{q},z\right).
\end{split}
\end{align}
Note that the left hand side is non-zero but proportional to the surface defect partition function even though the summation is taken over half-integers. We have checked this identitiy in series expansion in $\qe$ and $z^{-1}$.

Next, let us turn to the vortex string surface defect whose complexified FI parameter is in the domain $0<\vert z \vert<\vert \qe\vert<1$. The surface defect partition function is again decomposed into the classical part, the 1-loop part, and the non-perturbative part:
\begin{align}
\Psi_\b ^{R} (\mathbf{a},\mathbf{m},\ve_1,\ve_2;\qe,z) = \Psi_\b^{R, \text{classical}} (\mathbf{a},\mathbf{m},\ve_1,\ve_2;\qe,z)\Psi_\b ^{R,\text{1-loop}} (\mathbf{a},\mathbf{m},\ve_1,\ve_2) \Psi_\b ^{R,\text{non-pert}} (\mathbf{a},\mathbf{m},\ve_1,\ve_2;\qe,z),
\end{align}
where $\b \in \{1,2\}$ is the choice of the vacuum of the gauged linear sigma model on the $z_2$-plane. The classical part is given by
\begin{align}
\Psi^{R,\text{classical}}_\beta (\mathbf{a}, \mathbf{m},\ve_1,\ve_2 ; \qe,z) = z^{-\frac{m_{\beta+2}-m_{\bar\beta +2}}{2\ve_2}} \qe^{-\frac{a_1 ^2 +a_2 ^2}{2\ve_1 \ve_2}},
\end{align}
and the 1-loop part is given by
\begin{align}
\Psi^{R,\text{1-loop}}_\beta (\mathbf{a},\mathbf{m}, \ve_1,\ve_2) =   \frac{\prod_{\a,\a' =1,2} \Gamma_2 ( a_\a -a_{\a'}  ;\ve_1,\ve_2)  }{\prod_{\a=1,2}\prod_{i=1}^4 \Gamma_2 ( a_\a - m_i;\ve_1,\ve_2)} \frac{ \prod_{\a=1,2} \Gamma_1 (m_{\b+2} +\ve-a_\a ;\ve_2)}{\Gamma_1 ( m_{\b+2} -m_{\bar\b+2};\ve_2)}.
\end{align}
Finally, the non-perturbative part is given by
\begin{align}
\begin{split}
&\Psi_{\b}^{R,\text{non-pert} }(\mathbf{a},\mathbf{m},\ve_1,\ve_2;\qe,z) \\
&= \sum_{\boldsymbol\l} \qe ^{\vert \boldsymbol\l  \vert}   E \left[ \mathcal{T}_{A_1} ' \left[ \boldsymbol\l \right] \right]  \sum_{k=0} ^\infty \left(\frac{z}{\qe}\right)^k \prod_{l=1} ^{k} \frac{\EuScript{Y}' (-m_{\b+2} -\ve +l\ve_2) \left[ \boldsymbol\l \right]}{P_3 ' (-m_{\b+2} + \ve +l \ve_2)} \prod_{\Box \in \boldsymbol\l} \frac{-m_{\b+2} -\ve  -\ve_1 -c_\Box '}{-m_{\b+2} -\ve  -c_\Box '}.
\end{split}
\end{align}

We expect the full surface defect partition function to satisfy the blowup formula \eqref{eq:blowupsurf} in the absence of the flux $k=0$. In fact, we observe there is a slight deviation from this naive expectation, and the equality holds with an additional multiplicative prefactor:
\begin{align} \label{eq:vortexrightinteger}
\begin{split}
&\Psi_\b ^{R} (\mathbf{a}, \mathbf{m}, \varepsilon_1, \varepsilon_2 ; \mathfrak{q},z) \\ &= (1-\qe) ^{\frac{\ve_2 \left(2a_1 +2a_2 -2\ve -2\ve_1 -\sum_{i=1} ^4 m_i \right)}{\ve_1 (\ve_1 -\ve_2)}} \\
&\sum_{\substack{\mathbf{n} = (n_1, n_2) \in \mathbb{Z}^2 \\ n_1 +n_2 =0}}  \mathcal{Z} (\mathbf{a+n} \ve_1, \mathbf{m}+\ve_1,  \ve_1 , \ve_2-\ve_1 ; \qe) \Psi_\b ^{R} (\mathbf{a+n}\ve_2, \mathbf{m}+\ve_2,  \ve_1-\ve_2, \ve_2 ; \qe,z).
\end{split} 
\end{align}
Reducing this formula to the one for the non-perturbative part, we obtain
\begin{align}
\begin{split}
&\Psi^{R,\text{non-pert}} _\b (\mathbf{a}, \mathbf{m}, \varepsilon_1, \varepsilon_2 ; \mathfrak{q},z) \\ &= (1-\qe) ^{\frac{\ve_2 \left(2a_1 +2a_2 -2\ve -2\ve_1 -\sum_{i=1} ^4 m_i \right)}{\ve_1 (\ve_1 -\ve_2)}} \\
&\sum_{\substack{\mathbf{n} = (n_1, n_2) \in \mathbb{Z}^2 \\ n_1 +n_2 =0}} \qe^{\frac{1}{2} (n_1 ^2 +n_2 ^2)} L^{\mathbf{n},k=0} \prod_{\g=1,2} \frac{\Gamma \left( \frac{m_{\b+2} +\ve-a_\g}{\ve_2}  \right)}{\Gamma \left( \frac{m_{\b+2} +\ve -a_\g}{\ve_2} -n_\g  \right)} \\
& \quad\quad\mathcal{Z}^{\text{inst}} (\mathbf{a+n} \ve_1, \mathbf{m}+\ve_1,  \ve_1 , \ve_2-\ve_1 ; \qe) \Psi^{R,\text{non-pert}}_\b (\mathbf{a+n}\ve_2, \mathbf{m}+\ve_2,  \ve_1-\ve_2, \ve_2 ; \qe,z).
\end{split}
\end{align}
This blowup formula looks more natural when all the masses are re-defined to be anti-fundamentals. Let us define $\mathbf{m}' = \left( m_f ' \right)_{f=1} ^4= \left( m_f + \ve  \right)_{f=1} ^4$. Then the above blowup formula can be expressed as
\begin{align} \label{eq:vortexrightanti}
\begin{split}
&\Psi^{R,\text{non-pert}} _\b (\mathbf{a}, \mathbf{m}', \varepsilon_1, \varepsilon_2 ; \mathfrak{q},z) \\ &= (1-\qe) ^{\frac{\ve_2 \left(2a_1 +2a_2 +2\ve_2 -\sum_{i=1} ^4 m_i ' \right)}{\ve_1 (\ve_1 -\ve_2)}} \\
&\sum_{\substack{\mathbf{n} = (n_1, n_2) \in \mathbb{Z}^2 \\ n_1 +n_2 =0}} \qe^{\frac{1}{2} (n_1 ^2 +n_2 ^2)} {L'}^{\mathbf{n},k=0}  \prod_{\g=1,2} \frac{\Gamma \left( \frac{m_{\b+2}'-a_\g}{\ve_2}  \right)}{\Gamma \left( \frac{m_{\b+2}' -a_\g}{\ve_2} -n_\g  \right)} \\
& \quad\quad\mathcal{Z}^{\text{inst}} (\mathbf{a+n} \ve_1, \mathbf{m}',  \ve_1 , \ve_2-\ve_1 ; \qe) \Psi^{R,\text{non-pert}}_\b (\mathbf{a+n}\ve_2, \mathbf{m}',  \ve_1-\ve_2, \ve_2 ; \qe,z).
\end{split}
\end{align}
In particular, there is no shift in the new mass parameters. The $L'$ here is the ratio of bulk 1-loop contributions with anti-fundamentals,
\begin{align}
\begin{split}
{L'}^{\mathbf{n},k} (\mathbf{a},\mathbf{m},\ve_1,\ve_2):= \frac{{\mathcal{Z}'}^{\text{1-loop}} \left(\mathbf{a}+\mathbf{n} \ve_1 , \mathbf{m}'-\frac{k}{2} \ve_1 , \ve_1,\ve_2-\ve_1\right) {\mathcal{Z}'}^{\text{1-loop}} \left(\mathbf{a}+\mathbf{n} \ve_2 , \mathbf{m}'-\frac{k}{2} \ve_2 , \ve_1-\ve_2,\ve_2\right)}{{\mathcal{Z}'}^{\text{1-loop}} \left(\mathbf{a}  , \mathbf{m}' , \ve_1,\ve_2\right)},
\end{split}
\end{align}
where the 1-loop part of the bulk partition function with anti-fundamentals is
\begin{align}
{\mathcal{Z}'}^{\text{1-loop}} \left(\mathbf{a}  , \mathbf{m}' , \ve_1,\ve_2\right) =\frac{\prod_{\alpha,\beta=1,2 }\Gamma_2 (a_\alpha -a_\beta ; \ve_1 ,\ve_2)}{\prod_{\substack{\alpha=1,2 \\ i=1,\cdots, 4}}\Gamma_2 (m'_{i} -a_\a ;\ve_1, \ve_2)} .
\end{align}

For the case of the non-zero flux $k=1$, similarly, we have
\begin{align} \label{eq:blowupRhalf}
\begin{split}
&\Psi_\b ^{R} (\mathbf{a}, \mathbf{m}, \varepsilon_1, \varepsilon_2 ; \mathfrak{q},z) \\ &= (1-\qe) ^{1+\frac{\ve_2 \left(2a_1 +2a_2 -2\ve  -\sum_{i=1} ^4 m_i \right)}{\ve_1 (\ve_1 -\ve_2)}} \\
&\sum_{\substack{\mathbf{n} = (n_1, n_2) \in \left( \mathbb{Z} +\frac{1}{2} \right)^2 \\ n_1 +n_2 =0}}  \mathcal{Z} \left(\mathbf{a+n} \ve_1, \mathbf{m}+\frac{\ve_1}{2},  \ve_1 , \ve_2-\ve_1 ; \qe \right) \Psi_\b ^{R} \left(\mathbf{a+n}\ve_2, \mathbf{m}+\frac{\ve_2}{2},  \ve_1-\ve_2, \ve_2 ; \qe,z \right).
\end{split} 
\end{align}
In terms of the partition functions with anti-fundamentals, this blowup formula can also be written as
\begin{align}
\begin{split}
&\Psi_\b ^{R} (\mathbf{a}, \mathbf{m}', \varepsilon_1, \varepsilon_2 ; \mathfrak{q},z) \\ &= (1-\qe) ^{1+\frac{\ve_2 \left(2a_1 +2a_2 +2\ve  -\sum_{i=1} ^4 m'_i \right)}{\ve_1 (\ve_1 -\ve_2)}} \\
&\sum_{\substack{\mathbf{n} = (n_1, n_2) \in \left( \mathbb{Z} +\frac{1}{2} \right)^2 \\ n_1 +n_2 =0}}  \mathcal{Z} \left(\mathbf{a+n} \ve_1, \mathbf{m}'-\frac{\ve_1}{2},  \ve_1 , \ve_2-\ve_1 ; \qe \right) \Psi_\b ^{R} \left(\mathbf{a+n}\ve_2, \mathbf{m}'-\frac{\ve_2}{2},  \ve_1-\ve_2, \ve_2 ; \qe,z \right).
\end{split} 
\end{align}

\subsubsection{Analytic continuations and blowup formulas}
As we have learned in section \ref{subsec:surfacedef}, the surface defect partition functions are expressed as series expansions in certain convergence domains, either when they are engineered by orbifolding or partial higgsing. Accordingly, the blowup formulas for the surface defect partition functions that we have seen in the previous discussion were checked in specific convergence domains, where the very surface defect partition functions lie in. Meanwhile, we also have seen the analytic continuations across those convergence domains connect different surface defect partition functions through connection formulas. Thus a natural question that arises is whether the validity of the blowup formulas is not affected across different convergence domains.

The expectation is naturally that the blowup formulas in different convergence domains connect to each other by the connection formulas for the surface defect partition functions. Hence, this is also a non-trivial check on the validity of the blowup formulas for the surface defect partition functions, in the sense that the blowup formulas are grouped together by if and only conditions. In what follows, we explain that the naive expectation is indeed true.

Let us begin with the $\mathbb{Z}_2$-orbifold surface defect, whose expectation value converges in the domain $0<\vert \qe \vert<\vert z\vert<1$. When there is no flux through the exceptional divisor, the blowup formula reads
\begin{align}
\begin{split}
\Psi^{\mathbb{Z}_2 ,\text{non-pert}}_\beta &(\mathbf{a},\mathbf{m}, \ve_1, \ve_2; \qe, z ) =  \sum_{\substack{\mathbf{n} = (n_1, n_2) \in \mathbb{Z}^2 \\ n_1 +n_2 =0}} \qe^{\frac{1}{2} (n_1 ^2 + n_2 ^2 )} z^{\frac{1}{2} (n_{\bar\beta}-n_\beta)} L^{\mathbf{n},k=0} (\mathbf{a},\mathbf{m} ,\ve_1,\ve_2) \\
& \frac{\Gamma \left( \frac{a_\beta-a_{\bar\beta}}{\ve_2}+n_\beta-n_{\bar\beta} \right)}{\Gamma \left( \frac{a_\beta-a_{\bar\beta}}{\ve_2} \right)} \prod_{i=1,2} \frac{\Gamma \left( \frac{a_{\beta} - m_{i}}{\ve_2} \right)}{\Gamma \left( \frac{a_{\beta} -m_{i}}{\ve_2} + n_\beta  \right)} \\
& \mathcal{Z}^{\text{inst}} (\mathbf{a}+\mathbf{n} \ve_1, \mathbf{m}, \ve_1,\ve_2-\ve_1;\qe) \Psi_\beta ^{\mathbb{Z}_2,\text{non-pert}} (\mathbf{a} + \mathbf{n}\ve_2 , \mathbf{m} ,\ve_1-\ve_2,\ve_2;\qe,z).
\end{split}
\end{align}
As we have seen in section \ref{subsec:flowz2def}, we need to multiply an appropriate connection matrix to make use of the connection formula corresponding to the adiabatic flow of the surface defect on the $z_2$-plane.\footnote{In section \ref{subsec:flowz2def} we considered the analytic continuation of the intersecting surface defect partition functions. Here we have a surface defect only on the $z_2$-plane, but the connection formulas and the connection matrices for the analytic continuation are exactly the same. See \cite{JN2018}.} We multiply entries of the connection matrix $\mathbf{C}_\infty ^{(2)}$ to the blowup equation to yield
\begin{align}
\begin{split}
&\left( \mathbf{C}_\infty ^{(2)} (\mathbf{a},\mathbf{m},\ve_2) \right)_{\a\b}  \Psi^{\mathbb{Z}_2,\text{non-pert}} _\b (\mathbf{a} , \mathbf{m},\ve_1,\ve_2;\qe,z)  \\
&=  \sum_{\substack{\mathbf{n} = (n_1, n_2) \in \mathbb{Z}^2 \\ n_1 +n_2 =0}} \qe^{\frac{1}{2} (n_1 ^2 + n_2 ^2 )} z^{-\frac{1}{2}(n_\b - n_{\bar\b})}  L^{\mathbf{n},k=0} (\mathbf{a},\mathbf{m} ,\ve_1,\ve_2) \\
& \frac{\Gamma \left( \frac{a_\beta-a_{\bar\beta}}{\ve_2}+n_\beta-n_{\bar\beta} \right)}{\Gamma \left( \frac{a_\beta-a_{\bar\beta}}{\ve_2} \right)} \prod_{i=1,2} \frac{\Gamma \left( \frac{a_{\beta} - m_{i}}{\ve_2} \right)}{\Gamma \left( \frac{a_{\beta} -m_{i}}{\ve_2} + n_\beta  \right)}  \frac{\left( \mathbf{C}_\infty ^{(2)} (\mathbf{a},\mathbf{m},\ve_2) \right)_{\a\b} }{\left( \mathbf{C}_\infty ^{(2)} (\mathbf{a} +\mathbf{n} \ve_2,\mathbf{m},\ve_2) \right)_{\a\b} } \\
& \mathcal{Z}^{\text{inst}} (\mathbf{a}+\mathbf{n} \ve_1, \mathbf{m}, \ve_1,\ve_2-\ve_1;\qe) \\
& \left( \mathbf{C}_\infty ^{(2)} (\mathbf{a} +\mathbf{n}\ve_2,\mathbf{m},\ve_2) \right)_{\a\b}  \Psi^{\mathbb{Z}_2,\text{non-pert}} _\beta  (\mathbf{a} + \mathbf{n}\ve_2 , \mathbf{m} ,\ve_1-\ve_2,\ve_2;\qe,z).
\end{split}
\end{align}
A straightforward computation shows that the ratio of $\Gamma$-functions in the middle simplifies to
\begin{align}
 \frac{\Gamma \left( \frac{a_\beta-a_{\bar\beta}}{\ve_2}+n_\beta-n_{\bar\beta} \right)}{\Gamma \left( \frac{a_\beta-a_{\bar\beta}}{\ve_2} \right)} \prod_{i=1,2} \frac{\Gamma \left( \frac{a_{\beta} - m_{i}}{\ve_2} \right)}{\Gamma \left( \frac{a_{\beta} -m_{i}}{\ve_2} + n_\beta  \right)}  \frac{\left( \mathbf{C}_\infty ^{(2)} (\mathbf{a},\mathbf{m},\ve_2) \right)_{\a\b} }{\left( \mathbf{C}_\infty ^{(2)} (\mathbf{a} +\mathbf{n} \ve_2,\mathbf{m},\ve_2) \right)_{\a\b} } = \prod_{\g=1,2} \frac{\Gamma\left( \frac{a_\g -m_\a}{\ve_2} \right) }{\Gamma\left( \frac{a_\g -m_\a}{\ve_2} +n_\g \right)}.
\end{align}
In particular, this expression is independent of $\b$. The blowup formula thus becomes
\begin{align}
\begin{split}
& \left( \mathbf{C}_\infty ^{(2)} (\mathbf{a},\mathbf{m},\ve_2) \right)_{\a\b}  \Psi^{\mathbb{Z}_2,\text{non-pert}}_\b (\mathbf{a} , \mathbf{m},\ve_1,\ve_2;\qe,z)  \\
&=  \sum_{\substack{\mathbf{n} = (n_1, n_2) \in \mathbb{Z}^2 \\ n_1 +n_2 =0}} \qe^{\frac{1}{2} (n_1 ^2 + n_2 ^2 )} z^{-\frac{1}{2}(n_\b -n_{\bar\b})}  L^{\mathbf{n},k=0} (\mathbf{a},\mathbf{m} ,\ve_1,\ve_2)  \prod_{\g=1,2} \frac{\Gamma\left( \frac{a_\g -m_\a}{\ve_2} \right) }{\Gamma\left( \frac{a_\g -m_\a}{\ve_2} +n_\g \right)}  \\
& \quad\quad \mathcal{Z}^{\text{inst}} (\mathbf{a}+\mathbf{n} \ve_1, \mathbf{m}, \ve_1,\ve_2-\ve_1;\qe) \\
& \quad\quad \left( \mathbf{C}_\infty ^{(2)} (\mathbf{a} +\mathbf{n}\ve_2,\mathbf{m},\ve_2) \right)_{\a\b}  \Psi^{\mathbb{Z}_2,\text{non-pert}}_\beta  (\mathbf{a} + \mathbf{n}\ve_2 , \mathbf{m} ,\ve_1-\ve_2,\ve_2;\qe,z).
\end{split}
\end{align}
To make use of the connection formula , we need to multiply appropriate prefactors to the non-perturbative part of the surface defect partition functions on both sides and take the sum over $\b=1,2$. Then we may continue the complexified FI parameter to the domain $0<\vert \qe \vert <1<\vert z \vert$, where the left hand side becomes the vortex string surface defect partition functions. The ratio of additional prefactors just mentioned miraculously cancels the $z$ factor inside the summation, so that the sum over $\b=1,2$ on the right hand side also gives a vortex string surface defect partition function with the shifts in the arguments. Hence we arrive at the blowup formula for this surface defect partition function:
\begin{align}
\begin{split}
& \Psi^{L,\text{non-pert}}_\a (\mathbf{a},\mathbf{m}, \ve_1, \ve_2; \qe, z ) \\
&=  \sum_{\substack{\mathbf{n} = (n_1, n_2) \in \mathbb{Z}^2 \\ n_1 +n_2 =0}} \qe^{\frac{1}{2} (n_1 ^2 + n_2 ^2 )}  L^{\mathbf{n},k=0} (\mathbf{a},\mathbf{m} ,\ve_1,\ve_2) \prod_{\g=1,2} \frac{\Gamma \left( \frac{a_{\g} - m_{\a}}{\ve_2} \right)}{\Gamma \left( \frac{a_{\g} -m_{\a}}{\ve_2} + n_\g  \right)} \\
&\quad\quad\quad\quad\quad \mathcal{Z}^{\text{inst}} (\mathbf{a}+\mathbf{n} \ve_1, \mathbf{m}, \ve_1,\ve_2-\ve_1;\qe) \Psi_\a ^{L,\text{non-pert}} (\mathbf{a} + \mathbf{n}\ve_2 , \mathbf{m} ,\ve_1-\ve_2,\ve_2;\qe,z).
\end{split}
\end{align}
This is precisely the blowup formula \eqref{eq:blowupL} for the vortex string surface defect partition function in the domain $0<\vert \qe\vert <1<\vert z \vert$ suggested in the previous section.

Similarly, we can also analytically continue the blowup formula for the orbifold surface defect partition function to the domain $0<\vert z \vert <\vert \qe\vert <1$. This procedure is a bit more involved compared to the previous case due to a non-trivial shift in the Coulomb moduli. To make use of the connection formula, let us shift the Coulomb moduli and multiply the entries of the connection matrix $\mathbf{C}_0 ^{(2)}$ to the blowup formula \eqref{eq:blowuporbint}:
\begin{align}
\begin{split}
&\left(\mathbf{C}_0 ^{(2)} (\mathbf{a},\mathbf{m},\ve_2 )\right)_{\a\b} \Psi^{\mathbb{Z}_2,\text{non-pert}}_\beta (\mathbf{a} -\boldsymbol\d \ve_1,\mathbf{m}, \ve_1, \ve_2; \qe, z )\\
& =  \sum_{\substack{\mathbf{n} = (n_1, n_2) \in \mathbb{Z}^2 \\ n_1 +n_2 =0}} \qe^{\frac{1}{2} (n_1 ^2 + n_2 ^2 )} z^{\frac{1}{2} (n_{\bar\beta}-n_\beta)} L^{\mathbf{n},k=0} (\mathbf{a} -\boldsymbol\d \ve_1,\mathbf{m} ,\ve_1,\ve_2) \\
& \frac{\Gamma \left( \frac{a_\beta-a_{\bar\beta} -\ve_1 }{\ve_2}+n_\beta-n_{\bar\beta} \right)}{\Gamma \left( \frac{a_\beta-a_{\bar\beta} -\ve_1}{\ve_2} \right)} \prod_{i=1,2} \frac{\Gamma \left( \frac{a_{\beta} - m_{i} -\ve_1}{\ve_2} \right)}{\Gamma \left( \frac{a_{\beta} -m_{i}-\ve_1}{\ve_2} + n_\beta  \right)} \frac{\left(\mathbf{C}_0 ^{(2)} (\mathbf{a},\mathbf{m},\ve_2 )\right)_{\a\b}}{\left(\mathbf{C}_0 ^{(2)} (\mathbf{a} + (\mathbf{n} -\boldsymbol\d) \ve_2 ,\mathbf{m},\ve_2 )\right)_{\a\b}} \\
& \mathcal{Z}^{\text{inst}} (\mathbf{a}+(\mathbf{n}-\boldsymbol\d) \ve_1, \mathbf{m}, \ve_1,\ve_2-\ve_1;\qe) \\
& \left(\mathbf{C}_0 ^{(2)} (\mathbf{a} + (\mathbf{n} -\boldsymbol\d) \ve_2 ,\mathbf{m},\ve_2 )\right)_{\a\b} \Psi_\beta ^{\mathbb{Z}_2,\text{non-pert}} (\mathbf{a} + (\mathbf{n}-\boldsymbol\d)\ve_2 -\boldsymbol\d (\ve_1 -\ve_2) , \mathbf{m} ,\ve_1-\ve_2,\ve_2;\qe,z),
\end{split}
\end{align}
where $\boldsymbol\d = (\d_{\a\b})_{\a=1,2}$. Now a straightforward computation shows that
\begin{align}
\begin{split}
&\frac{L^{\mathbf{n}, k=0} (\mathbf{a} -\boldsymbol\d \ve_1,\mathbf{m},\ve_1,\ve_2)}{L^{{\mathbf{n} -\boldsymbol\d}, k=0} (\mathbf{a} -\boldsymbol\d \ve_1,\mathbf{m},\ve_1,\ve_2)} \frac{\Gamma \left( \frac{a_\beta-a_{\bar\beta} -\ve_1 }{\ve_2}+n_\beta-n_{\bar\beta} \right)}{\Gamma \left( \frac{a_\beta-a_{\bar\beta} -\ve_1}{\ve_2} \right)} \prod_{i=1,2} \frac{\Gamma \left( \frac{a_{\beta} - m_{i} -\ve_1}{\ve_2} \right)}{\Gamma \left( \frac{a_{\beta} -m_{i}-\ve_1}{\ve_2} + n_\beta  \right)} \frac{\left(\mathbf{C}_0 ^{(2)} (\mathbf{a},\mathbf{m},\ve_2 )\right)_{\a\b}}{\left(\mathbf{C}_0 ^{(2)} (\mathbf{a} + (\mathbf{n} -\boldsymbol\d) \ve_2 ,\mathbf{m},\ve_2 )\right)_{\a\b}} \\
&=  \frac{\prod_{\g=1,2}\Gamma \left( \frac{m_{\a+2} +\ve -a_\g}{\ve_2}  \right)}{\Gamma\left( \frac{m_{\a+2} +\frac{\ve_2}{2} +\ve_1 -a_\b}{\ve_2} - n_\b +\frac{1}{2} \right) \Gamma\left( \frac{m_{\a+2} +\frac{\ve_2}{2} +\ve_1 -a_{\bar\b}}{\ve_2} - n_{\bar\b} -\frac{1}{2} \right)} \\
&=  \prod_{\g=1,2} \frac{\Gamma \left( \frac{m_{\a+2} +\ve -a_\g}{\ve_2}  \right)}{\Gamma\left( \frac{m_{\a+2} +\frac{\ve_2}{2} +\ve_1 -a_\g}{\ve_2} - n_\g ' \right) },
\end{split}
\end{align}
where we introduced half-integers $n_\b ' = n_\b -\frac{1}{2}$ and $n_{\bar\b} ' = n_{\bar\b} +\frac{1}{2}$. Note that this expression becomes independent of $\b$. Now we multiply the prefactor introduced in section \ref{sec:orbsurfd} and take the sum over $\b=1,2$ in the both sides of the blowup equation. The left hand side becomes the vortex string surface defect partition function by the analytic continuation. On the right hand side, the $z$ factor inside the sum miraculously cancels due to the ratio of prefactors, and the sum over $\b=1,2$ also yields the vortex string surface defect partition function with the shifted arguments. All in all, we get
\begin{align}
\begin{split}
&\Psi_\a ^{R,\text{non-pert}} (\mathbf{a}, \mathbf{m},\ve_1,\ve_2 ;\qe,z)\\
&= (1-\qe)^{1+\frac{\ve_2 \left( 2a_1+2a_2 -2\ve -\sum_{i=1} ^4 m_i  \right)}{\ve_1 (\ve_1-\ve_2)}} \\
&  \sum_{\substack{\mathbf{n} = (n_1, n_2) \in  \left( \mathbb{Z} +\frac{1}{2} \right)^2 \\ n_1 +n_2 =0}} \qe^{\frac{1}{2} (n_1 ^2 +n_2 ^2) -\frac{1}{4}} L^{\mathbf{n},k=1} (\mathbf{a},\mathbf{m},\ve_1,\ve_2)   \prod_{\g=1,2} \frac{\Gamma \left( \frac{m_{\a+2} +\ve -a_\g}{\ve_2}  \right)}{\Gamma\left( \frac{m_{\a+2} +\frac{\ve_2}{2} +\ve_1 -a_\g}{\ve_2} - n_\g  \right) } \\
& \quad\quad\quad \mathcal{Z}^{\text{inst}} \left(\mathbf{a}+\mathbf{n} \ve_1,\mathbf{m}+\frac{\ve_1}{2} ,\ve_1 ,\ve_2-\ve_1;\qe \right) \Psi_\a ^{R,\text{non-pert}} \left( \mathbf{a}+\mathbf{n}\ve_2, \mathbf{m}+\frac{\ve_2}{2} ,\ve_1-\ve_2,\ve_2;\qe,z \right).
\end{split}
\end{align}
This is precisely the blowup formula \eqref{eq:blowupRhalf} for the vortex string surface defect partition function in the domain $0<\vert z \vert<\vert \qe \vert <1$ with non-zero flux. Amazingly, we recovered the blowup formula with non-zero flux (half-integer sum) from the blowup formula with zero flux (integer sum). The prefactor in front also comes out correctly as suggested in the previous section.

We can alternatively start from the blowup formula \eqref{eq:blowuporbhalf} for the orbifold surface defect partition function with non-zero flux, and analytically continue to the domain $0<\vert \qe \vert< 1 <\vert z \vert$ or $0<\vert z \vert<\vert \qe \vert< 1$. A straightforward computation similar to the one just described shows that it analytically continues to the blowup formulas \eqref{eq:vortexhalf} and \eqref{eq:vortexrightinteger} for the vortex string surface defect partition functions, respectively. Note in particular we recover a blowup formula with zero flux \eqref{eq:vortexrightinteger} (integer sum) from a blowup formula with non-zero flux \eqref{eq:blowuporbhalf} (half-integer sum).

\section{Isomonodromic tau functions:\\Connecting the NS limit and the self-dual limit} \label{sec:taufunction}
In this section, we finally derive the main conjecture of \cite{GIL2012} relating the isomonodromic tau functions to the self-dual limit ($\ve_1=-\ve_2$) of the gauge theory partition functions. In particular, we show that the NS limit ($\ve_2 \to 0$) of the blowup formula for the surface defect partition functions leads to the conjectured statement. In this sense, the blowup formula connects the NS limit and the self-dual limit of the gauge theory partition function, thereby explaining the mysterious fact that the same isomonodromic tau function emerges in both limits. 

\subsection{Painlev\'{e} VI tau function}
As we have studied in section \ref{subsec:blowupsurf}, the blowup formula reads
\begin{align}
\begin{split}
&\Psi (a, \mathbf{m}, \ve_1, \ve_2 ; \qe,z)  = \sum_{n \in \mathbb{Z}} \mathcal{Z}( a+n \ve_1, \mathbf{m}, \ve_1, \ve_2-\ve_1 ; \qe) \Psi (a+n\ve_2, \mathbf{m}, \ve_1-\ve_2, \ve_2; \qe,z),
\end{split}
\end{align}
for any value of $z$. We are omitting a prefactor which for $z$ in the domain $0<\vert z \vert<\vert \qe \vert<1$ approaches $1$ in the NS limit $\ve_2 \to 0$. We have seen in section \ref{subsec:HJeq} that the ${\ve}_2 \to 0$ asymptotics of the vev of the surface defect is
\begin{align}
\Psi (a, \mathbf{m}, \ve_1, \ve_2 ; \qe,z) = \exp \left[ \frac{\ve_1}{\ve_2} S (a,\mathbf{m},\ve_1;\qe,z) + \cdots \right].
\end{align}
When we take the limit $\ve_2\to 0$ to the blowup formula, the singular terms $( \exp \mathcal{O} (\ve_2 ^{-1}) )$ cancel each other, but due to the $\ve_1$-shifts in the arguments the regular terms leave a non-trivial equation. We are led to
\begin{align} \label{eq:tau}
e^{S + \ve_1 \frac{\partial S}{\partial \ve_1}} = \sum_{n \in \mathbb{Z}} e^{n \ve_1 \frac{\partial S}{\partial a}}\mathcal{Z}( a+n\ve_1 , \mathbf{m},\ve_1,-\ve_1 ; \qe).
\end{align}
Note that on the right hand side we have an infinite sum of $SU(2)$ gauge theory partition functions on the self-dual $\Omega$-background with the integer shifts in the Coulomb modulus. Since the $\Omega$-background parameter $\ve_1$ only plays the role of the mass scale, we define the following massless parameters
\begin{align}
\alpha := \frac{a}{\ve_1}, \quad \theta_0:= \frac{m_3 -m_4}{2 \ve_1}, \quad \theta_\qe:= \frac{m_3 +m_4}{2 \ve_1}, \quad \theta_1:= \frac{m_1 +m_2}{2 \ve_1}, \quad  \theta_\infty:= \frac{m_1-m_2}{2 \ve_1}.
\end{align}
We denote $\mathcal{Z}( \a+n , \boldsymbol\theta ; \qe) := \mathcal{Z} (a + n\ve_1 , \mathbf{m} , \ve_1,-\ve_1;\qe)$.

The equation \eqref{eq:tau} is valid for any $z$. To relate this equation to the isomonodromic tau function, we consider the isomonodromic flow $z(\qe)$ which preserves the monodromy of the associated Fuchsian system. In other words, we consider the symplectomorphism generated by $\widetilde{S} (\alpha, z, \boldsymbol\theta;\qe)$,
\begin{align}
p = \frac{\p \widetilde{S}}{\p z}, \quad \beta = \frac{\p \widetilde{S}}{\p \a}.
\end{align}
As we have seen in section \ref{sec:monod}, $\alpha$ and $\beta$ parametrize the monodromy data of the Fuchsian system. The isomonodromic flow $z(\qe)$ and $p(\qe)$ are thus determined by requiring $\alpha$ and $\beta$ to be constants of motion. By taking the time derivative to the second equation, we get
\begin{align} \label{eq:eom1}
\begin{split}
0&=\frac{d \b}{d \qe} =\left. \left(  \frac{\p}{\p \a} \frac{\p \widetilde{S}}{\p \qe} \right) \right\vert_{z=z(\qe)} + \left. \frac{\p ^2 \widetilde{S} }{\p \a \p z}\right\vert_{z=z(\qe)} \frac{dz(\qe)}{d \qe}  \\ 
&= \left. \frac{\p^2 \widetilde{S}}{\p \a \p z} \right\vert_{z=z(\qe)}  \left( \left. -\frac{\p H^+(z,p;\qe)}{\p p}\right\vert_{\substack{z=z(\qe) \\ p=p(\qe)}} + \frac{dz(\qe)}{d\qe} \right),
\end{split}
\end{align}
where we have used $\frac{\p \widetilde{S}}{\p \qe} = -H^+\left( z, \frac{\p \widetilde{S}}{\p z}; \qe \right)$ in the second line. Also we can take the time derivative to the first equation to obtain
\begin{align}
\begin{split}
\frac{d p(\qe)}{d\qe} &= \left. \left( \frac{\p}{\p z} \frac{\p \widetilde{S}}{\p \qe} \right) \right\vert_{z=z(\qe)} + \left. \frac{\p ^2 \widetilde{S}}{\p z^2} \right\vert_{z=z(\qe)} \frac{dz(\qe)}{d\qe} \\
&= - \left. \frac{\p H^+(z,p;\qe)}{\p z} \right\vert_{\substack{z=z(\qe) \\ p=p(\qe)}},
\end{split}
\end{align}
where we have used $\frac{\p \widetilde{S}}{\p \qe} = -H^+\left( z, \frac{\p \widetilde{S}}{\p z}; \qe \right)$ in the second line. In other words, the isomonodromic flow $z(\qe)$ and $p(\qe)$ are solutions to the Hamiltonian equations of motion, as they should be. The constants of motion $\alpha$ and $\beta$ can be regarded as the integration constants of these equations.

Now we restrict the equation \eqref{eq:tau} to $z=z(\qe)$. Then, noting that $\frac{\p S}{\p \a} =  \frac{\p \widetilde{S}}{\p \a}   =\beta  $ since the prefactor for the effective twisted superpotential does not depend on the Coulomb moduli $\a=\frac{a}{\ve_1}$, we have
\begin{align} \label{eq:tauu}
\tau(\qe) := e^{\left. \left(S + \ve_1 \frac{\partial S}{\partial \ve_1}\right)\right\vert_{\substack{z=z(\qe)}} } = \sum_{n \in \mathbb{Z}} e^{n \beta}\mathcal{Z}( \a+n , \boldsymbol\theta  ; \qe),
\end{align}
where $\a$ and $\b$ are now constants which parametrize the monodromies of the associated Fuchsian system. We claim that the left hand side of this equation can indeed be identified as the Painlev\'{e} VI tau function.

Before we present the proof of the statement, let us examine the expression written on the right hand side. The gauge theory partition function $\mathcal{Z} (\a , \boldsymbol\th ; \qe)$ is the product of the classical part, the 1-loop part, and the instanton part:
\begin{align}
\mathcal{Z} (\a , \boldsymbol\th ; \qe) = \mathcal{Z}^{\text{cl}} (\a ; \qe) \mathcal{Z}^{\text{1-loop}} (\a , \boldsymbol\th ) \mathcal{Z}^{\text{inst}} (\a , \boldsymbol\th ; \qe).
\end{align}
The classical part is simply
\begin{align}
\mathcal{Z}^{\text{cl}} (\alpha ; \qe) = \qe^{\alpha^2},
\end{align}
while the 1-loop part is given as
\begin{align}
\begin{split}
\mathcal{Z}^{\text{1-loop}} (\alpha, \boldsymbol\theta) &= \prod_{\pm} \frac{ \Gamma_2 (\pm2 \alpha \ve_1;\ve_1,-\ve_1)}{\prod_{i=1} ^4 \Gamma_2 ( \pm \a \ve_1- m_i ; \ve_1, -\ve_1)} \\
&= \frac{1}{(2\pi) ^{3-2\th_\qe -2\th_1}} \prod_{\pm} \frac{\prod_{\epsilon =\pm} G(1\pm \a -\th_\qe + \epsilon \th_0) G(1 \pm \a -\th_1 +\epsilon \th_\infty) }{G(1\pm 2\a)},
\end{split}
\end{align}
where we used \eqref{eq:barnesg} in the second equality. The constant prefactor in front is not very important, and can be absorbed into the definition of $\tau(\qe)$. Lastly the instanton part is given as
\begin{align}
\begin{split}
&\mathcal{Z}^{\text{inst}} (\alpha ,\boldsymbol\th ;\qe) \\
&= \sum_{\l, \mu} \qe^{\vert \l \vert+ \vert \mu \vert} \prod_{(i,j) \in \l} \frac{((\alpha +i-j -\th_1)^2 -\th_\infty^2)((\alpha+i-j -\th_\qe)^2 -\th_0^2)}{h^2 _\l (i,j) (-2\alpha+1 -i-j+\mu_i+ \l_j ^t )^2} \\
&\quad\quad\quad\quad\quad\,\;\; \prod_{(i,j) \in \mu} \frac{((-\alpha +i-j -\th_1)^2 -\th_\infty^2)((-\alpha+i-j -\th_\qe)^2 -\th_0^2)}{h^2 _\mu (i,j) (2\alpha+1 -i-j+ \l_i +\mu_j ^t )^2},
\end{split}
\end{align}
where $\l$ and $\mu$ are Young diagrams, $\l^t$ is the transpose of $\l$, and $h_\l (i,j) := \l_i - j +1 -i +\l_j ^t$ is the hook length of $(i,j) \in \l$.

Now we prove that the left hand side of \eqref{eq:tauu} satisfies the defining relation for the tau function. In other words, we take the derivative $\frac{d}{d\qe} \log \tau(\qe)$ and show that it indeed reproduces the right hand side of the GIL relation. First, note that there is a slight difference between the generating function $\widetilde{S}$ and the asymptotics $S$, namely,
\begin{align}
\widetilde{S} = S + \Delta S,
\end{align}
where the difference $\Delta S$ varies according to the convergence domain:
\begin{align} \label{eq:extras}
\begin{split}
\Delta S^L &= (\th_0 + \th_\qe+\th_1) \log z + 2 \th_1 \log \left( 1 - \frac{1}{z} \right) + (\th_0 + \th_\qe)^2 \log \qe \\
&-(\th_0+\th_\qe +\th_1+ \th_\infty )(\th_0+\th_\qe +\th_1- \th_\infty )  \log (1-\qe),  \\
\Delta S^{\mathbb{Z}_2} &= (\th_0 + \th_\qe) \log z +(\th_0 + \th_\qe) ^2 \log \qe \\
&-(\th_0+\th_\qe +\th_1 +\th_\infty)(\th_0+\th_\qe +\th_1 -\th_\infty) \log (1-\qe),
\end{split}
\end{align}
and thus we have
\begin{align} \label{eq:dqops}
\frac{d}{d\qe} \left. \left( S + \ve_1 \frac{\p S}{\p \ve_1} \right) \right\vert_{z=z(\qe)} = \frac{d}{d\qe} \left. \left( \widetilde{S} + \ve_1 \frac{\p \widetilde{S}}{\p \ve_1} \right) \right\vert_{z=z(\qe)} - \frac{d}{d\qe} \left. \left( \Delta S + \ve_1 \frac{\p \Delta S}{\p \ve_1} \right) \right\vert_{z=z(\qe)}.
\end{align}
The first term can be computed as
\begin{align}
\begin{split}
\frac{d \widetilde{S} (\a , z(\qe) ,\boldsymbol\th;\qe)}{d\qe} &= \left.\frac{\p \widetilde{S}}{\p \qe} \right\vert_{z=z(\qe)} +   \left.\frac{\p \widetilde{S}}{\p z} \right\vert_{z=z(\qe)} \frac{d z(\qe)}{d\qe} \\
&=-H^+ (z(\qe),p(\qe);\qe) + p(\qe) \left.\frac{\p H^+(z,p;\qe)}{\p p} \right\vert_{\substack{z=z(\qe) \\ p=p(\qe)}} \\
&=\left.\left[ \left( p\frac{\p}{\p p} -1  \right)H^+(z,p;\qe)\right]  \right\vert_{\substack{z=z(\qe) \\ p=p(\qe)}},
\end{split}
\end{align}
where we have used the equation of motion \eqref{eq:eom1} for the second equality. Also, the second term in \eqref{eq:dqops} can be computed as
\begin{align}
\begin{split}
\frac{d}{d\qe} \left.\left(\ve_1 \frac{\p \widetilde{S}}{\p \ve_1}  \right)\right\vert_{z=z(\qe)} &= \left. \left( \ve_1 \frac{\p}{\p \ve_1}  \frac{\p \widetilde{S}}{\p \qe} \right) \right\vert_{z=z(\qe)}  + \left.  \ve_1 \frac{\p^2 \widetilde{S}}{\p \ve_1 \p z }  \right\vert_{z=z(\qe)}  \frac{dz(\qe)}{d\qe} \\
& =- \left.\left[ \ve_1 \frac{\p}{\p \ve_1} H^+\left( z,\frac{\p \widetilde{S}}{\p z};\qe \right) \right] \right\vert_{z=z(\qe)}+ \left.  \ve_1 \frac{\p^2 \widetilde{S}}{\p \ve_1 \p z }  \right\vert_{z=z(\qe)}  \frac{dz(\qe)}{d\qe} \\
&=  - \ve_1  \left.\frac{\p H^+(z,p;\qe)}{\p \ve_1}  \right\vert_{\substack{z=z(\qe) \\ p=p(\qe)}} + \left.  \ve_1 \frac{\p^2 \widetilde{S}}{\p \ve_1 \p z }  \right\vert_{z=z(\qe)} \left( \frac{dz(\qe)}{d\qe} -\left. \frac{\p H^+(z,p;\qe)}{\p p} \right\vert_{\substack{z=z(\qe) \\ p=p(\qe)}} \right) \\
&= - \ve_1  \left.\frac{\p H^+(z,p;\qe)}{\p \ve_1}  \right\vert_{\substack{z=z(\qe) \\ p=p(\qe)}},
\end{split}
\end{align}
where we used the equation of motion \eqref{eq:eom1} in the last line. Finally, a simple computation shows that
\begin{align}
\frac{d}{d\qe} \left. \left( \Delta S + \ve_1 \frac{\p \Delta S}{\p \ve_1} \right) \right\vert_{z=z(\qe)} = -\frac{(\th_0 + \th_\qe)^2}{\qe} -\frac{(\th_0+\th_\qe +\th_1+ \th_\infty )(\th_0+\th_\qe +\th_1- \th_\infty )}{1-\qe},
\end{align}
for both expressions in \eqref{eq:extras}. All in all, we have
\begin{align}
\begin{split}
\frac{d}{d\qe} \log \tau(\qe) &= \left.\left[ \left( p\frac{\p}{\p p} -\ve_1 \frac{\p}{\p \ve_1}-1  \right)H^+(z,p;\qe)\right]  \right\vert_{\substack{z=z(\qe) \\ p=p(\qe)}} \\
& +\frac{(\th_0 + \th_\qe)^2}{\qe} + \frac{(\th_0+\th_\qe +\th_1+ \th_\infty )(\th_0+\th_\qe +\th_1- \th_\infty )}{1-\qe}.
\end{split}
\end{align}
Note that the Painlev\'{e} VI Hamiltonian $H^+(z,p;\qe)$ is decomposed into a term quadratic in $p$, a term linear in $p$, and a $p$-independent term. The quadratic term changes the sign due to the $p \p_p -1$ operation. The $p$-independent term also changes the sign due to the $-\ve_1 \p_{\ve_1} -1$ operation, since it is quadratic in $\theta$-parameters. At last, the $p$-linear term also changes the sign because this term is also linear in $\theta$-parameters. Therefore, we obtain
\begin{align} \label{eq:tauderived}
\frac{d}{d\qe} \log \tau(\qe) = H^+(z(\qe),p(\qe);\qe) +\frac{(\th_0 + \th_\qe)^2}{\qe} + \frac{(\th_0+\th_\qe +\th_1+ \th_\infty )(\th_0+\th_\qe +\th_1- \th_\infty )}{1-\qe}.
\end{align}
Thus we have proven that $\tau(\qe)$ is indeed the Painlev\'{e} VI tau function. The terms on the right hand side other than the Hamiltonian are only rational functions of time $\qe$, and therefore could have been absorbed into the ambiguity of the definition of the tau function itself. In \cite{NikBlowup} a general argument about the tau-functions non-stationary integrable systems with Hamiltonians quadratic in momenta is presented. 

\paragraph{Remark} In the domain $0<\vert z \vert< \vert \qe \vert <1$, we cannot directly use the blowup formula \eqref{eq:vortexrightinteger} due to the shift in the fundamental masses. Instead, we can re-define the masses to the ones for the anti-fundamentals and use the blowup formula \eqref{eq:vortexrightanti}, where there is no shift for the masses. Even though the fundamentals and anti-fundamentals appear differently in the gauge theory partition function, they do in the self-dual limit of the $\Omega$-background. Hence, by taking the limit $\ve_2 \to 0$ to this blowup formula we recover the same expression for the tau function,
\begin{align}
\tau(\qe) = \sum_{n \in \mathbb{Z}} e^{n \beta}\mathcal{Z}( \a+n , \boldsymbol\theta  ; \qe),
\end{align}
where the $\qe$-derivative gives
\begin{align} \label{eq:anothertau}
\begin{split}
\frac{d}{d\qe} \log \tau(\qe) &= {H'}^+(z(\qe),p(\qe);\qe) \\
& -\frac{z(\qe)(z(\qe)-\qe)(z(\qe)-1)}{\qe(\qe-1)} p(\qe) \left( \frac{1}{z(\qe)-\qe} +\frac{1}{z(\qe)-1}\right) +\frac{2z(\qe) (\th_0+\th_\qe+\th_1-1)}{\qe(\qe-1)} \\
& +\frac{(\th_0 + \th_\qe)^2}{\qe} + \frac{(\th_0+\th_\qe +\th_1+ \th_\infty )(\th_0+\th_\qe +\th_1- \th_\infty )}{1-\qe}.
\end{split}
\end{align}
It is important to note that the Hamiltonian is still different from $H^+(z,p;\qe)$ due to the fact that we have used anti-fundamental masses. In particular, the Hamiltonian here is
\begin{align} \label{eq:anotherham}
\begin{split}
{H'}^+ (z,p;\qe) &= \frac{z(z-\qe)(z-1)}{\qe (\qe-1)} p\left( p-\frac{2\th_0}{z} -\frac{2\th_\qe-1}{z-\qe}-\frac{2\th_1-1}{z-1} \right) \\
& +\frac{z(\th_0+\th_\qe+\th_1+\th_\infty-1)(\th_0+\th_\qe+\th_1-\th_\infty-1)}{\qe(\qe-1)}.
\end{split}
\end{align}
Then $(z(\qe),p(\qe))$ is a solution to the Hamiltonian equation of motion, i.e., isomonodromic flow generated by this Hamiltonian. Compared to the Hamiltonian $H(w,p_w;\qe)$, there is half-integer shifts in the $\th$-parameters,
\begin{align}
\begin{split}
&\th_1 \longrightarrow \th_1 - \frac{1}{2} \\
&\th_\infty \longrightarrow \th_\infty -\frac{1}{2}.
\end{split}
\end{align}

\subsection{Okamoto transformations} \label{sec:ok}
The relation that we derived in \eqref{eq:tauderived} is not precisely in the form written down in \cite{GIL2013}. This is because the isomonodromic flow $z(\qe)$ that we obtained from the gauge theory is not identical to the isomonodromic flow considered there. Here, we explain how these two distinct isomonodromic flows are related to each other by an Okamoto transformation (namely, a B\"{a}cklund transformation for Painlev\'{e} VI, see \cite{NikBlowup} for more discussion), and how this procedure directly recovers the exact form of the GIL relation from \eqref{eq:tauderived}\footnote{in \cite{NikBlowup} a slightly different route, using the orbifold surface defect, is taken, leading to the GIL relation in a more direct way}.

\subsubsection{Okamoto transformation and the Painlev{\'e} VI tau function}

We have seen in section \ref{sec:genfun} that the complexified FI parameter $z(\qe)$ of the surface defect obeys Painlev{\'e} VI when the monodromies of the corresponding Fuchsian system are constant. The main problem is the monodromy parameters which define this isomonodromic flow $z(\qe)$ are slightly different from the monodromy parameters which define the isomonodromic flow, say, $w(\qe)$ considered in \cite{GIL2013}. In fact, from $w(\qe)$ to $z(\qe)$ the $\th$-parameters change as
\begin{align} \label{eq:thshift}
\begin{split}
&\th_q \longrightarrow \th_q + \frac{1}{2} \\
&\th_\infty \longrightarrow \th_\infty -\frac{1}{2}.
\end{split}
\end{align}
This can be easily checked by directly comparing the Painlev\'{e} VI Hamiltonians which define the flows $z(\qe)$ and $w(\qe)$. To recover the GIL relation as written in \cite{GIL2013}, we need to find the transformation rule between $(z(\qe), p(\qe))$ and $(w(\qe),p_w(\qe))$, and re-express the Hamiltonian $H(z(\qe),p(\qe);\qe)$ in terms of $(w(\qe),p_w(\qe))$.

The transformation of Painlev\'{e} system with respect to the shift \eqref{eq:thshift} was studied in \cite{Okamoto1986}. It corresponds to one of the Okamoto transformations of Painlev\'{e} VI. The full group of Okamoto transformations is isomorphic to the affine Weyl group of $F_4$. Its subgroup corresponding to
the $\widehat{D_4}$ can be seen in the quiver description of the moduli space of flat $SL(2)$-connections on the $4$-punctured sphere (it also related to the
$Spin(8)$ $R$-symmetry group of the $SU(2)$ gauge theory with $4$ fundamental hypers). 

Below we describe how the relevant Okamoto transformation is accounted for in our expression of the Painlev\'{e} tau function.

Let us be given with a solution $(w(\qe),p_w(\qe))$ of the Painlev\'{e} VI system:
\begin{align}
\frac{dw(\qe)}{d\qe} = \frac{\p H}{\p p_w}, \quad \frac{dp_w(\qe)}{d\qe} = - \frac{\p H}{\p w},
\end{align}
where
\begin{align}
\begin{split}
H (w,p_w;\qe) &= \frac{w(w-\qe)(w-1)}{\qe (\qe-1)} p_w\left( p_w-\frac{2\th_0}{w} -\frac{2\th_\qe-1}{w-\qe}-\frac{2\th_1}{w-1} \right) \\
& +\frac{w(\th_0+\th_\qe+\th_1+\th_\infty)(\th_0+\th_\qe+\th_1-\th_\infty-1)}{\qe(\qe-1)}.
\end{split}
\end{align}
We define an auxiliary function $h(\qe)$:
\begin{align} \label{eq:h}
\begin{split}
h(\qe):&=\qe(\qe-1) H(w(\qe),p_w(\qe);\qe) -\qe \left( \th_0 + \th_1 \right)^2 \\
&-\frac{1}{2} \left( \th_0 ^2 - \th_1 ^2 +2 \th_0 ( 2\th_\qe-1) +(\th_\qe +\th_\infty)(\th_\qe-\th_\infty-1) \right).
\end{split}
\end{align}
Then it was proven in \cite{Okamoto1986} that $h(\qe)$ satisfies the following nonlinear ordinary differential equation:
\begin{align} \label{eq:heq}
\begin{split}
&\frac{d h}{d\qe} \left[ \qe(1-\qe) \frac{d^2 h}{d\qe^2} \right]^2 + \left[ \frac{dh}{d\qe} \left( 2h-(2\qe-1) \frac{dh}{d\qe} \right) + (\th_0^2 -\th_1 ^2 )(\th_\qe^2 -\th_\infty^2 -\th_\qe-\th_\infty) \right]^2 \\
&= \left( \frac{dh}{d\qe} +(\th_0+\th_1)^2 \right)\left( \frac{dh}{d\qe} +(\th_0-\th_1)^2 \right)\left( \frac{dh}{d\qe} +(\th_\qe+\th_\infty)^2 \right)\left( \frac{dh}{d\qe} +(\th_\qe-\th_\infty -1)^2 \right).
\end{split}
\end{align}
In other words, we can say a solution $(w(\qe),p_w(\qe))$ to Painlev\'{e} VI provides a solution $h(\qe)$ to \eqref{eq:heq}. In fact, the converse is also true: given a solution $h(\qe)$ to \eqref{eq:heq}, we can construct solutions $w(\qe)$ and $p_w(\qe)$ to Painlev\'{e} VI as rational functions of $h$, $\frac{dh}{d\qe}$, and $\frac{d^2 h}{d\qe^2}$ \cite{Okamoto1986}. In this sense, we have a one-to-one birational correspondence $\Gamma$ between the solution spaces of the two equations:
\begin{align}
\Gamma(h(\qe)) = (w(\qe),p_w(\qe)).
\end{align}

Let us now consider the shift \eqref{eq:thshift}. We apply this shift to \eqref{eq:heq} and find that the following $h^+ (\qe)$ satisfies this shifted equation \cite{Okamoto1986},
\begin{align} \label{eq:h+}
h^+  = h - w(w-1) p_w +(\th_0+\th_\qe+\th_1+\th_\infty)w - \frac{1}{2}(2\th_0 +\th_\qe +\th_\infty).
\end{align}
By applying the birational correspondence $\Gamma$ to $h^+$, we obtain a solution $(z(\qe),p(\qe))$ to Painlev\'{e} VI with the shift \eqref{eq:thshift}. Since $h^+$ is written in terms of $w$ and $p_w$, we obtain a birational transformation from $(w(\qe),p_w(\qe))$ to $(z(\qe),p(\qe))$ accordingly.

This birational transformation is quite lengthy and complicated, so it will not be explicitly written here. Nevertheless, we can easily obtain the expression of the new Hamiltonian $H^+(z(\qe),p(\qe);\qe)$ in terms of the original variables $w(\qe)$ and $p_w(\qe)$, by combining \eqref{eq:h} and \eqref{eq:h+}:
\begin{align}
H^+(z(\qe),p(\qe);\qe) = H(w(\qe),p_w(\qe);\qe)-\frac{w(\qe) (w(\qe)-1)}{\qe(\qe-1)} p_w(\qe) + \frac{w(\qe)(\th_0+\th_\qe+\th_1+\th_\infty)}{\qe(\qe-1)}.
\end{align}
Substituting this to the relation \eqref{eq:tauderived} derived in the previous section, we finally obtain
\begin{align} \label{eq:tauderived2}
\begin{split}
\frac{d}{d\qe} \log \tau(\qe)&= H(w(\qe),p_w(\qe);\qe)-\frac{w(\qe) (w(\qe)-1)}{\qe(\qe-1)} p_w(\qe) + \frac{w(\qe)(\th_0+\th_\qe+\th_1+\th_\infty)}{\qe(\qe-1)} \\
& +\frac{(\th_0 + \th_\qe)^2}{\qe} + \frac{(\th_0+\th_\qe +\th_1+ \th_\infty )(\th_0+\th_\qe +\th_1- \th_\infty )}{1-\qe}.
\end{split}
\end{align}
This is precisely the GIL relation written in \cite{GIL2013}.

\paragraph{Remark} In the domain $0<\vert z \vert<\vert \qe \vert <1$, we have another Hamiltonian ${H'}^+ (z,p;\qe)$ \eqref{eq:anotherham}. Just as what we have done above, we may re-express $z(\qe)$, $p(\qe)$, and ${H'}^+ (z(\qe),p(\qe);\qe)$ in terms of $w(\qe)$ and $p_w (\qe)$ by taking account of the Okamoto transformation from $(w,p_w)$ to $(z,p)$, which involves the half-integer shifts in $\th$-parameters,
\begin{align}
\begin{split}
&\th_1 \longrightarrow \th_1 - \frac{1}{2} \\
&\th_\infty \longrightarrow \th_\infty -\frac{1}{2}.
\end{split}
\end{align}
Then we may substitute this expression into \eqref{eq:anothertau}, recovering the GIL expression \eqref{eq:tauderived2} for the isomonodromic tau function.

\subsubsection{The Okamoto maps of  monodromy data} \label{sec:okmon}
In making the Okamoto transformation from $(w(\qe),p_w(\qe))$ to $(z(\qe),p(\qe))$ corresponding to the shift \eqref{eq:thshift}, we have not yet specified which new solution a given solution is mapped to under the transformation. In other words, we only argued the function $h^+ (\qe)$ \eqref{eq:h+} provides a solution to the equation \eqref{eq:heq} with the shifted $\th$-parameters, but it is a particular solution in the two-dimensional solution space. When the birational correspondence $\Gamma$ is applied, this ambiguity is passed to the integration constants $\alpha$ and $\beta$ for the isomonodromic flow $(z(\qe),p(\qe))$.
 
The question can be rephrased in the context of the Riemann-Hilbert correspondence as follows. The Riemann-Hilbert map sends the solution $(w(\qe),p_w(\qe))$ of the Painlev\'{e} VI system  to a point in the moduli space  $\EuScript{M}_\qe (\boldsymbol\th)$ of $SL(2)$ flat connections on the four punctured sphere, with the conjugacy classes for small loops around the punctures fixed in terms of the $\th$-parameters. The Okamoto transformation $(z(\qe),p(\qe))$ of $(w(\qe),p_w(\qe))$ is a solution of a new Painlev\'{e} VI system. The latter is sent, by the Riemann-Hilbert map, to an element of the $SL(2)$-monodromy space $\EuScript{M}_\qe \left( \th_0, \th_\qe+\frac{1}{2}, \th_1,\th_\infty -\frac{1}{2} \right)$. It is similar  to the original monodromy space but the shifts in the $\th$-parameters as \eqref{eq:thshift} imply these moduli spaces are not isomorphic as symplectic varieties. The problem is to find the element in this monodromy space obtained from the procedure described so far, i.e., to study how the push-forward of the Okamoto transformation with respect to the Riemann-Hilbert map acts on the monodromy space. 

It is important to properly answer this question to precisely identify the meaning of $\a$ and $\b$ appearing on the right hand side of the GIL relation,
\begin{align}
\tau(\qe)=  \sum_{n \in \mathbb{Z}} e^{n \beta}\mathcal{Z}( \a+n , \boldsymbol\theta  ; \qe).
\end{align}
Recall from section \ref{sec:monodromy} that $\a$ and $\b$ parametrize the monodromies of the Fuchsian system associated to $(z(\qe),p(\qe))$ by \eqref{eq:acycle} and \eqref{eq:bcycle}, namely,
\begin{align}
&\Tr \, M_A = - 2 \cos 2\pi \a
\end{align}
and
\begin{align}
\begin{split}
\text{Tr}\, M_B &= \frac{\left( -\cos 2\pi \th_\infty +\cos 2\pi \th_1 \right) \left( \cos 2\pi \th_0 - \cos 2\pi \th_\qe \right)}{2 \sin ^2 \pi \a} \\
& - \frac{\left( \cos 2\pi \th_\infty +\cos 2\pi \th_1 \right) \left( \cos 2\pi \th_0 + \cos 2\pi \th_\qe \right)}{2 \cos ^2 \pi \a} \\
& + \sum_{\pm} 4 \frac{\prod_{\epsilon=\pm} \sin \pi (\mp \a - \th_\qe +\epsilon \th_0) \sin \pi (\mp \a -\th_1 +\epsilon \th_\infty) }{\sin ^2 2\pi \a}  e^{\pm\b} .
\end{split}
\end{align}
The question is how the monodromies of the Fuchsian system associated to $(w(\qe),p_w(\qe))$ are parametrized by $\a$ and $\b$. This is indeed a non-trivial problem, but in \cite{CL2007} it was shown that the answer is in fact very simple.\footnote{For more works on the action of Okamoto transformations on the monodromy space, see also \cite{IIS2003, DM2003, LT2008}} We just have to undo the half-integer shifts \eqref{eq:thshift} in the above formulas, yielding
\begin{align}
&\Tr \, M_A  = -2 \cos 2\pi \a
\end{align}
and
\begin{align}
\begin{split}
\text{Tr}\, M_B &= \frac{\left( \cos 2\pi \th_\infty +\cos 2\pi \th_1 \right) \left( \cos 2\pi \th_0 + \cos 2\pi \th_\qe \right)}{2 \sin ^2 \pi \a} \\
& + \frac{\left( \cos 2\pi \th_\infty -\cos 2\pi \th_1 \right) \left( \cos 2\pi \th_0 - \cos 2\pi \th_\qe \right)}{2 \cos ^2 \pi \a} \\
& - \sum_{\pm} 4 \frac{\prod_{\epsilon=\pm} \cos \pi (\mp \a - \th_\qe +\epsilon \th_0) \cos \pi (\mp \a -\th_1 +\epsilon \th_\infty) }{\sin ^2 2\pi \a}  e^{\pm\b} .
\end{split}
\end{align}
These formulas precisely match with the definitions of $\a$ and $\b$ parameters in \cite{GIL2013}. Hence we complete the derivation of the GIL relation.

\section{Discussion} \label{sec:discussion}
We have derived the conjectural expression for the isomonodromic tau function as an infinite sum of $\EuScript{N}=2$ gauge theory partition functions by taking the NS limit of the blowup formula for the surface defect partition function, given that the free energy of the $\EuScript{N}=2$ gauge theory with a surface defect is the Hamilton-Jacobi potential for the isomonodromic flow. Our method nicely connects the two independent approaches to the isomonodromic problem, one associated to the self-dual limit and the other associated to the NS limit of the $\Omega$-background. It also provides a physical intuition for the origin of the gauge theoretical expression of the isomonodromic tau function.

The obvious desirable generalizations of our story would be higher rank quiver gauge theories on the supersymmetric side, and higher rank isomonodromy problems on higher genus Riemann surfaces with more punctures. The relevant geometry and the conjectural generalization of the GIL formula are described in  \cite{NikBlowup}. Let us outline some issues with these generalizations. 

First of all, for the general Riemann surface the supersymmetric side is most likely non-Lagrangian class $S$ theory. For genus zero with $2$ regular and $p+1$ minimal punctures, for $SL(N)$ flat connections, the supersymmetric side is the quiver gauge theory with 
linear quiver with $p$ nodes, with the gauge group $SU(N)^{p}$. For genus one with $p$ minimal punctures one gets the $SU(N)^p$ theory corresponding to the affine quiver with $p$ nodes.

For example, already in the $SU(2)^p$ case,  when the Riemann surface is the sphere with $p+3$ punctures, we need $p$ insertions of the surface defects, whose complexified FI parameters with their conjugate momenta give coordinates $(z_i , p_i)_{i=1} ^p$ on the moduli space
${\mathcal{M}}^{\rm flat}_{\boldsymbol\th}$ of flat $SL(2, {\BC})$-connections. Their expectation value conjecturally satisfies $p$ differential equations in both the bulk gauge couplings and the complexified FI parameters, whose NS limit gives the Hamilton-Jacobi equations for the isomonodromic flows. Finally the blowup formula for the surface defect partition function would involve a sum over a $p$-dimensional lattice of the self-dual partition function of the bulk theory at the shifted Coulomb moduli. 

Another interesting direction is the extension to the higher rank theory (see the appendices $D$ and $F$ in \cite{NikBlowup}). The simplest example is the $A_2$ theory with the Riemann sphere with two regular punctures at $0$ and $\infty$ and two minimal punctures at $\qe$ and $1$, meaning that the eigenvalues of $A_0, A_\infty \in \mathfrak{sl}(3)$ are all distinct, while  the eigenvalues of $A_\qe , A_1 \in \mathfrak{sl}(3)$ are maximally degenerate (see the appendices $D$ and $F$ in \cite{NikBlowup}). The dimension of the reduced moduli space is $4$. With a proper choice of the pair of pants decomposition, the corresponding class $\mathcal{S}$ theory is the $SU(3)$ gauge theory with six hypermultiplets. Then we may insert a $\mathbb{Z}_3$-orbifold surface defect. Two fractional couplings, say, $z_1$ and $z_2$, and their conjugate momenta (the topological charges of the surface theory)  would provide a local Darboux coordinate system on the reduced moduli space of flat connections. 
The orbifold surface defect partition function satisfies the Knizhnik-Zamolodchikov equation \cite{NT}. In the ${\ve}_{2} \to 0$ limit it reduces to the Hamilton-Jacobi equation  \cite{Reshetikhin} for the isomonodromic deformation of the $\mathfrak{sl}(3)$ meromorphic connection with regular (first-order) poles, with two regular and two minimal residues. Such connection can be mapped, in a standard fashion, to a third-order ordinary differential equation with regular singularities. The isomonodromic flow can be, in turn, also mapped to a third-order ordinary differential equation in $\qe$. However, to exhibit all deformations we need more parameters. On the supersymmetric gauge theory side the additional parameter, say ${\rm log}({\qe}_{3})$, comes from the ${\Tr} {\phi}^3$ (formal) deformation of the microscopic prepotential (see \cite{Losev:2003py}). However, we don't quite understand the geometry of this deformation in the two dimensional context.  
Informally, we should study the deformation of $\qe_3$ as well as of $\qe$ that preserves the monodromy data, except that it is not guaranteed that with $\qe_3$ turned on the relevant object still can be viewed as a meromorphic connection. The Hamilton-Jacobi potential would be properly generalized with $\qe_3$ and this is still expected to be the free energy of the surface defect theory with an appropriate inclusion of $\qe_3$ in gauge theoretical manner. The same applies to the tau function as an infinite sum of the bulk gauge theory partition functions.

Another aspect to be investigated is the S-duality. Strictly speaking, the GIL expression of the tau function only allows an expansion around $\qe=0$. To achieve expansions around other critical points, $\qe=1$ and $\qe=\infty$, we need similar expressions as infinite sum of gauge theory partition functions, where the gauge coupling $\qe$ is replaced by $1-\qe$ or $\qe^{-1}$. We choose different pants decomposition of the four-punctured sphere accordingly, so that the new $({\alpha}, {\beta})$ coordinates which appear in the GIL-like formulas would differ from the original ones, related to each other by canonical transformations as discussed in \cite{NRS2011}. In the gauge theory context, this amounts to going to the S-dual frame. It would be nice to see whether blowup formulas for surface defect partition functions with such S-duality transformations would still provide the gauge theoretical expressions for the isomonodromic tau functions expanded in the corresponding domains of $\qe$. Indeed, the simpler versions of the blowup formulas were used in \cite{Vafa:1994tf} in tests of the S-duality of the ${\mathcal{N}}=4$ theory. 

We believe the novel blowup formulas for the partition functions of the gauge theory in the presence of surface defects deserve further analysis by themselves. Their implications in the view of the refined topological string and the quantum toroidal algebra are unclear as of yet. In the geometric engineering of $\EuScript{N}=2$ gauge theories \cite{KKV}, the insertion of half-BPS surface defects corresponds to incorporating open strings \cite{DGH}. In the point of view of the algebraic engineering \cite{BFH}, where the gauge theory partition function is represented as a correlation function of operators which intertwine representations of quantum toroidal algebra, the insertion of the surface defect can be realized by properly extending the quantum toroidal algebra \cite{BJ}. It would be nice to formulate the blowup formula in these contexts and study its implications.

\appendix
\section{Partition functions of $\EuScript{N}=2$ supersymmetric quiver gauge theories} \label{appA}
We give a brief review on the partition functions of the $\EuScript{N}=2$ quiver gauge theories. For more details on this subject, see \cite{Nekrasov_BPS1} for example.

 For an oriented graph $\gamma$, we denote the sets of its vertices and edges and $\text{Vert}_\gamma$ and $\text{Edge}_\gamma$, respectively. We define $s,t:\text{Edge}_\gamma \to \text{Vert}_\gamma$ as the maps which send an edge to its source and target, respectively. For each vertex we assign two integers,
\begin{align}
\mathbf{n} = (\mathpzc{n}_\mathbf{i})_{\mathbf{i} \in \text{Vert}_\gamma} \in \left( \mathbb{Z}^{>0} \right) ^{\text{Vert}_\gamma}, \quad \mathbf{m} = (\mathpzc{m}_\mathbf{i})_{\mathbf{i} \in \text{Vert}_\gamma} \in \left( \mathbb{Z}^{\geq 0} \right) ^{\text{Vert}_\gamma}.
\end{align}
The $\EuScript{N}=2$ quiver gauge theory associated to $\gamma$ is the four-dimensional $\EuScript{N}=2$ supersymmetric gauge theory, whose gauge group is 
\begin{align} \label{globalgauge}
G_{g} = \bigtimes_{\mathbf{i} \in \text{Vert}_\gamma} U(\mathpzc{n}_\mathbf{i}),
\end{align}
and whose flavor group is
\begin{align} \label{flavor}
G_f = \left( \bigtimes_{\mathbf{i} \in \text{Vert}_\gamma} U(\mathpzc{m}_\mathbf{i}) \times U(1) ^{\text{Edge}_\gamma} \right) \Bigg/ U(1) ^{\text{Vert}_\gamma}.
\end{align}
Here the overall $U(1)^{\text{Vert}_\gamma}$ transformation has been mod out due to the gauge symmetry,
\begin{align}
(u_\mathbf{i})_{\mathbf{i} \in \text{Vert}_\gamma} : \left( (g_\mathbf{i})_{\mathbf{i} \in \text{Vert}_\gamma} , ( u_\mathbf{e})_{\mathbf{e} \in \text{Edge}_\gamma} \right) \mapsto \left( (u_\mathbf{i} g_\mathbf{i})_{\mathbf{i} \in \text{Vert}_\gamma} , (u_{s(\mathbf{e})} u_\mathbf{e} u_{t(\mathbf{e})} ^{-1})_{\mathbf{e} \in \text{Edge}_\gamma} \right).
\end{align}
The field contents of the theory are the following: the vector multiplets $\boldsymbol{\Phi} = (\Phi_\mathbf{i})_{\mathbf{i} \in \text{Vert}_\gamma}$ in the adjoint representation of $G_g$, the fundamental hypermultiplets $\boldsymbol{Q}_{\text{fund}} = (Q_\mathbf{i})_{\mathbf{i} \in \text{Vert}_\gamma}$ in the fundamental representation of $G_g$ and the antifundamental representation of $G_f$, and finally the bifundamental hypermultiplets $\boldsymbol{Q}_{\text{bifund}} = (Q_\mathbf{e})_{\mathbf{e} \in \text{Edge}_\gamma}$ in the bifundamental representation $(\overline{n_{s(\mathbf{e})}} ,n_{t(\mathbf{e})})$ of $G_g$. The $\EuScript{N}=2$ supersymmetric action is then fixed up to the gauge couplings,
\begin{align}
\mathfrak{q}_\mathbf{i} = \text{exp}(2 \pi i \tau_\mathbf{i}) \quad \left( \tau_\mathbf{i} = \frac{\vartheta_\mathbf{i}}{2\pi} + \frac{4\pi i}{g_\mathbf{i} ^2} \right), \quad \mathbf{i} \in \text{Vert}_\gamma,
\end{align}
and the masses of the hypermultiplets,
\begin{align} \label{mass}
&\textbf{\textit{m}} = ((\mathbf{m}_\mathbf{i})_{\mathbf{i} \in \text{Vert}_\gamma}, (m_\mathbf{e})_{ \mathbf{e} \in  \text{Edge}_\gamma}), \nonumber \\
&\mathbf{m}_\mathbf{i} = \text{diag}(m_{\mathbf{i},1} , \cdots, m_{\mathbf{i},\mathpzc{m}_\mathbf{i}}) \in \text{End}(\mathbb{C}^{\mathpzc{m}_\mathbf{i}}), \quad m_\mathbf{e} \in \mathbb{C}.
\end{align}
The global symmetry group of the theory is 
\begin{align}
H = G_g \times G_f \times G_{\text{rot}},
\end{align}
where $G_g$ \eqref{globalgauge} is the group of global gauge symmetry, $G_f$ \eqref{flavor} is the group of flavor symmetry, and $G_{\text{rot}} = SO(4)$ is the group of the Lorentz symmetry. We turn on equivariant parameters for the maximal torus $T_H \subset H$. The equivariant parameters for $G_g$ is the vacuum expectation values of the complex scalars,
\begin{align}
\langle \Phi_\mathbf{i} \rangle = \mathbf{a}_\mathbf{i}, \quad  \mathbf{a}_\mathbf{i} = \text{diag}(a_{\mathbf{i},1}, \cdots, a_{\mathbf{i},\mathpzc{n}_\mathbf{i}}) \in \text{End}(\mathbb{C}^{\mathpzc{n}_\mathbf{i}}), \quad \mathbf{i} \in \text{Vert}_\gamma.
\end{align}
The equivariant parameters for $G_f$ is the masses of the hypermultiplets \eqref{mass}. Finally the equivariant parameters for $G_{\text{rot}}$ is the $\Omega$-deformation parameters $\varepsilon_1, \varepsilon_2$. The partition function of the theory is a function of these parameters $(\mathbf{a}, \textbf{\textit{m}}, \boldsymbol{\varepsilon}) \in \text{Lie}(T_H)$. In expressing the partition function, we abuse our notation and denote the vector spaces and their $T_H$-equivariant characters in the same letters. Hence we write
\begin{align}
N_{\mathbf{i}} = \sum_{\alpha =1} ^{\mathpzc{n}_{\mathbf{i}}} e^{\beta a_{\mathbf{i},\alpha}}, \quad M_{\mathbf{i}} = \sum_{f=1} ^{\mathpzc{m}_{\mathbf{i}}} e^{\beta m_{\mathbf{i}, f}}.
\end{align}
It is helpful to use the following notation for abbreviated expressions,
\begin{align}
\begin{split}
&q_{i} \equiv e^{\beta \varepsilon_{i}}, \quad P_{i} \equiv 1-q_{i} \quad i=1,2, \\
&q_{12} \equiv q_1 q_2,\quad P_{12} \equiv (1-q_1)(1-q_2).
\end{split}
\end{align}
The partition function factors into the classical, one-loop, and the instanton parts:
\begin{align}
\mathcal{Z}(\mathbf{a}, \boldsymbol{m}, \boldsymbol{\varepsilon}, \mathfrak{q}) = \mathcal{Z}^{\text{classical}}\; \mathcal{Z}^{\text{1-loop}} \; \mathcal{Z}^{\text{inst}}.
\end{align}
The classical part is given by
\begin{align}
 \mathcal{Z}^{\text{classical}}(\mathbf{a}, \boldsymbol{\varepsilon}, \mathfrak{q}) = \prod_{\mathbf{i} \in \text{Vert}_\gamma} \mathfrak{q}_{\mathbf{i} } ^{ -\frac{1}{2\varepsilon_1 \varepsilon_2} \sum_{\alpha=1} ^{\mathpzc{n}_{\mathbf{i}}} a_{\mathbf{i}, \alpha} ^2 }.
\end{align}
The one-loop part is given by
\begin{align} \label{1loop}
\begin{split}
&\mathcal{Z}^{\text{1-loop}}(\mathbf{a}, \boldsymbol{m}, \boldsymbol{\varepsilon}) \\
& = E \left[ \frac{1}{(1-e^{-\beta \varepsilon_1} )(1-e^{-\beta \varepsilon_2})} \left( \sum_{\mathbf{i} \in \text{Vert}_\gamma } (M_{\mathbf{i}} -N_{\mathbf{i}})N_{\mathbf{i}} ^* + \sum_{\mathbf{e} \in \text{Edge}_\gamma} e^{\beta m_{\mathbf{e}}} N_{t(\mathbf{e})} N_{s(\mathbf{e})} ^* \right) \right],
\end{split}
\end{align}
where the $E$-symbol is defined by
\begin{align} \label{epsilon}
E\left[ \cdots \right] \equiv \exp \left[ \frac{d}{ds} \Bigg\vert_{s=0} \frac{1}{\Gamma(s)} \int_0 ^\infty d\beta \beta^{s-1} [\cdots] \right],
\end{align}
which converts a character into a product of weights. In particular, the $E$-symbol regularizes an infinite product of weights such as \eqref{1loop} by the Barnes double gamma function,
\begin{align}
\Gamma_2 (x;\varepsilon_1,\varepsilon_2) \equiv \exp \left[ -\frac{d}{ds} \Bigg\vert_{s=0} \frac{1}{\Gamma(s)} \int_0 ^{\infty} d\beta \beta^{s-1} \frac{ e^{-\beta x}}{(1-e^{-\beta \varepsilon_1})(1-e^{-\beta \varepsilon_2})} \right].
\end{align}

The instanton part $\mathcal{Z}^{\text{inst}}$ is obtained by a $T_H$-equivariant integral over the instanton moduli space. Given the vector of the instanton charges $\mathbf{k} = (k_\mathbf{i})_{\mathbf{i} \in \text{Vert}_\gamma} \in \mathbb{Z}^{\geq 0}$, the total framed noncommutative instanton moduli space of the quiver gauge theory for $\gamma$ is
\begin{align}
\EuScript{M}_\gamma (\mathbf{n}, \mathbf{k}) \equiv \bigtimes_{\mathbf{i} \in \text{Vert}_\gamma} \EuScript{M}(\mathpzc{n}_\mathbf{i} , k_\mathbf{i}) \label{quivmodul},
\end{align}
where $\EuScript{M} (\mathpzc{n}_\mathbf{i} , k_\mathbf{i})$ is the ADHM moduli space
\begin{align}
&\EuScript{M} (n , k)  = \begin{cases} \quad \quad B_{1,2}  : K \rightarrow K, \\ I : N \rightarrow K, J : K \rightarrow N\end{cases} \Bigg\vert \begin{rcases} \left[ B_1 , B_2   \right] + I J=0, \quad \quad \quad \quad \quad   \\ [B_1 , {B_1 } ^\dagger ] + [B_2  , {B_2 } ^ \dagger] + I {I} ^\dagger - {J} ^\dagger J = \zeta \end{rcases} \Bigg/ U(k). \\
&(N = \mathbb{C}^n, K = \mathbb{C}^k) \nonumber
\end{align}
Solving the real moment map equation $[B_1 , {B_1 } ^\dagger ] + [B_2  , {B_2 } ^ \dagger] + I {I} ^\dagger - {J} ^\dagger J = \zeta$ and dividing by the compact $U(k)$ is equivalent to imposing the stability condition and dividing by the complex group $GL(k)$,
\begin{align} \label{adhmmod}
&\EuScript{M} (n , k)  = \begin{cases} \quad \quad B_{1,2}  : K \rightarrow K, \\ I : N \rightarrow K, J : K \rightarrow N\end{cases} \Bigg\vert \begin{rcases} \left[ B_1 , B_2   \right] + I J=0,  \quad\quad \\\quad  K = \mathbb{C} [B_1, B_2]\: I (N) \quad \end{rcases} \Bigg/ GL(k).
\end{align}
The $T_H$-equivariant integration over the instanton moduli space \eqref{quivmodul} localizes on the set of fixed points of $T_H$-action, $\EuScript{M}_\gamma (\mathbf{n}, \mathbf{k})^{T_H}$, which is the set of colored partitions $\boldsymbol{\lambda}=((\lambda^{(\mathbf{i}, \alpha)})_{\alpha=1} ^{\mathpzc{n}_\mathbf{i}})_{\mathbf{i} \in \text{Vert}_\gamma}$, where each $\lambda^{(\mathbf{i}, \alpha)}$ is a partition,
\begin{align} \lambda^{(\mathbf{i}, \alpha)} = \left( \lambda^{(\mathbf{i},\alpha)} _1 \geq \lambda^{(\mathbf{i}, \alpha)}_2 \geq \cdots \geq \lambda^{(\mathbf{i},\alpha)}_{l(\lambda^{(\mathbf{i},\alpha)})} > \lambda^{(\mathbf{i}, \alpha)}_{l(\lambda^{(\mathbf{i},\alpha)})+1} = \cdots = 0   \right),
\end{align}
with the size $\vert \lambda^{(\mathbf{i}, \alpha)} \vert = \sum_{i=1} ^{l(\lambda^{(\mathbf{i}, \alpha)})} \lambda^{(\mathbf{i},\alpha)} _i = k_{\mathbf{i}, \alpha}$ constrained by $k_\mathbf{i} = \sum_\alpha k_{\mathbf{i}, \alpha} = \vert \boldsymbol{\lambda} ^{(\mathbf{i})} \vert$. At each fixed point $\boldsymbol{\lambda}$, the vector space $K_{\mathbf{i}}$ carrys a representation of $T_H$ with the weights given by the formula
\begin{align}
K_{\mathbf{i}} [\boldsymbol{\lambda}] = \sum_{\alpha=1} ^{\mathpzc{n}_\mathbf{i}} \sum_{\Box \in \lambda^{(\mathbf{i}, \alpha)}} e^{\beta c_{\square}},
\end{align}
where we have defined the content of the box,
\begin{align} 
c_{\square} = a_{\mathbf{i}, \alpha} + \varepsilon_1 (i-1) + \varepsilon_2 (j-1) \quad \text{for} \quad \Box = (i,j) \in \lambda^{(\mathbf{i},\alpha)} \iff 1 \leq j \leq \lambda_i ^{(\mathbf{i},\alpha)}.
\end{align}
The tangent bundle and the matter bundle comprise the character
\begin{align}
\mathcal{T}[\boldsymbol{\lambda}] &= \sum_{\mathbf{i} \in \text{Vert}_\gamma} \left( N_\mathbf{i} K_\mathbf{i} ^*+ q_{12} N_\mathbf{i} ^* K_\mathbf{i} - P_{12} K_\mathbf{i} K_\mathbf{i} ^* -M_\mathbf{i} ^* K_\mathbf{i} \right) \nonumber \\
&  - \sum_{\mathbf{e} \in \text{Edge}_\gamma} e^{\beta m_\mathbf{e}} (N_{t(\mathbf{e})} K_{s(\mathbf{e})} ^* +q_{12} N_{s(\mathbf{e})} ^*  K_{t(\mathbf{e})} - P_{12} K_{t(\mathbf{e})} K_{s(\mathbf{e})} ^*)  \label{charac},
\end{align}
assoicated to each fixed point $\boldsymbol{\lambda} \in \EuScript{M}_\gamma (\mathbf{n}, \mathbf{k})^{T_H}$. At last the instanton part of the partition function is evaluated by
\begin{align}
\mathcal{Z}^{\text{inst}} ( \mathbf{a}; \textbf{\textit{m}}; \boldsymbol{\varepsilon} ; \mathfrak{q}) = \sum_{\boldsymbol{\lambda}} \prod_{\mathbf{i} \in \text{Vert}_\gamma} \mathfrak{q}_\mathbf{i} ^{\vert \boldsymbol{\lambda}^{(\mathbf{i})} \vert} \; E \left[ \mathcal{T}[\boldsymbol{\lambda}] \right], \label{instpart}
\end{align}
where we have used the $E$-symbol \eqref{epsilon}. Note that the one-loop part and the instanton part can be combined into
\begin{align}\label{1loopinst}
\begin{split}
&\mathcal{Z}^{\text{1-loop}}(\mathbf{a}, \boldsymbol{m}, \boldsymbol{\varepsilon})  \: \mathcal{Z}^{\text{inst}} ( \mathbf{a}; \textbf{\textit{m}}; \boldsymbol{\varepsilon} ; \mathfrak{q}) \\
&=\sum_{\boldsymbol{\lambda}} \prod_{\mathbf{i} \in \text{Vert}_\gamma} \mathfrak{q}_\mathbf{i} ^{\vert \boldsymbol{\lambda}^{(\mathbf{i})} \vert} \; E\left[ \frac{1}{(1-e^{-\beta \varepsilon_1})(1-e^{-\beta \varepsilon_2})} \left( \sum_{\mathbf{i} \in \text{Vert}_\gamma } (M_{\mathbf{i}} -S_{\mathbf{i}})S_{\mathbf{i}} ^* + \sum_{\mathbf{e} \in \text{Edge}_\gamma} e^{\beta m_{\mathbf{e}}} S_{t(\mathbf{e})} S_{s(\mathbf{e})} ^* \right) \right],
\end{split}
\end{align}
with the character $S_{\mathbf{i}} \equiv N_\mathbf{i} - P_{12} K_{\mathbf{i}}$.

The regularized characteristic polynomials of the adjoint scalars form important chiral observables, called the $\EuScript{Y}$-observables, are defined by
\begin{align}
\EuScript{Y}_{\mathbf{i}}(x) \equiv x^{\mathpzc{n}_\mathbf{i}} \exp \sum_{l=1} ^\infty -\frac{1}{lx ^l} \text{Tr}\: \Phi_\mathbf{i} ^l \vert_0,
\end{align}
Their expressions at the fixed point $\boldsymbol{\lambda}$ are written as
\begin{align} \label{yobs}
\EuScript{Y}_\mathbf{i} (x)[\boldsymbol{\lambda}] = \prod_{\alpha=1} ^{\mathpzc{n}_\mathbf{i}} \left( (x-a_{\mathbf{i},\alpha}) \prod_{\square \in \lambda^{(\mathbf{i},\alpha)}} \frac{(x- c_\square -\varepsilon_1)(x-c_\square-\varepsilon_2)}{(x-c_\square)(x-c_\square-\varepsilon)}  \right).
\end{align}
which shows that upon the regularization, the instanton contribution makes the polynomials into rational functions of the auxiliary variable $x$. 
The $\EuScript{Y}$-observable can be simply written as
\begin{align}
\EuScript{Y}_\mathbf{i} (x)[\boldsymbol{\lambda}] = \beta^{-\mathpzc{n}_\mathbf{i}}\: E [-e^{\beta x} S_\mathbf{i} ^*].
\end{align}
Note that the $\EuScript{Y}$-observables are the generating functions for the chiral observables
\begin{align}
\begin{split}
\EuScript{O}_{\mathbf{i},k} [\boldsymbol{\lambda}] &\equiv \text{Tr} \: \Phi_{\mathbf{i}} ^k \vert_0 [\boldsymbol{\lambda}] \\
& =  \sum_{\alpha=1} ^{\mathpzc{n}_\mathbf{i}} \left[ a_{\mathbf{i},\alpha} ^k + \sum_{\Box \in \lambda^{(\mathbf{i},\alpha)}} \left( (c_{\Box} + \varepsilon_1)^k + ( c_{\Box} + \varepsilon_2)^k - c_{\Box} ^k - (c_{\Box} + \varepsilon)^k \right) \right].
\end{split} 
\end{align}
The $qq$-characters for the quiver gauge theories are given as certain Laurent polynomials of the $\EuScript{Y}$-observables.

\section{1-loop part of the partition functions} \label{app:1loop}
The 1-loop parts of the partition functions are regularized by using the Barnes double gamma function,
\begin{align}
\Gamma_2 (x;\ve_1,\ve_2) := \exp \left[ \left.-\frac{d}{ds} \right\vert_{s=0} \frac{1}{\Gamma(s)} \int_0 ^\infty d \b \b^{s-1} \frac{ e^{-\b x} }{(1- e^{-b \ve_1})(1-e^{-b \ve_2}) }\right],
\end{align}
and also the gamma function
\begin{align}
\Gamma_1 (x;\ve_2) := \exp \left[ \left. - \frac{d}{ds} \right\vert_{s=0} \frac{1}{\Gamma(s)} \int_0 ^\infty d\b \b^{s-1} \frac{e^{\b x}}{1-e^{-\b\ve_2}} \right] = \frac{\sqrt{2\pi /\ve_2}}{\ve_2 ^{\frac{x}{\ve_2}} \Gamma\left( \frac{x}{\ve_2}\right)}.
\end{align}
In particular, the 1-loop part of the partition function of the $SU(2)$ gauge theory with four fundamental hypermultiplets is given by
\begin{align}
\mathcal{Z}^{\text{1-loop}} = \prod_\pm \frac{\Gamma_2 (0;\ve_1,\ve_2)\Gamma_2 (\pm 2a ;\ve_1,\ve_2)  }{\prod_{i=1}^4 \Gamma_2 (\pm a - m_i;\ve_1,\ve_2)}.
\end{align}
If the theory is coupled to the surface defect on $z_2$-plane, the gauged linear sigma model living on the surface defect also contributes to the 1-loop part of the partition function. Depending on the complexified FI parameter $z$ of the sigma model, the gauge and the matter contents of the sigma model vary and thus the 1-loop part of the partition function changes accordingly. For example, in the domain $0<\vert \qe \vert< \vert z \vert <1 $,
\begin{align}
\Psi^{\text{1-loop}} _\pm = \prod_{\xi=\pm} \frac{\Gamma_2 (0;\ve_1,\ve_2)\Gamma_2 ( \xi 2a  ;\ve_1,\ve_2)  }{\prod_{i=1}^4 \Gamma_2 ( \xi a  - m_i;\ve_1,\ve_2)} \frac{ \prod_{i=1,2} \Gamma_1 (\pm a - m_i ;\ve_2)}{\Gamma_1 (\pm 2 a;\ve_2)}.
\end{align}
Accordingly, the 1-loop part of the effective twisted superpotneital is computed as
\begin{align}
S_\pm ^{\text{1-loop}} = \lim_{\ve_2 \to 0} \frac{\ve_2}{\ve_1} \log \Psi^{\text{1-loop}} _\pm 
\end{align}
Using the following identities regarding the NS limits of the gamma functions,
\begin{align}
\begin{split}
&\frac{\p}{\p x} \left( \lim_{\ve_2 \to 0} \ve_2 \log \Gamma_2 (x;\ve_1,\ve_2) \right) = - \log \Gamma_1 (x;\ve_1)\\
&\frac{\p}{\p x} \left( \lim_{\ve_2 \to 0} \ve_2 \log \Gamma_1 (x;\ve_2) \right) = - \log x,
\end{split}
\end{align}
we obtain
\begin{subequations}
\begin{align}
&\frac{\p S_+ ^{\text{1-loop}}}{\p \a} = \log \left[  \frac{\Gamma \left( \frac{ 2a}{\ve_1} \right)^2}{\Gamma \left( -\frac{2a}{\ve_1} \right)^2} \prod_{i=1} ^4 \frac{\Gamma \left( - \frac{a +m_i}{\ve_1} \right)}{ \Gamma \left( \frac{a - m_i}{\ve_1} \right)} \frac{(2 a)^2}{\prod_{i=1,2} (a - m_i )} \right] \\
&\frac{\p S_- ^{\text{1-loop}}}{\p \a} = \log \left[   \frac{\Gamma \left( \frac{ 2a}{\ve_1} \right)^2}{\Gamma \left( -\frac{2a}{\ve_1} \right)^2} \prod_{i=1} ^4 \frac{\Gamma \left( - \frac{a +m_i}{\ve_1} \right)}{ \Gamma \left( \frac{a - m_i}{\ve_1} \right)}  \frac{\prod_{i=1,2} (-a - m_i )}{(2a)^2} \right].
\end{align}
\end{subequations}
This is precisely the relation \eqref{eq:1loopdrv} that we used to absorb the 1-loop part into the twisted superpotential $\widetilde{S}$.

In section \ref{sec:blowup}, we have defined the function $L^{\mathbf{n},k} (\mathbf{a},\mathbf{m},\ve_1,\ve_2)$ by a ratio of the 1-loop part of the bulk partition function,
\begin{align}
\begin{split}
L^{\mathbf{n},k} (\mathbf{a},\mathbf{m},\ve_1,\ve_2):= \frac{\mathcal{Z}^{\text{1-loop}} \left(\mathbf{a}+\mathbf{n} \ve_1 , \mathbf{m}+\frac{k}{2} \ve_1 , \ve_1,\ve_2-\ve_1\right) \mathcal{Z}^{\text{1-loop}} \left(\mathbf{a}+\mathbf{n} \ve_2 , \mathbf{m}+\frac{k}{2} \ve_2 , \ve_1-\ve_2,\ve_2\right)}{\mathcal{Z}^{\text{1-loop}} \left(\mathbf{a}  , \mathbf{m} , \ve_1,\ve_2\right)}.
\end{split}
\end{align}
A straightforward computation shows that 
\begin{align}
L^{\mathbf{n},k} (\mathbf{a}, \mathbf{m},\ve_1,\ve_2) = \frac{\prod_{\a,\b=1} ^N s^{n_\a -\frac{k}{2} } (m_\b -a_\a ,\ve_1,\ve_2)}{\prod_{\a ,\b =1} ^N s^{n_\a -n_\b} (a_\b-a_\a,\ve_1,\ve_2) },
\end{align}
where
\begin{align}
s^n (x,\ve_1,\ve_2) = \begin{cases} \prod_{ \substack{i,j \geq 0 \\ i+j\leq n-1}} (x- i \ve_1 -j \ve_2) \quad\quad\quad\quad\quad\quad\quad\quad\,\,\,\, n>0 \\ \prod_{\substack{i,j \geq 0 \\ i+j \leq -n-2} } (x+(i+1)\ve_1+(j+1)\ve_2) \quad\quad\quad n<-1 \\ 1  \quad\quad\quad\quad\quad\quad\quad\quad\quad\quad\quad\quad\quad\quad\quad\quad\quad\quad\,\,\,\, n=0 \,\, \text{or}\,\, 1 \end{cases}
\end{align}
In particular, this is a rational function of equivariant parameters.

\section{Double gamma function and Barnes G-function}
We define the double gamma function $\Gamma_2 (x;\ve_1,\ve_2)$ by
\begin{align}
\Gamma_2 (x;\ve_1,\ve_2) = \exp \left[ - \left. \frac{d}{ds} \right\vert_{s=0} \frac{1}{\Gamma(s)} \int_0 ^\infty d \b \b ^{s-1} \frac{e^{-\b x}}{(1-e^{-\b \ve_1})(1-e^{-\b \ve_2}) } \right].
\end{align}
In the self-dual limit of the $\Omega$-background $\ve_2=-\ve_1$, it is easy to show that
\begin{align}
\Gamma_2 (x;\ve_1, -\ve_1) \, \dot{=} \, \exp\left[  \left. \frac{d}{ds} \right\vert_{s=0} \sum_{n_1, n_2 \geq 0} \frac{1}{(1+\frac{x}{\ve_1} +n_1+n_2)^s}  \right],
\end{align}
where the equality is regarded as a regularization of the right hand side. From this relation we identify
\begin{align} \label{eq:barnesg}
\Gamma_2 (x;\ve_1, -\ve_1) = \frac{(2\pi) ^{\frac{1}{2} + \frac{x}{2\ve_1}}}{G \left(1+ \frac{x}{\ve_1} \right)},
\end{align}
where $G(x)$ is the usual Barnes G-function.

\bibliographystyle{JHEP}

\begin{thebibliography}{10}



\bibitem{agt}
L.~Alday, D.~Gaiotto and Y.~Tachikawa, \emph{{Liouville Correlation Functions
  from Four-dimensional Gauge Theories}},
  \href{http://dx.doi.org/10.1007/s11005-010-0369-5}{\emph{Lettt.Math.Phys.}
  {\bfseries 91} (2010) 167--197},
  [\href{https://arxiv.org/abs/0906.3219}{{\ttfamily 0906.3219}}].
  
   
\bibitem{ising1} E.~Barouch, B.~McCoy, C.~Tracy, T.T.~Wu,  \emph{Spin-spin correlation functions for the two dimensional Ising model: exact theory in the scaling region}, Physical Review {\bf B13} (1976) 316-374\ .
  
  \bibitem{BFH} J-E.~Bourgine, M.~Fukuda, K.~Harada, Y.~Matsuo, and R-D.~Zhu \emph{{(p,q)-webs of DIM representations, 5d N=1 instanton partition functions and qq-characters}},
  \href{http://dx.doi.org/10.1007/JHEP11(2017)034}{\emph{JHEP}
  {\bfseries 11} (2017) 034},
  [\href{https://arxiv.org/abs/1703.10759}{{\ttfamily 1703.10759}}].

\bibitem{BJ} J-E.~Bourgine and S.~Jeong \emph{{New quantum toroidal algebras from 5D $\mathcal{N} = 1$ instantons on orbifolds}},
  \href{http://dx.doi.org/10.1007/JHEP05(2020)127}{\emph{JHEP}
  {\bfseries 05} (2020) 127},
  [\href{https://arxiv.org/abs/1906.01625}{{\ttfamily 1906.01625}}].
  
  \bibitem{CV1992} 
  S.~Cecotti and C.~Vafa,
  \emph{Ising Model and $\mathcal{N}=2$ Supersymmetric Theories},
  Commun.Math.Phys. {\bf 157}, 139--178 (1993)
  doi:10.1007/BF02098023
  [hep-th/9209085]. 
  
   \bibitem{FZ} 
  V.A.~Fateev and A.B.~Zamolodchikov, \emph{Operator Algebra and 
  Correlation Functions in the Two-Dimensional Wess-Zumino} $SU(2) \times SU(2)$ \emph{Chiral Model}, Sov. J. Nucl. Phys. 43 (1986) 657-664
  
  
\bibitem{GIL2012}
O.~Gamayun, N.~Iorgov and O.~Lisovyy, \emph{{Conformal field theory of
  Painlev{\'e} VI}},
  \href{http://dx.doi.org/10.1007/JHEP10(2012)038}{\emph{JHEP} {\bfseries 10}
  (2012) 038}, [\href{https://arxiv.org/abs/1207.0787}{{\ttfamily 1207.0787}}].

 \bibitem{Losev:2003py} 
  A.~S.~Losev, A.~Marshakov and N.~A.~Nekrasov,
  \emph{Small instantons, little strings and free fermions},
  In, M.~Shifman (ed.) et al.: \emph{From fields to strings, Ian Kogan Memorial volume}, vol. 1, pp. 581-621
  [hep-th/0302191].

\bibitem{NRS2011}
N.~Nekrasov, A.~Rosly and S.~Shatashvili, \emph{{Darboux coordinates, Yang-Yang
  functional, and gauge theory}},
  \href{http://dx.doi.org/10.1016/j.nuclphysbps.2011.04.150}{\emph{Nucl.Phys.Proc.Suppl.}
  {\bfseries 216} (2011) 69--93},
  [\href{https://arxiv.org/abs/1103.3919}{{\ttfamily 1103.3919}}].

\bibitem{ref:nekwit}
N.~Nekrasov and E.~Witten, \emph{{The Omega Deformation, Branes, Integrability,
  and Liouville Theory}},
  \href{http://dx.doi.org/10.1007/JHEP09(2010)092}{\emph{JHEP} {\bfseries 09}
  (2010) 92}, [\href{https://arxiv.org/abs/1002.0888}{{\ttfamily 1002.0888}}].

\bibitem{JN2018}
S.~Jeong and N.~Nekrasov, \emph{{Opers, surface defects, and Yang-Yang
  functional}},  to appear in Adv. Theor. Math. Phys., \href{https://arxiv.org/abs/1806.08270}{{\ttfamily
  1806.08270}}.

\bibitem{LLNZ2013}
A.~Litvinov, S.~Lukyanov, N.~Nekrasov and A.~Zamolodchikov, \emph{{Classical
  Conformal Blocks and Painleve VI}},
  \href{http://dx.doi.org/10.1007/JHEP07(2014)144}{\emph{JHEP} {\bfseries 07}
  (2014) 144}, [\href{https://arxiv.org/abs/1309.4700}{{\ttfamily 1309.4700}}].


\bibitem{ny}
H.~Nakajima and K.~Yoshioka, \emph{{Instanton counting on blowup. I.
  4-dimensional pure gauge theory}},
  \href{http://dx.doi.org/10.1007/s00222-005-0444-1}{\emph{Invent.Math.}
  {\bfseries 162} (2005) 313--355},
  [\href{https://arxiv.org/abs/math/0306198}{{\ttfamily math/0306198}}].
  
  
\bibitem{Nekrasov:2002qd} 
N.~Nekrasov,
  \emph{Seiberg-Witten prepotential from instanton counting},
  Adv.\ Theor.\ Math.\ Phys.\  {\bf 7}, no. 5, 831 (2003)
  doi:10.4310/ATMP.2003.v7.n5.a4
  [hep-th/0206161].


  
  \bibitem{Nekrasov:2003rj} 
  N.~Nekrasov and A.~Okounkov,
  \emph{Seiberg-Witten theory and random partitions},
  Prog.\ Math.\  {\bf 244}, 525 (2006)
  doi:10.1007/0-8176-4467-9$\_$15
  [hep-th/0306238].
  
  \bibitem{NekLisbon}
  N.~Nekrasov, \emph{Localizing gauge theories}, 14th International
  Congress on Mathematical Physics 2003, Lisbon, J.C.~Zambrini (Ed.), 
  World Scientific (2006)
  
  \bibitem{NN2004}
$\sim\sim\sim$, \emph{On the BPS/CFT correspondence},  Lecture at the University of Amsterdam string theory group seminar, Feb. 3, 2004 . \\
 $\sim\sim\sim$, \emph{2d CFT-type equations from
4d gauge theory}, Lecture at the IAS conference ``Langlands Program and Physics",
March 8-10, 2004\\
$\sim\sim\sim$, \emph{Supersymmetric gauge theories and quantization of integrable systems}, Lecture at the Strings' 2009 conference,
{\tt http://strings2009.roma2.infn.it/talks/Nekrasov$\_$Strings09.pdf}

\bibitem{Nekrasov_BPS1}
N.~Nekrasov, \emph{{BPS/CFT correspondence: non-perturbative Dyson-Schwinger
  equations and $qq$-characters}},
  \href{http://dx.doi.org/10.1007/JHEP03(2016)181}{\emph{JHEP} {\bfseries 03}
  (2016) 181}, [\href{https://arxiv.org/abs/1512.05388}{{\ttfamily
  1512.05388}}]. \\
$\sim\sim\sim$, \emph{{BPS/CFT correspondence II: Instantons at crossroads, Moduli
  and Compactness Theorem}},
  \href{https://arxiv.org/abs/1608.07272}{{\ttfamily 1608.07272}} \\
$\sim\sim\sim$, \emph{{BPS/CFT Correspondence III: Gauge Origami partition
  function and $qq$-characters}},
  \href{http://dx.doi.org/10.1007/s00220-017-3057-9}{\emph{Commun.Math.Phys.}
  {\bfseries 358} (2018) 863--894},
  [\href{https://arxiv.org/abs/1701.00189}{{\ttfamily 1701.00189}}].

\bibitem{Nekrasov_BPS45}
$\sim\sim\sim$, \emph{{BPS/CFT correspondence IV: sigma models and defects in
  gauge theory}},
  \href{http://dx.doi.org/10.1007/s11005-018-1115-7}{\emph{Lett.Math.Phys.}
  {\bfseries 109} (2019) 579--622},
  [\href{https://arxiv.org/abs/1711.11011}{{\ttfamily 1711.11011}}].\\
$\sim\sim\sim\sim$, \emph{{BPS/CFT correspondence V: BPZ and KZ equations from
  $qq$-characters}},  \href{https://arxiv.org/abs/1711.11582}{{\ttfamily
  1711.11582}}.
  
  \bibitem{NikBlowup}
  N.~Nekrasov, \emph{Blowups  in BPS/CFT correspondence, 
and Painlev{\'e} VI},   \href{https://arxiv.org/abs/2007.03646}{{\ttfamily
  2007.03646}}.

\bibitem{Nekrasov:2009rc}
N.~A.~Nekrasov and S.~L.~Shatashvili,
\emph{Quantization of Integrable Systems and Four Dimensional Gauge Theories},
doi:10.1142/9789814304634$\_$0015
[arXiv:0908.4052 [hep-th]].


\bibitem{Jeong2017}
S.~Jeong and X.~Zhang, \emph{{BPZ equations for higher degenerate fields and
  non-perturbative Dyson-Schwinger equations}},
  \href{https://arxiv.org/abs/1710.06970}{{\ttfamily 1710.06970}}.

\bibitem{GIL2013}
O.~Gamayun, N.~Iorgov and O.~Lisovyy, \emph{{How instanton combinatorics solves
  Painlev\'{e} VI, V and III's}},
  \href{http://dx.doi.org/10.1088/1751-8113/46/33/335203}{\emph{J. Phys. A:
  Math. Theor.} {\bfseries 46} (2013) 335203},
  [\href{https://arxiv.org/abs/1302.1832}{{\ttfamily 1302.1832}}].


\bibitem{gai1}
D.~Gaiotto, \emph{{N=2 dualities}},
  \href{http://dx.doi.org/10.1007/JHEP08(2012)034}{\emph{JHEP} {\bfseries 08}
  (2012) 034}, [\href{https://arxiv.org/abs/0904.2715}{{\ttfamily 0904.2715}}].

\bibitem{Jeong2017}
S.~Jeong, \emph{{Splitting of surface defect partition functions and integrable systems}},
  \href{http://dx.doi.org/10.1016/j.nuclphysb.2018.12.007}{\emph{Nucl.Phys.B} {\bfseries 938}
  (2019) 775--806}, [\href{https://arxiv.org/abs/1709.04926}{{\ttfamily
  1709.04926}}].

\bibitem{JZ2019}
S.~Jeong and X.~Zhang, \emph{{A note on chiral trace relations from
  $qq$-characters}},
  \href{http://dx.doi.org/10.1007/JHEP04(2020)026}{\emph{JHEP} {\bfseries 04}
  (2020) 026}, [\href{https://arxiv.org/abs/1910.10864}{{\ttfamily
  1910.10864}}].

\bibitem{GFPP2016}
J.~Gomis, B.~Floch, Y.~Pan, and W.~Peelaers, \emph{{Intersecting Surface Defects and Two-Dimensional CFT}},
  \href{http://dx.doi.org/10.1103/PhysRevD.96.045003}{\emph{Phys.Rev.D} {\bfseries 96}
  (2017) 045003}, [\href{https://arxiv.org/abs/1610.03501}{{\ttfamily
  1610.03501}}].

\bibitem{PP2016}
Y.~Pan, and W.~Peelaers, \emph{{Intersecting Surface Defects and Instanton Partition Functions}},
  \href{http://dx.doi.org/10.1007/JHEP07(2017)073}{\emph{JHEP} {\bfseries 07}
  (2017) 073}, [\href{https://arxiv.org/abs/1612.04839}{{\ttfamily
 1612.04839}}].

\bibitem{ILT} N. ~Iorgov, O. ~Lisovyy, and J. ~Teschner, \emph{{Isomonodromic tau-functions from Liouville conformal blocks}},
  \href{http://dx.doi.org/10.1007/s00220-014-2245-0}{\emph{Comm.Math.Phys.}
  {\bfseries 336} (2015) 671--294},
  [\href{https://arxiv.org/abs/1401.6104}{{\ttfamily 1401.6104}}].

\bibitem{Gav} P. ~Gavrylenko, \emph{{Isomonodromic $\tau$-functions and ${W}_N$ conformal blocks}},
  \href{http://dx.doi.org/10.1007/JHEP09(2015)167}{\emph{JHEP}
  {\bfseries 09} (2015) 167},
  [\href{https://arxiv.org/abs/1505.00259}{{\ttfamily 1505.00259}}].

\bibitem{Tes} J. ~Teschner, \emph{{Classical conformal blocks and isomonodromic deformations}},
  [\href{https://arxiv.org/abs/1707.07968}{{\ttfamily 1707.07968}}].

\bibitem{LN2017}
M.~Lencs{\'e}s and F.~Novaes, \emph{{Classical Conformal Blocks and Accessory
  Parameters from Isomonodromic Deformations}},
  \href{http://dx.doi.org/10.1007/JHEP04(2018)096}{\emph{JHEP} {\bfseries 04}
  (2018) 096}, [\href{https://arxiv.org/abs/1709.03476}{{\ttfamily
  1709.03476}}].

\bibitem{GavIL} P. ~Gavrylenko, N. ~Iorgov, and O. ~Lisovyy \emph{{Higher rank isomonodromic deformations and W-algebras}},
  \href{http://dx.doi.org/10.1007/s11005-019-01207-6}{\emph{Lett.Math.Phys.}
  {\bfseries 110} (2020) 327-364},
  [\href{https://arxiv.org/abs/1801.09608}{{\ttfamily 1801.09608}}].

\bibitem{CPT} I. ~Coman, E. ~Pomoni, and J. ~Teschner \emph{{From quantum curves to topological string partition functions}},
  [\href{https://arxiv.org/abs/1811.01978}{{\ttfamily 1811.01978}}].

\bibitem{BMGT} G. Bonelli, F. ~Del Monte, P. ~Gavrylenko, and A. ~Tanzini \emph{{N=2* gauge theory, free fermions on the torus and Painlev\'{e} VI}},
  \href{http://dx.doi.org/10.1007/s00220-020-03743-y}{\emph{Comm.Math.Phys.}
  {\bfseries 377} (2020) 1381-1419},   [\href{https://arxiv.org/abs/1901.10497}{{\ttfamily 1901.10497}}].


  
  
\bibitem{NT} N.~Nekrasov, O.~Tsymbalyuk, \emph{Surface defects in gauge theory and Knizhnik-Zamolodchikov equation}, to appear

\bibitem{JLN} S.~Jeong, N.~Lee, and N.~Nekrasov, to appear


\bibitem{Okamoto1986}
K.~Okamoto, \emph{{Studies on the Painlev\'{e} equations I.-Sixth Painlev\'{e}
  equation $\text{P}_{\text{VI}}$}},
  \href{http://dx.doi.org/10.1007/BF01762370}{\emph{Ann. Math. Pura Appl.}
  {\bfseries 146} (1986) 337--381}.
  
  
  \bibitem{Reshetikhin}
  N.~Reshetikhin, \emph{The Knizhnik-Zamolodchikov System as a Deformation of the Isomonodromy Problem}, Lett.~Math.~Phys. {\bf 26} (1992) 167-177 .
  
  \bibitem{Ribault:2005wp}
S.~Ribault and J.~Teschner,
\emph{$H_{+}^{3}$-WZNW correlators from Liouville theory},
JHEP \textbf{06}, 014 (2005)
doi:10.1088/1126-6708/2005/06/014
[arXiv:hep-th/0502048 [hep-th]].
  
  
  \bibitem{SJM} M.~Sato, T.~Miwa, M.~Jimbo, \emph{Holonomic quantum fields I-V}, Publ. RIMS Kyoto Univ. {\bf 14} (1978), 223-267; {\bf 15} (1979), 201-278, 577-629, 871-972; {\bf 16} (1980), 531-584. 
  
  

\bibitem{CL2007}
S.~Cantat and F.~Loray, \emph{{Holomorphic dynamics, Painlev\'{e} VI equation
  and Character Varieties}},  \href{https://arxiv.org/abs/0711.1579}{{\ttfamily
  0711.1579}}.

\bibitem{IIS2003}
M.~Inaba, K.~Iwasaki and M.~Saito, \emph{{B\"{a}cklund Transformations of the
  Sixth Painlev\'{e} Equation in Terms of Riemann-Hilbert Correspondence}},
  \href{http://dx.doi.org/10.1155/S1073792804131310}{\emph{Internat. Math. Res.
  Notices} {\bfseries 1} (2004) 1--30},
  [\href{https://arxiv.org/abs/math/0309341}{{\ttfamily math/0309341}}].

\bibitem{DM2003}
B.~Dubrovin and M.~Mazzocco, \emph{{Canonical structure and symmetries of the
  Schlesinger equations}},
  \href{http://dx.doi.org/10.1007/s00220-006-0165-3}{\emph{Comm. Math. Phys.}
  {\bfseries 271} (2007) 289--373},
  [\href{https://arxiv.org/abs/math/0311261}{{\ttfamily math/0311261}}].

\bibitem{LT2008}
O.~Lisovyy and Y.~Tykhyy, \emph{{Algebraic solutions of the sixth Painlev\'{e}
  equation}},  \href{https://arxiv.org/abs/0809.4873}{{\ttfamily 0809.4873}}.


 
  

\bibitem{KKV} S.~Kats, A.~Klemm, and C.~Vafa, \emph{{Geometric Engineering of Quantum Field Theories}},
  \href{http://dx.doi.org/10.1016/S0550-3213(97)00282-4}{\emph{Nucl.Phys.B}
  {\bfseries 497} (1997) 173--195},
  [\href{https://arxiv.org/abs/hep-th/9609239}{{\ttfamily hep-th/9609239}}].
  
  
\bibitem{Klemm:1996bj} 
  A.~Klemm, W.~Lerche, P.~Mayr, C.~Vafa and N.~P.~Warner,
  \emph{Self-dual strings and  ${\mathcal{N}}=2$ supersymmetric field theory},
  Nucl.\ Phys.\ B {\bf 477}, 746 (1996)
  doi:10.1016/0550-3213(96)00353-7
  [hep-th/9604034].  
  
   
 
  
  \bibitem{Vafa:1994tf} 
  C.~Vafa and E.~Witten,
  \emph{A Strong coupling test of S-duality},
  Nucl.\ Phys.\ B {\bf 431}, 3 (1994)
  doi:10.1016/0550-3213(94)90097-3
  [hep-th/9408074]. 
 
 \bibitem{Witten5} E.~Witten, \emph{Solutions of four-dimensional field theories via $M$-theory},
  Nucl.\ Phys.\ B {\bf 500}, 3 (1997)
  doi:10.1016/S0550-3213(97)00416-1
  [hep-th/9703166].
  
\bibitem{DGH} T.~Dimofte, S.~Gukov, and L.~Hollands, \emph{{Vortex Counting and Lagrangian 3-manifolds}},
  \href{http://dx.doi.org/10.1007/s11005-011-0531-8}{\emph{Lettt.Math.Phys.}
  {\bfseries 98} (2011) 225--287},
  [\href{https://arxiv.org/abs/1006.0977}{{\ttfamily 1006.0977}}].



  \bibitem{Wyllard:2009hg}
N.~Wyllard,
\emph{$A_{N-1}$ conformal Toda field theory correlation functions from conformal N = 2 SU(N) quiver gauge theories},
JHEP \textbf{11} (2009), 002
doi:10.1088/1126-6708/2009/11/002
[arXiv:0907.2189 [hep-th]].

\bibitem{BPZ}
A. A. Belavin, A. M. Polyakov, A. B. Zamolodchikov, \emph{Infinite conformal symmetry in two-dimensional quantum field theory}, \emph{Nucl. Phys. B} {\bf 247} (1984) 333-380.




\bibitem{ZW}
A.~B.~Zamolodchikov, \emph{Infinite extra symmetries in two-dimensional conformal quantum field theory}, Theor. ~Math.~Physics (in Russian), {\bf 65} (3) (1985) 347?359, ISSN 0564-6162, MR 0829902
  
\bibitem{Zamolodchikov:1995aa} 
  A.~B.~Zamolodchikov and A.~B.~Zamolodchikov,
  \emph{Structure constants and conformal bootstrap in Liouville field theory},
  Nucl.\ Phys.\ B {\bf 477}, 577 (1996)
  doi:10.1016/0550-3213(96)00351-3
  [hep-th/9506136].  
\end{thebibliography}
\providecommand{\href}[2]{#2}\begingroup\raggedright\endgroup

\end{document}